\documentclass[%
 reprint,
superscriptaddress,
 amsmath,amssymb,
 aps,
]{revtex4-2}

\usepackage{graphicx}
\usepackage{dcolumn}
\usepackage{bm}
\usepackage{lipsum} 
\usepackage{physics,amsmath}
\usepackage[dvipsnames]{xcolor}
\usepackage{hyperref}

\def\Tfor#1{\underline{\underline{\underline{\underline{#1}}}}}
\def\Tthr#1{\underline{\underline{\underline{#1}}}}
\def\mat#1{\underline{\underline{#1}}}
\def\vec#1{{\underline{#1}}}

\def\mat#1{\underline{\underline{#1}}}
\def\vec#1{{\underline{#1}}}

\begin{document}

\preprint{APS/123-QED}

\title{Fluctuation-Response Theory of Non-Equilibrium Complex Fluids}

\author{Ryota Takaki}
\email{ryota@pks.mpg.de}
\affiliation{%
 Max Planck Institute for the Physics of Complex Systems 
}%

\author{Frank Jülicher}%
\email{julicher@pks.mpg.de}
\affiliation{%
 Max Planck Institute for the Physics of Complex Systems 
}%
\affiliation{Center for Systems Biology Dresden, Dresden, Germany}
\affiliation{Cluster of Excellence Physics of Life, TU Dresden, Dresden, Germany}




\date{\today}

\begin{abstract}
A fundamental challenge in soft matter physics is to describe materials, such as the living cytoplasm and tissues, that are simultaneously active, chemically driven, and exhibit long-lasting memory of mechanical stresses.  Here, we construct a generalized hydrodynamic framework at finite wavevectors and frequencies that is applicable to non-equilibrium fluids with memory. 
By leveraging stationary correlation identities, we derive a generalized linear response theory for non-equilibrium steady states. This framework serves as a formal extension of Onsager's regression hypothesis beyond thermal equilibrium. Our approach provides a direct pathway to derive transport coefficients from steady-state fluctuations without the traditional Mori–Zwanzig projection-operator formalism, generalizing the Green–Kubo relations to non-equilibrium systems.
As a corollary, we derive two model-free variants of the non-equilibrium fluctuation-response relation for non-Markovian dynamics.
These generalized relations explicitly capture the non-equilibrium circulating currents-an out-of-equilibrium signature that is invisible to conventional scalar formulations or frameworks that treat degrees of freedom independently.
Applying our theory to chemically driven active fluids reveals the emergence of active viscoelastic memory, wherein chemical reaction cycles dynamically renormalize the macroscopic viscous response. Strikingly, this active memory can induce negative storage modulus at finite frequencies, a behavior absent in ordinary viscoelastic fluids. Our first-principles framework rigorously extends linear rheology to non-equilibrium systems and provides a foundation for understanding non-Markovian dynamics across a broad range of biological and synthetic active matter.

\end{abstract}

\maketitle

\section{Introduction}

Biological soft matter exhibits complex behaviors arising from both its intrinsic material properties and the self-generated forces produced as a result of metabolic activity~\cite{marchetti2013hydrodynamics,bowick2022symmetry}. These self-generated forces, or active stresses, are, for example, generated by molecular motors that consume adenosine triphosphate (ATP)~\cite{julicher1997modeling,iino2020introduction,kolomeisky2007molecular,mugnai2020theoretical}. The interplay between mechanical and chemical processes, i.e., chemo–mechanical coupling, is a defining feature of living systems and underlies force generation, remodeling, and pattern formation across scales~\cite{lipowsky2008chemomechanical,lipowsky2000universal,toyabe2011thermodynamic,takaki2019kinesin,takaki2022information,recho2019theory,naganathan2017mechanochemical,mizuno2007nonequilibrium}.

The dynamical behavior of soft matter can be characterized through response and correlation functions. In systems at thermal equilibrium, the fluctuation-dissipation theorem relates linear response functions to spontaneous correlation functions ~\cite{kubo1966fluctuation,kubo2012statistical}.
Early generalizations of fluctuation-response relation beyond thermal equilibrium were developed by Agarwal~\cite{agarwal1972fluctuation}. Since then, a growing body of work has developed response theories and fluctuation-response relations for a broad range of nonequilibrium dynamics, including non-Markovian stochastic processes~\cite{hanggi1977time}, Langevin dynamics~\cite{harada2005equality,harada2006energy,
chun2026fluctuationresponsetheorynonequilibriumlangevin},
and general Markov processes~\cite{seifert2010fluctuation,baiesi2009fluctuations,
baiesi2009nonequilibrium,zheng2025unified}. For Markovian steady states, a generalized fluctuation-dissipation theorem has also been derived from the Hatano-Sasa identity~\cite{hatano2001steady,prost2009generalized}. Related response identities and fluctuation-response bounds have
been developed for Markovian networks~\cite{aslyamov2025nonequilibrium,bao2024nonequilibrium,
kwon2025fluctuation}. Further developments address nonequilibrium
field theories~\cite{Nardini2017,johnsrud2025fluctuation},
hydrodynamically interacting many-body systems
\cite{golestanian2025hydrodynamically},
and hydrodynamic transport and active matter
\cite{han2021fluctuating,chun2021nonequilibrium,
kirkpatrick2025fluctuation}. General inequalities constraining nonequilibrium response have been derived for broad classes of perturbations
\cite{dechant2020fluctuation,dechant2026finite}, and complementary
response and generalized Green-Kubo formulations based on
integration through transients have been developed for driven
colloidal and active systems~\cite{fuchs2002theory,fuchs2009mode,dal2019linear}.

Starting from Onsager theory~\cite{onsager1931reciprocal,onsager1931reciprocal2}, nonequilibrium hydrodynamics descriptions of active matter can be developed~\cite{prost2015active,julicher2018hydrodynamic,mayer2010anisotropies,alert2020physical}.

The theory captures general features of fluid-like biological structures such as the actomyosin cortex and epithelial layers~\cite{prost2015active,julicher2018hydrodynamic,mayer2010anisotropies,alert2020physical}. 
Many biological materials, from dense tissues to protein condensates and chromosomes, operate in a physical regime where slow structural relaxation and solid-like responses dominate over biologically relevant timescales~\cite{kirkpatrick2015colloquium,maier2018stress,bi2016motility,angelini2011glass,jawerth2020protein,takaki2025active,takaki2023theory,takaki2024sequence}. In such systems, stress and fluxes depend on the history of deformation and chemical activity, requiring a theory that naturally includes memory effects.

The observation of complex dynamics in biological material has led to growing interest in active glasses and dense active matter, where high component density and activity interact to produce rich emergent phenomena~\cite{janssen2019active,sadhukhan2024perspective,berthier2019glassy,henkes2020dense,paoluzzi2024flocking,flenner2016nonequilibrium,mandal2020extreme}. While extensions of microscopic glass theories—such as mode-coupling and random first-order transition approaches—have illuminated aspects of these systems~\cite{liluashvili2017mode,nandi2018random,mandal2020extreme,paul2023dynamical}, a general continuum framework that (i) is applicable to fluids in far-from-equilibrium, (ii) is valid at finite wavelength and frequency (spatio-temporal memory), and (iii) naturally encodes chemo–mechanical feedback, has remained elusive. Here we develop a generalized hydrodynamic framework for active viscoelastic fluids that addresses these needs. 

This paper is organized as follows. In Section~\ref{sec:correlation}, we derive a general equation for correlation functions in a steady state and average responses, given in Eqs.~(\ref{eq:dpsi/dt}) and~(\ref{eq:dA/dt}), respectively, which provide the foundation for this work. We further present general fluctuation-response relations for nonequilibrium steady states with non-Markovian dynamics, Eqs.~\eqref{eq:FDT_chi_physical_main} and \eqref{eq:Harada-sasa_like_physical}.
We then proceed to the applications of Eq.~\eqref{eq:dA/dt} to construct constitutive equations for progressively more complex systems: in Section \ref{sec:stress_CE}, to complex fluids in non-equilibrium steady states, see Eqs.~(\ref{eq:CEQ_fluid})-(\ref{eq:eta}); in Section~\ref{sec:chemical_reaction} to chemical reaction networks, see Eq.~(\ref{eq:CE_chem}); and in Section~\ref{sec: fluids_chemical} to non-equilibrium fluids driven by internal chemical activity, see Eqs.~(\ref{eq:CE_general_shear})-(\ref{eq:coefficients_upsilon}), where we highlight a chemo–mechanical feedback that we term \textit{Active Viscoelastic Memory}. We conclude and summarize our results in the discussion.


\section{Fluctuation-response relations for steady states \label{sec:correlation}}
\subsection{Properties of correlation functions}
In this section, we derive our central fluctuation–response relation [Eq.~\eqref{eq:dA/dt}] applicable to steady states, including non-equilibrium, which serves as the foundation for the subsequent sections for the constitutive equations of non-equilibrium fluids.

We consider steady state correlation functions in ergodic systems.
Under the assumption of ergodicity, steady state correlation functions can be defined equivalently as averages over a single long-time trajectory or over a phase-space ensemble, see Appendix~\ref{App:correlation_def}.  
This framework is applicable to systems that relax within a finite time scale but allows for slow relaxation processes.

Consider a set of dynamical variables which in general take complex values, $ \vec{A} = (A_1, A_2, \ldots)^\top$\,, where $^\top$ denotes the transposed. The underline denotes the rank of tensors, see Appendix~\ref{App:tcensor_underline}.  We define fluctuations around the steady state by
\begin{equation}
\delta\vec{A}(t)\equiv \vec{A}(t)-\langle \vec{A}\rangle\quad,
\end{equation}
where $\expval{\cdot}$ is the ensemble or an average over a single long-time trajectory. 
Since our theory concerns fluctuations, all steady-state correlation
functions are understood as connected correlations.  Equivalently, we may
choose mean-subtracted variables so that $\langle \vec{A}\rangle=0$, without loss of
generality, and henceforth use $\vec{A}(t)$ to denote the fluctuation $\delta\vec{A}(t)$\,.

We denote the steady state dynamic correlation matrix of the variables using dyadic product (see Appendix \ref{App:dyadic}),
\begin{equation} 
\label{eq:psi(t)}
\mat{\psi}(t) \equiv \expval{\vec{A}(t)  \vec{A}^\dagger}\quad, 
\end{equation}
where we use the notation $ \vec{A} \equiv \vec{A}(t=0)$ such that the absence of the time argument implies time $t=0$, and the dagger, $\dagger$, denotes hermitian conjugate, i.e. transposed complex conjugate. 
The time translational invariance of the steady state correlation function implies (see Appendix~\ref{sec:derivative_correlation}),
\begin{equation} 
\label{eq:d2/dt2psi}
\frac{d^2}{dt^2}\expval{\vec{A}(t) \vec{A}^\dagger} =- \expval{\vec{\dot{A}}(t)  \vec{\dot{A}}^\dagger}_{} \quad,
\end{equation}
where the dots denote time derivatives.
We define dynamic correlation functions
\begin{equation} 
\label{eq:phi_omega}
\mat{\phi}(t) \equiv \expval{\vec{{\dot{A}}}(t) \vec{\dot{A}}^\dagger} \quad.
\end{equation}
We also introduce the static correlation functions
\begin{equation} 
\label{}
\mat{g} \equiv \expval{\vec{A} \, \vec{A}^\dagger}  \quad \text{and}\quad \mat{\omega} \equiv   \expval{\vec{\dot{A}} \, \vec{A}^\dagger} \quad.
\end{equation}  
Here, the static correlation $\mat{g}$ is Hermitian, while the reactive frequency matrix $\mat{\omega}$\ is anti-Hermitian. When the variables in $\vec{A}$ are real valued, $\mat{\omega}$ becomes an antisymmetric matrix. Consequently, the trace of $\mat{\omega}$ vanishes. Therefore, for a single real variable, it follows that $\mat{\omega} = 0$\,.  Note that 
for Langevin equations with white noise, $\mat{\psi}(t)$ has a cusp at $t=0$, so that the time derivative is not well-defined.
We define 
\begin{equation} 
\label{}
\mat{\omega} \equiv \frac{1}{2} \qty(\frac{d}{dt}\, \mat{\psi}(t)\bigg|_{t=0^+} +\frac{d}{dt}\,\mat{\psi}(t)\bigg|_{t=0^-} )\quad ,
\end{equation}
which ensures that the reactive frequency matrix is anti-hermitian. 

If all components of $\vec{A}$ have the same time-reversal signature, then time-reversal symmetry implies $\mat{\omega}=\mat{0}$ (see Appendix~\ref{sec:symmetry}). Therefore, in this case a non-vanishing $\mat{\omega}$ is a clear signature of a non-equilibrium steady state with broken time-reversal symmetry. By contrast, if $\vec{A}$ contains variables with opposite time-reversal signatures, a non-vanishing $\mat{\omega}$ can arise even at equilibrium.

We would like to obtain a dynamical equation for the time dependent correlation function $\mat{\psi}(t)$. With this aim, we use the Laplace transform of Eq.~(\ref{eq:d2/dt2psi}), which reads 
\begin{equation} 
\label{eq: SecondD_in_Laplace}
s^2 \mat{\widehat{\psi}}(s)-s \mat{g}-\mat{\omega}= -\mat{\widehat{\phi}}(s) \quad,
\end{equation}
where the hat denotes the Laplace transform and $s$ is the Laplace variable, see Appendix~\ref{sec:convention}. 
The following identity holds for general correlation functions in a steady state: 
\begin{equation} 
\begin{split}
\label{eq:identity}
 &s\mat{\widehat{\psi}}(s)-\mat{g} = \\
 &-\left[
    \mat{I}
    -
    \frac{1}{s}
    \qty( \mat{\widehat{\phi}}(s)-\mat{\omega})
    \cdot
    \mat{g}^{-1}
\right]^{-1}
\cdot
\qty( \mat{\widehat{\phi}}(s)-\mat{\omega})
\cdot
\mat{g}^{-1}\cdot \mat{\widehat{\psi}}(s) \quad .
\end{split}
\end{equation}
This can be readily checked, applying Eq.~\eqref{eq: SecondD_in_Laplace} to the right-hand side of Eq.~\eqref{eq:identity}, which becomes equal to the left-hand side of the equation. 
Returning to the time domain, we obtain an exact non-linear relation for the dynamical evolution of the correlation function,
\begin{equation} 
\begin{split}
\label{eq:dpsi/dt}
\frac{d}{dt}\,\mat{\psi}(t) = &- \int_0^t dt' \mat{K}(t-t')\cdot \mat{\psi}(t') \quad.
\end{split}
\end{equation}
The matrix $\mat{K}(t-t')$ represents causal memory kernels which vanish for $t< t'$\,.  The Laplace-space representation of $\mat{K}$ reads
\begin{equation} 
\label{eq:K_0}
\mat{\widehat{K}}(s) \equiv  \left[
    \mat{I}
    -
    \frac{1}{s}
    \qty( \mat{\widehat{\phi}}(s)-\mat{\omega})
    \cdot
    \mat{g}^{-1}
\right]^{-1}
\cdot
\qty( \mat{\widehat{\phi}}(s)-\mat{\omega})
\cdot
\mat{g}^{-1}\quad.
\end{equation}
This formulation generalizes the approach of Ref.~\cite{berne1966calculation}, derived for equilibrium systems, to multiple variables and to non-equilibrium steady states. 
From Eq.~\eqref{eq: SecondD_in_Laplace}, it follows that 
\begin{equation} 
\label{eq:omega=phi0}
 \mat{\widehat{\phi}}(s=0) =\mat{\omega} \quad ,
\end{equation}
provided that the correlation $\mat{\psi}(t)$ remains finite at long time. 
Therefore we can write the memory kernel of Eq.~\eqref{eq:K_0} as 
\begin{equation} 
\label{eq:K}
\mat{\widehat{K}}(s) =  \left[
    \mat{I}
    -
    \frac{1}{s}\qty(
    \Delta\mat{\widehat{\phi}}(s)\cdot\mat{g}^{-1})
\right]^{-1}
\cdot \qty(
\Delta\mat{\widehat{\phi}}(s)
\cdot \mat{g}^{-1}) \quad,
\end{equation}
defining the transient part of $\widehat{\phi}(s)$,
\begin{equation} 
\begin{split}
\label{eq:delphi}
\Delta \mat{\widehat{\phi}}(s) \equiv \mat{\widehat{\phi}}(s) -\mat{\widehat{\phi}}(0)\quad ,
\end{split}
\end{equation}
where $\mat{\widehat{\phi}}(0)$ denotes evaluation at $s=0$. 
Throughout, we write the Laplace-space argument $s=0$ explicitly, while the real-time argument $t=0$ is suppressed.
When all components of $\vec{A}$ have the same time-reversal signature, a non-vanishing value of $\mat{\widehat{\phi}}(0)$ is an indicator of a non-equilibrium steady state. 
Note that $\mat{\psi}(t)$ is Hermitian and $\mat{\omega} = 0$ when time-reversal symmetry holds and time-reversal signature of the variables is the same. Non-zero $\mat{\omega} =\mat{\widehat{\phi}}(0)$ induces a non-Hermitian component in $\mat{\psi}(t)$ once time-reversal symmetry is broken for variables with identical time-reversal signature.

From Eq.~\eqref{eq:dpsi/dt}, we can derive the relaxation of the mean
following an infinitesimal perturbation of the steady-state distribution
whose relative change is spanned by the chosen dynamical variables; see
Appendix~\ref{App:correlation_deriv}. To linear order, the perturbed mean
obeys
\begin{equation} 
\label{eq:dA/dt}
\frac{d}{dt}  \expval{\vec{A}(t)} = - \int_0^t dt' \mat{K}(t-t')\cdot \expval{\vec{A}(t')}\quad.
\end{equation}
Here the mean $\expval{\vec{A}(t)}$ and $\expval{\vec{A}(t')}$ are taken over perturbed ensemble whereas the memory kernel $\mat{K}(t)$ is evaluated in the reference (unperturbed) steady state.  
The time derivative $d\expval{\vec{A}(t)}/dt$ describes the relaxation of the perturbed mean $\langle \vec{A}(t)\rangle$ and the memory kernel $\mat{K}(t)$ governs how this perturbation relaxes over time.

In Appendix~\ref{App:Harmonic}-\ref{App:UDHO}, we illustrate the formalism developed in this section with an overdamped two-dimensional rotating harmonic oscillator and underdamped harmonic oscillator.

\subsection{Generalized fluctuation-response relations}
Using Eq.~\eqref{eq:dA/dt}, we derive two variant forms of the fluctuation-response relation in Appendix~\ref{sec:FDT}, both of which are applicable to non-equilibrium steady states with non-Markovian dynamics. The first addresses an additive kinematic perturbation, which results in the analogous form to the Harada-Sasa relation~\cite{harada2005equality}, while the second addresses a state perturbation, corresponding to the conventional fluctuation-dissipation theorem~\cite{kubo1966fluctuation}. Here we discuss the conventional fluctuation-dissipation form and leave the kinematic form in the Appendix~\ref{sec:FDT}, Eq.~\eqref{eq:Harada-sasa_like_physical}. Notably, these generalized relations are completely model-free and do not depend on the specific underlying equations of motion.

We consider an infinitesimal perturbation of the steady state by a generalized force $\vec{\lambda}$ that shift the steady states average from $\expval{\vec A}=0$ to $\mat{g}\cdot\vec{\lambda}$. 
This response is governed by the same memory kernel of Eq.~\eqref{eq:dA/dt} and reads
\begin{equation}
\label{eq:dA_h_main}
\frac{d}{dt}\expval{\vec{A}(t)}
=
-\int_0^t dt'\,
\mat{K}(t-t')
\cdot
\qty[
\expval{\vec{A}(t')}
-
\mat{g}\cdot\vec{\lambda}(t')
] \;.
\end{equation}
Here, $\expval{\vec{A}(t')}$ denotes the change of the mean relative to
the unperturbed steady state.
The corresponding causal response matrix $\mat{\chi}(t)$ is defined by
\begin{equation}
\label{eq:response}
\expval{\vec{A}(t)}
=
\int_0^t dt'\,
\mat{\chi}(t-t')
\cdot
\vec{\lambda}(t') \quad,
\end{equation}
where $\mat{\chi}(t)=\mat{0}$ for $t<0$. The generalized force $\vec{\lambda}$ need not coincide with an energy-conjugate force. The connection between the generalized force and energy-conjugate force is derived in the Appendix~\ref{sec:response-dissipation}.

The response is related to the
memory kernel by
\begin{equation}
    \mat{\widehat{\chi}}(s)
    =
    \left[
        s \mat I+\mat{\widehat K}(s)
    \right]^{-1}
    \cdot
    \mat{\widehat K}(s)\cdot \mat{g}
    \quad,
\end{equation}
see Eqs.~\eqref{eq:dA_h}-\eqref{eq:response_state}.
The symmetric and antisymmetric parts of the response matrix are defined as
\begin{align}
\label{eq:cai_S_A_main}
\mat{\chi}^{\rm S}(t)  \equiv \frac{1}{2} \qty(\mat{\chi}(t)  +   \mat{\chi}(t)^\top   ) \quad;  \\[2mm] 
\quad \mat{\chi}^{\rm A}(t)  \equiv \frac{1}{2} \qty(\mat{\chi}(t) -   \mat{\chi}(t)^\top ) \quad.
\end{align}


The fluctuation-response relation generalizing the fluctuation-dissipation theorem to non-equilibrium, is given in Fourier space by
\begin{equation}
    \label{eq:FDT_chi_physical_main}
    \mat{\widetilde{\psi}}(\omega)  =- \frac{2}{\omega}\mathrm{Im}\qty[ \mat{\widetilde{\chi}}^{\mkern1mu  \rm S}(\omega)]+ \frac{2i}{\omega}\mathrm{Re}\qty[ \mat{\widetilde{\chi}}^{\mkern1mu  \rm A}(\omega)]  \quad,
\end{equation}
see the detailed derivation in Appendix~\ref{sec:state_pert}.
Here $\mat{\widetilde{\psi}}(\omega)$ is the Fourier transform of the dynamical correlation matrix $\mat{\psi}(t)$.
Since stationarity implies $\mat{\psi}(-t)=\mat{\psi}(t)^{\dagger}$, the Fourier spectrum satisfies $\mat{\widetilde{\psi}}(\omega)=\mat{\widetilde{\psi}}(\omega)^{\dagger}$ for each frequency $\omega$.
For given correlations $\mat{\widetilde{\psi}}(\omega)$, Eq. (\ref{eq:FDT_chi_physical_main}) determines both the imaginary part of $\mat{\widetilde\chi}^{\mkern1mu  \rm S}(\omega)$ and the real part
of $\mat{\widetilde\chi}^{ \mkern1mu \rm A}(\omega)$, the missing real part of $\mat{\widetilde\chi}^{\mkern1mu  \rm S}(\omega)$ and  the imaginary part of $\mat{\widetilde\chi}^{\mkern1mu  \rm A}(\omega)$ are then determined by causality via Kramers-Kronig relations~\cite{toll1956causality,de1926theory}.

In equilibrium, for variables with identical time-reversal signatures without space-dependence, time-reversal symmetry implies the reciprocity condition $\mat{\widetilde{\psi}}(\omega)=\mat{\widetilde{\psi}}(\omega)^\top$. Consequently, the antisymmetric components vanish, yielding $\mat{\widetilde{\chi}}^{\mkern1mu \rm A}(\omega) = \mat{0}$, and Eq.~\eqref{eq:FDT_chi_physical_main} reduces to the equilibrium fluctuation-dissipation theorem, see Appendix~\ref{sec:state_pert}.
In equilibrium for variables with unequal time reversal signatures without space-dependence, the cross-correlation spectrum become antisymmetric and purely imaginary contribution in Eq.~\eqref{eq:FDT_chi_physical_main}. This leads to an antisymmetric linear response in this case of non-dissipative coupling. When the chosen variables are spatial fields, the generalized fluctuation-response relations in Eqs.~\eqref{eq:FDT_chi_physical_main} and~\eqref{eq:Harada-sasa_like_physical} remain valid, but their equilibrium interpretation must also take into account the spatial-inversion signatures of the variables in addition to their time-reversal signatures.


Crucially, the multivariable structure of these relations explicitly represents non-equilibrium circulating currents, which remain  invisible to scalar formulations of the fluctuation-response relation~\cite{harada2005equality,Nardini2017}, as well as to multivariable extensions that treat degrees of freedom independently~\cite{Nardini2017,harada2006energy}. Specifically, for variables with identical time-reversal signature, broken detailed balance generates non-equilibrium circulating probability fluxes that manifest directly as non-zero antisymmetric (non-reciprocal) response matrices $\mat{\widetilde{\chi}}^{\rm A}(\omega)$. By contrast, in any strictly one-variable system, circulating flux does not exist and
the correlation matrix $\mat{\widetilde{\psi}}(\omega)$ becomes a real valued scalar, forcing $\mat{\widetilde{\chi}}^{\rm A}(\omega) = 0$. In this one-variable scenario, the non-equilibrium circulating flux is absent, and the non-equilibrium activity may be hidden in the static covariance $\mat{g}$, often interpreted as an effective temperature $T_{\rm eff}$.

In the subsequent sections, we employ Eq.~(\ref{eq:dA/dt}) with the memory kernel Eq.~(\ref{eq:K}) to derive the constitutive equations for complex fluids in non-equilibrium steady states, fluctuations in chemical reactions, and, finally, non-equilibrium complex fluids driven by chemical reactions.

\section{Constitutive equation for non-equilibrium complex fluids \label{sec:stress_CE}}
In this section, we derive the constitutive equation for the stress in non-equilibrium complex fluids using the formalism introduced in Section~\ref{sec:correlation}.  The detailed derivation of this section is found in the Appendix~\ref{App:CE_j}.
We adopt an index-free notation for tensor contractions to simplify the presentation. Because this can obscure the order of contractions, Appendix~\ref{App:contractions} summarizes the conventions we use. For completeness, the detailed derivations in the Appendices present the contractions explicitly using indexes.

\subsection{Momentum density and conservation law}
We begin by defining the relevant dynamical variable and its associated conservation law.
We introduce the momentum density for a fluid composed of $N$ particles, 
\begin{equation} 
\label{eq:MC_real}
\vec{j}(\vec{r},t) = \sum_{i=1}^N m_i \,{\vec{v_i}}\, \delta(\vec{r}-\vec{r_i}(t))\quad,
\end{equation}
where $m_i$, $\vec{r_i}$, and $\vec{v_i}$ denote the mass, position, and velocity  of particle $i$, respectively.
Eq.~\eqref{eq:MC_real} is written in Fourier space,
\begin{equation} 
\label{}
\vec{j}(\vec{q},t) = \sum_{i=1}^N m_i\, \vec{v_i} \,\exp(i \vec{q}\cdot \vec{r_i}(t)) \quad.
\end{equation}
 The momentum density follows the conservation law,
\begin{equation} 
\label{eq:j_conserv}
\partial_t \, \vec{j}(\vec{q},t) = -i\, \mat{\sigma}(\vec{q},t)\cdot \vec{q}\quad ,
\end{equation}
where $\mat{\sigma}(\vec{q},t)$ is stress tensor given by the Irving–Kirkwood expression~\cite{irving1950statistical},
\begin{equation} 
\begin{split}
\label{eq:}
\mat{\sigma}(\vec{q},t) = &- \sum_{i=1}^{N} m_i \vec{v_i}\, \vec{v_i}^\top \exp(i \vec{q}\cdot \vec{r_i}) \\
&- \frac{1}{2} \sum_{i \neq j}^N  \vec{F_{ij}}\,\vec{r_{ij}}^\top\,   g(i\vec{q}\cdot \vec{r_{ij}})  \exp(i \vec{q} \cdot \vec{r_j})\quad.
\end{split}
\end{equation}
Here $\vec{r_{ij}} = \vec{r_i}-\vec{r_j}$ is the relative position vector and $\vec{F_{ij}}$ is the pair-wise force acting between particles. The function $g(x)$ is defined as $g(x) = (e^x-1)/x$.

The momentum density is related to the macroscopic strain rate to linear order by 
\begin{equation} 
\label{eq:strate}
\mat{\gamma}(\vec{q},t) = -i\,  \vec{j}(\vec{q},t)\,\vec{q}^\top  /\rho_0 \quad ,
\end{equation}
where $\rho_0$ is the average mass density of the system. 

\subsection{Constitutive equation of stress \label{sec:CE_j}}
We derive the constitutive equation of stress using Eq.~(\ref{eq:dA/dt}) by choosing the fluctuation of momentum density as a dynamical variable,
\begin{equation}
\label{eq:A_MD}
\vec{A} = \vec{j}(\vec{q},t)\quad .
\end{equation}
The static correlation $\mat{g}$ is given by
\begin{equation}
\label{eq:}
\mat{g}(\vec{q})= \expval{\vec{j}(\vec{q}) \, \vec{j}^\dagger(\vec{q})}\quad.
\end{equation}
For equilibrium isotropic systems, equipartition implies
\begin{equation}
\label{eq:Kubo_j}
 \expval{\vec{j}(\vec{q}) \, \vec{j}^\dagger(\vec{q})}_{\rm eq} = \rho_0 V k_BT \mat{I} \quad , 
\end{equation}
where $V$ is the system volume, $k_B$ is the Boltzmann constant, and $T$ is the temperature.
Motivated by this equilibrium limit, we may interpret $\mat{g}(\vec q)$ as defining a
non-equilibrium, wavelength-dependent effective temperature matrix, which reduces to the
thermodynamic temperature at equilibrium.

We first compute the term $\Delta\mat{{\widehat{\phi}}}(\vec{q},s) \cdot \mat{g}^{-1}$ in the memory kernel $\mat{\widehat{K}}$ [Eq.~(\ref{eq:K})]. The flux correlation matrix in Laplace space is 
\begin{equation}
\label{eq:phi_j}
    \mat{\widehat{\phi}}(\vec{q},s) = \langle \widehat{\dot{\vec{j}}}(\vec{q},s) \, {\dot{\vec{j}}}^\dagger(\vec{q}) \rangle \quad,
\end{equation}
and the transient part is defined as $ \Delta \mat{\widehat{\phi}}(\vec{q},s)=\mat{\widehat{\phi}}(\vec{q},s)-\mat{\widehat{\phi}}(\vec{q},0)$.
We define the normalized stress correlation tensor,
\begin{equation}
\label{eq:N_mech}
 \Tfor{ \widehat{N}}(\vec{q},s) \equiv \rho_0 \,  \expval{\mat{\widehat{\sigma}}(\vec{q},s)\, \mat{\sigma}^\dagger(\vec{q})} \cdot \mat{g}^{-1} \quad. 
\end{equation}
Using the conservation law, Eq.~\eqref{eq:j_conserv}, to write the time-derivative of the momentum density using the stress in Eq.~\eqref{eq:phi_j}, we obtain, 
\begin{equation}
\label{eq:qNq}
\Delta \mat{{\widehat{\phi}}}(\vec{q},s) \cdot \mat{g}^{-1}= \vec{q}\cdot \Delta \widehat{\Tfor{N}}(\vec{q},s) \cdot\vec{q}\,/\rho_0 \quad , 
\end{equation}
where the transient part of the normalized stress correlation function is defined as
\begin{equation} 
\label{eq:deltaN}
\Delta \widehat{\Tfor{N}}(\vec{q},s) \equiv  \widehat{\Tfor{N}}(\vec{q},s)-  \widehat{\Tfor{N}}(\vec{q},0)\quad .
\end{equation}
Substituting Eq.~\eqref{eq:qNq} into the memory kernel $\mat{\widehat{K}}(\vec{q},s)$ and using Eq.~\eqref{eq:dA/dt}, we derive the following equality involving the stress perturbation in Fourier-Laplace space:  
\begin{widetext}
\begin{equation} 
\begin{split}
\label{eq:main_eqality}
\qty( \expval{{\widehat{\mat{\sigma}}}(\vec{q},s)} - \qty[\mat{I}  - \frac{1}{\rho_0 s}\vec{q}\cdot \Delta \widehat{\Tfor{N}}(\vec{q},s) \cdot \vec{q}]^{-1} \cdot\Delta \Tfor{\widehat{N}}(\vec{q},s):  \expval{{\widehat{\mat{\gamma}}}(\vec{q},s)} ) \cdot \vec{q}=0 \quad.
\end{split}
\end{equation}
\end{widetext}

Eq.~(\ref{eq:main_eqality}) determines the linear response function up to a  divergence-free contribution if the correlation functions are known. Specifically, it dictates the relationship between the stress response and stress correlation functions for the longitudinal and shear components of the stress tensor. These components are parallel to the wavevector $\vec{q}$\,, or coupling the parallel and perpendicular directions. They are the only components that affect the evolution of the momentum density, see Eq.~(\ref{eq:j_conserv}), and are therefore fully determined by Eq.~(\ref{eq:main_eqality}). The stress components perpendicular to $\vec{q}$ (the transverse direction) do not affect the evolution of the momentum density and are therefore left undetermined by Eq.~(\ref{eq:main_eqality}). This ambiguity is analogous to the gauge freedom in electromagnetism; while the linear response function for the hydrodynamic components is well-defined, the components associated with the purely transverse stress cannot be unambiguously determined. 
Related aspects concerning the ambiguity between the stress correlation and elastic modulus have been discussed in~\cite{semenov2024general,grimm2025stress}.

To obtain a constitutive relation for all components, we fix the gauge. As detailed in Appendix~\ref{sec:gauge}, this leads to
the constitutive equation
\begin{equation} 
\label{eq:CEQ_fluid}
\expval{\mat{\widehat{\sigma}}(\vec{q},s)} =\Tfor{\widehat{\eta}}(\vec{q},s) : \expval{\mat{\widehat{\gamma}} (\vec{q},s)}   \ ,
\end{equation}
with the generalized viscosity tensor, 
\begin{equation} 
\begin{split}
\label{eq:eta}
\Tfor{\widehat{\eta}}(\vec{q},s)\equiv  
 \qty(\mat{I}  - \frac{1}{\rho_0 s}\vec{q}\cdot \Delta \widehat{\Tfor{N}}(\vec{q},s) \cdot \vec{q} )^{-1} \cdot \Delta \widehat{\Tfor{N}}(\vec{q},s) \; .
\end{split}
\end{equation} 
Eqs.~(\ref{eq:CEQ_fluid})-(\ref{eq:eta}) represent the constitutive equation of stress for fluids in finite wavelength and frequency. This applies to non-equilibrium steady states as well as to equilibrium. 
The generalized viscosity in Eq.~(\ref{eq:eta}) extends the expression for the frequency and wave-dependent shear viscosity for equilibrium fluids, obtained by Evans \cite{j2007statistical,evans1981equilibrium}, to non-equilibrium steady states.

This construction extracts the dynamic, frequency dependent,  stress response kernel from the stress correlation, provided the components of $\Tfor{\widehat N}(\vec q,s)$ are accessible (e.g.\ from simulation or experiment). It therefore fixes both shear and dynamic longitudinal transport coefficients, including dissipative and non-dissipative (reactive) parts due to possible antisymmetric chiral components.
However, the static isotropic part of the stress such as the thermodynamic pressure at equilibrium is not a transport coefficient. It is therefore not determined by our transient response kernels constructed from momentum density. 
To treat the static compressional response, the mass density must be included among the dynamical variables. This allows one to separate the static response from the remaining dynamic viscous response. We carry out this extended construction in Appendix~\ref{App:mass_momentum}.

Time-reversal symmetry at equilibrium implies $ \vec{q}\cdot \widehat{\Tfor{N}}(\vec{q},0) \cdot \vec{q}=\mat{0}$ for finite wavelength. For systems in a non-equilibrium steady state, however,  $ \vec{q}\,\cdot \widehat{\Tfor{N}}(\vec{q},0) \cdot \vec{q}\neq \mat{0}$ in general. This can be understood by noting that $\mat{{\widehat{\phi}}}(\vec{q},0) =  \mat{\omega}$ with
\begin{equation}
\label{eq:omega_N}
\mat{\omega} \cdot \mat{g}^{-1}= \vec{q}\cdot \widehat{\Tfor{N}}(\vec{q},0) \cdot\vec{q}/\rho_0 \quad ,
\end{equation}
where $\mat{\omega}=0$ at equilibrium by time reversal symmetry.

The connection to conventional viscous transport coefficient is recovered in the hydrodynamic limit at equilibrium. The Green-Kubo relation for viscosity is obtained by taking the long-wavelength limit ($q \to 0$) of the correlation function, followed by the zero-frequency limit of its Laplace transform ($s \to 0$), which yields a finite result. This is detailed in Appendix~\ref{sec:Green-Kubo}.

\subsection{Isotropic fluids in a two-dimensions \label{sec:iso_fluid_2D}}
The generalized viscosity in Eq.~\eqref{eq:eta} is computed from stress-stress correlation functions. These correlations can be obtained from simulation or experiment. Alternatively, one could determine them self‐consistently using a mode‐coupling–type approach~\cite{gotze2009complex} or using microscopic models which allow analytical computation. Here as an example and for the comparison purpose to the chemically driven fluids in the later section (Section~\ref{sec: fluids_chemical}), we compute the generalized viscosity for isotropic achiral fluids in two dimensions using a specific form of the correlation function. We consider in two dimensions, without loss of generality, the $y$-axis to be aligned with the direction of $\vec{q}$, i.e., $\vec{q} = (0,q)$. 
In this reference frame, isotropy and the absence of chirality impose strong constraints on material tensors (correlations and transport coefficients): for any even-rank material tensor, components with an odd number of a given index must vanish~\cite{wittmer2023correlations}. For example, $T_{xy}=0$ and $T_{xyyy}=0$. 
We emphasize that this symmetry constraint applies to material properties not to the strain rate tensor $\widehat{\gamma}_{\alpha\beta}$.
Independently, the choice of coordinate frame implies $q_x=0$, and therefore
$\widehat{\gamma}_{\alpha x}(\vec q,s)= -i q_x j_\alpha(\vec q,s)/\rho_0=0$, i.e.\ $\widehat{\gamma}_{xx}(\vec q,s)=\widehat{\gamma}_{yx}(\vec q,s)=0$, while $\widehat{\gamma}_{xy}(\vec q,s)$ can remain nonzero because it involves $q_y$.
Applying these considerations to Eq.~\eqref{eq:CEQ_fluid}, we can write the constitutive equation as 
\begin{equation} 
\label{eq:CE_mech_simplified}
\expval{\widehat{\sigma}_{xy}(\vec{q},s)} = \widehat{\eta}(\vec{q},s)\expval{\widehat{\gamma}_{xy}(\vec{q},s)}\quad; 
\end{equation}
\begin{equation} 
\label{}
\expval{\widehat{\sigma}_{yy}(\vec{q},s)} =  \widehat{\zeta} (\vec{q},s)\expval{\widehat{\gamma}_{yy}(\vec{q},s)}\quad,
\end{equation}
where the generalized viscosities are given by
\begin{equation} 
\label{eq:eta_hat_def}
\widehat{\eta}(\vec{q},s) = \frac{\Delta \widehat{N}^{}_{xyxy}\rho_0 s}{\rho_0 s - q^2\Delta \widehat{N}^{}_{xyxy}}\quad;
\end{equation}
and 
\begin{equation} 
\begin{split}
\label{eq:zeta_hat_def}
\widehat{\zeta}(\vec{q},s) = \frac{\Delta \widehat{N}^{}_{yyyy}\rho_0 s}{\rho_0 s - q^2\Delta \widehat{N}^{}_{yyyy}}\quad.
\end{split}
\end{equation}

Motivated by Refs.~\cite{klochko2018long,maier2018stress,maier2017emergence} where the stress correlation functions were calculated by a projection operator method and by the fluctuation-dissipation relation at equilibrium, we consider the transient parts of the stress of the simple form 
\begin{equation} 
\label{eq:DN_perp}
\Delta \widehat{N}_{xyxy}(\vec{q},s)=\frac{\eta^\perp \rho_0 s }{\eta^\perp q^2 + \rho_0 s}\quad, 
\end{equation}
and
\begin{equation} 
\label{eq:DN_para}
\Delta \widehat{N}_{yyyy}(\vec{q},s)=\frac{\eta^\parallel \rho_0 s }{\eta^\parallel q^2 + \rho_0 s} \quad.
\end{equation}
We thus obtain the rheology of a Newtonian fluid with the constant shear and bulk viscosities $\eta^{\perp}$ and $\eta^{\parallel}$, respectively, such that we have in the time domain  
\begin{equation}
\label{eq:newton_simple}
{\eta}(\vec{q},t)=\eta^{\perp}\delta(t)\quad,
\qquad
{\zeta}(\vec{q},t)=\eta^{\parallel}\delta(t)\quad.
\end{equation}
This minimal scenario will serve as a reference for the chemically driven case analyzed in Section~\ref{sec: simplified_active}-\ref{sec:example:active}.

\section{Constitutive equation for chemical reactions \label{sec:chemical_reaction} }
Chemical reactions generate thermodynamic driving forces for various biological processes, including the active stresses produced by molecular motors. In this section, we apply Eq.~(\ref{eq:dA/dt}) to the chemical potential differences arising from such reactions. We begin this section by introducing the fundamental concepts of chemical reactions such as chemical potentials and reaction rates.
\subsection{Chemical potentials and reactions rates}
We consider a mixture of $M$ chemical species. The chemical potential of species $a$ at position $\vec{r}$ and time $t$, $\mu_a(\vec{r},t)$, is defined using the number density of species $a$, $n_a$, and free energy density $f(n_1,...,n_M)$,
\begin{equation} 
\label{eq:}
\mu_a(\vec{r},t) \equiv \frac{\partial f}{\partial n_a(\vec{r},t)}\quad. 
\end{equation}
We consider chemical reactions $I=1,...,I_{\rm max}$ which are described by the stoichiometric coefficient $\nu^I_a$. Here $\nu^I_a$ is the number of molecule species of $a$ removed by one event of the reaction $I$. When $\nu^I_a > 0$, species $a$ acts as a substrate and its number decreases; conversely, when $\nu^I_a < 0$, species $a$ is a product and its number increases.
We define the chemical potential difference  between reactants and products in the chemical reaction $I$: 
\begin{equation} 
\label{eq:Deltamu}
\Delta \mu_I(\vec{r},t) \equiv \nu^I_a \mu_a(\vec{r},t)\quad.
\end{equation}
The repeated indices imply the summation.  Here $\Delta \mu_I$ is also called reaction Gibbs free energy. The rate of number density change for species $a$ is denoted $r_a = dn_a/dt$, which satisfies 
$r_a = -\nu_a^I r_I$, where $r_I$ is the net reaction rate of the reaction $I$. Taking the time derivative of Eq.~(\ref{eq:Deltamu}) leads to

\begin{equation} 
\label{eq:dmu/dt_0}
\frac{d \Delta \mu_I(\vec{r},t)}{dt} =\nu_a^I \qty( \frac{\partial n_b}{\partial t}\frac{\partial \mu_a}{\partial n_b})
=- r_J    \nu_a^I   \frac{\partial \mu_a}{\partial n_b}     \nu_b^J \quad .
\end{equation}
We define 
\begin{equation} 
\label{eq:R_r}
R_I(\vec{r},t) \equiv r_J(\vec{r},t) \kappa_{IJ}(\vec{r},t)\quad,
\end{equation}
with  the susceptibility $\kappa_{IJ}$,
\begin{equation} 
\label{eq:def_kappa}
\kappa_{IJ}(\vec{r},t)  \equiv  \nu_a^I   \frac{\partial \mu_a (\vec{r},t)}{\partial n_b(\vec{r},t)}     \nu_b^J \quad.
\end{equation}
Therefore Eq.~(\ref{eq:dmu/dt_0}) can be written as
\begin{equation} 
\label{eq:dmu/dt}
\frac{d \Delta \mu_I(\vec{r},t)}{dt} = -R_I(\vec{r},t)\quad. 
\end{equation}
This evolution equation for $\Delta \mu_I(\vec{r},t)$ plays a role for chemical reactions analogous to the momentum conservation law [Eq.~(\ref{eq:j_conserv})] for fluids.

We consider a spatially homogeneous unperturbed steady state of the multicomponent chemical mixture.  Therefore the susceptibility matrix $\mat{\kappa}$ is treated as constant in space and time [Eq.~\eqref{eq:def_kappa}].
This does not preclude spatiotemporal fluctuations of $\Delta \mu_I(\vec r,t)$. 
As we show in Appendix~\ref{App:CE_mu}, the susceptibility and the static fluctuations of the chemical potentials are related in equilibrium as
\begin{equation} 
\label{eq:kubo_mu}
\expval{\Delta \vec{\mu}(\vec{q})\, \Delta \vec{\mu}^\dagger(\vec{q})}_{\rm eq} =V k_BT  \,\mat{\kappa}  \quad ,
\end{equation}
where $\Delta \vec{\mu}\equiv (\Delta \mu_1,\Delta \mu_2,.... )^\top$. 
Eq.~(\ref{eq:kubo_mu}) represents the static Kubo relation for the chemical potential fluctuations, analogous to Eq.~\eqref{eq:Kubo_j} for equilibrium fluids. 

\subsection{Constitutive equation for chemical reaction rates}
We now derive the constitutive equation for the reaction rates. Similar to the previous section, we begin from Eq.~(\ref{eq:dA/dt}). In this case, we consider the dynamical variables using the chemical potential differences, 
\begin{equation} 
\label{eq:}
\vec{A}= \Delta \vec{\mu}(\vec{q},t)\quad  .
\end{equation}
The static matrix $\mat{g}$ is given by 
\begin{equation} 
\label{eq:}
\mat{g}(\vec{q}) = \expval{\Delta \vec{\mu}(\vec{q})\, \Delta\vec{\mu}^\dagger (\vec{q}) }\quad.
\end{equation}
The flux correlation in Laplace space is 
\begin{equation} 
\label{eq:phi_chem}
\mat{\widehat{\phi}}(\vec{q},s) = \expval{\vec{\widehat{{R}}} (\vec{q},s)\, \vec{{R}}^\dagger(\vec{q})} \quad.
\end{equation}
We define the normalized correlation,
\begin{equation} 
\label{eq:Nmu}
\mat{\widehat{N}}(\vec{q},s) \equiv \mat{\widehat{\phi}}(\vec{q},s)  \cdot \mat{g}^{-1}\quad.
\end{equation}

Applying Eq.~(\ref{eq:dmu/dt}) to the left-hand side of Eq.~(\ref{eq:dA/dt}) and substituting the Eq.~\eqref{eq:Nmu} into the memory kernel, we obtain the constitutive equation in Fourier-Laplace space 
\begin{equation} 
\label{eq:}
\expval{\vec{\widehat{R}}(\vec{q},s)}=  \widehat{\mat{K}}(\vec{q},s) \cdot \expval{\Delta \vec{\widehat{\mu}}(\vec{q},s)}\quad. \\
\end{equation}
This can be transformed to the constitutive equation for the chemical reaction rate $r$,
\begin{equation} 
\label{eq:CE_chem}
\expval{\vec{\widehat{r}}(\vec{q},s)}=  \mat{\kappa}^{-1}\cdot \widehat{\mat{K}}(\vec{q},s) \cdot \expval{\Delta \vec{\widehat{\mu}}(\vec{q},s)}\quad, 
\end{equation}
with the kernel given by 
\begin{equation} 
\label{eq:KIJ}
\mat{\widehat{K}}(\vec{q},s) =\left[\mat{I} -\frac{1}{s}\Delta \mat{\widehat{N}}(\vec{q},s) \right]^{-1}\cdot\, \Delta \mat{\widehat{N}}(\vec{q},s)\quad,
\end{equation}
where $\Delta \mat{\widehat{N}}(\vec{q},s)= \mat{\widehat{N}}(\vec{q},s)-\mat{\widehat{N}}(\vec{q},0)$.
Eqs.~(\ref{eq:CE_chem})-(\ref{eq:KIJ}) form the constitutive  equation for chemical reaction rates, analogous to the constitutive equation for the stress obtained in Eqs.~(\ref{eq:CEQ_fluid})-(\ref{eq:eta}).

\section{Non-equilibrium complex fluids driven by chemical reactions \label{sec: fluids_chemical}}
In this section, we consider complex fluids whose behavior is governed by internal chemical reactions. Such systems are driven into non-equilibrium steady states by the continuous energy consumption of their individual constituents. Our formalism unifies the mechanical response of a fluid with the chemical reactions that drive its activity, systematically yielding the constitutive equations for chemically driven active fluids. The detailed derivations in this section may be found in Appendix~\ref{App: CE_j_mu_detail}.
\subsection{General form of constitutive equation}
Building upon the previous considerations, we now derive the constitutive equation for an active fluid driven by chemical reactions, beginning with the selection of the appropriate dynamical variables:
\begin{equation} 
\label{eq:variable_j_mu}
\vec{A} = \qty(\vec{j}(\vec{q},t)\,,\ \Delta \vec{\mu}(\vec{q},t))^\top \,.
\end{equation}
For simplicity, we assume that the chemical variables are not statically correlated with the macroscopic directional current in the steady state. Importantly, the fluid itself need not be isotropic. This implies their steady-state cross-correlation vanishes,
\begin{equation} 
\label{}
\expval{\vec{j}(\vec{q})\, \Delta {\vec{\mu}}^{\dagger}(\vec{q})} = \expval{\Delta \vec{\mu}(\vec{q})\, \vec{j}^\dagger(\vec{q})}=\mat{0}\quad.
\end{equation}
This condition means that the static correlation matrix $\mat{g}$ is block diagonal and thus $\mat{g}^{-1}$ is as well block diagonal.
Therefore we have the static correlation matrix,
\begin{equation} 
\begin{split}
\label{}
\mat{g}(\vec{q}) 
=
 \begin{pmatrix}
\mat{g}^{jj} & 0 \\ \\
0 & \mat{g}^{\mu \mu}
\end{pmatrix}\quad,
\end{split}
\end{equation}
where 
\begin{equation}
    \mat{g}^{jj} \equiv \expval{\vec{j}(\vec{q})\,\vec{j}^\dagger(\vec{q})} \quad; \quad \mat{g}^{\mu \mu} \equiv \expval{\Delta \vec{\mu}(\vec{q})\,\Delta \vec{\mu}^\dagger(\vec{q})} \quad.
\end{equation}

The flux correlation matrix in Laplace space is given by
\begin{equation} 
\begin{split}
\label{}
\mat{\widehat{\phi}}(\vec{q},s) &=  \begin{pmatrix}
\expval{\widehat{\dot{\vec{j}}}(\vec{q},s)\, \dot{\vec{j}}^{\dagger}(\vec{q})} & \expval{\widehat{\dot{\vec{j}}}(\vec{q},s)\, \Delta \dot{\vec{\mu}}^{\dagger}(\vec{q})}\\ \\
\expval{\Delta \widehat{\dot{\vec{\mu}}}(\vec{q},s)\, \dot{\vec{j}}^\dagger(\vec{q})}& \expval{\Delta \widehat{\dot{\vec{\mu}}}(\vec{q},s)\,  \Delta \dot{\vec{\mu}}^{\dagger}(\vec{q})}
\end{pmatrix} \quad.
\end{split}
\end{equation}
Using the momentum conservation Eq.~\eqref{eq:j_conserv} and Eq.~\eqref{eq:dmu/dt}, the matrix product $\mat{\widehat{\phi}}(\vec{q},s) \cdot \mat{g}^{-1}$ can be written in
the compact form
\begin{equation} 
\label{eq:phi_j_mu}
\mat{\widehat{\phi}}(\vec{q},s)\cdot \mat{g}^{-1} =
 \begin{pmatrix}
 \vec{q}\cdot \Tfor{\widehat{N}}^{\sigma \sigma}(\vec{q},s) \cdot\vec{q}/\rho_0 & i \vec{q}\cdot \Tthr{\widehat{N}}^{\sigma R}(\vec{q},s)/\rho_0\\ \\
- \Tthr{\widehat{N}}^{R\sigma}(\vec{q},s) \cdot i\vec{q}/\rho_0&  \mat{\widehat{N}}^{RR}(\vec{q},s) /\rho_0
\end{pmatrix}\;.
\end{equation}
Here the normalized correlation tensors are defined as
\begin{align} 
\label{eq:correlations}
&\Tfor{\widehat{N}}^{\sigma \sigma} (\vec{q},s)  \equiv \rho_0  \cdot \expval{\mat{\widehat{\sigma}}(\vec{q},s)\, \mat{\sigma}^{\dagger}(\vec{q})} \cdot \qty(\mat{g}^{jj})^{-1}\quad ,\\[2mm]
&\Tthr{\widehat{N}}^{\sigma R} (\vec{q},s) \equiv \rho_0 \expval{\widehat{\mat{\sigma}}(\vec{q},s)\, \vec{R}^{\dagger}(\vec{q})} \cdot \qty(\mat{g}^{\mu \mu})^{-1}\quad ,\\[2mm] 
&\Tthr{\widehat{N}}^{R \sigma} (\vec{q},s) \equiv\rho_0 \expval{\widehat{\vec{R}}(\vec{q},s)\, \mat{\sigma}^{\dagger}(\vec{q})}\cdot \qty(\mat{g}^{jj})^{-1} \quad ,\\[2mm]
&\mat{\widehat{N}}^{R R} (\vec{q},s)  \equiv  \rho_0 \expval{\widehat{\vec{R}}(\vec{q},s)\, \vec{R}^{\dagger}(\vec{q})}\cdot \qty(\mat{g}^{\mu \mu})^{-1} \quad. 
\end{align}
The correlations $\Tfor{\widehat{N}}^{\sigma \sigma}$ and $\mat{\widehat{N}}^{RR}$ are analogous to Eq.~(\ref{eq:N_mech}) (stress correlation) and Eq.~(\ref{eq:phi_chem}) (reaction rate correlation), respectively. The correlations $\Tthr{\widehat{N}}^{\sigma R}$ and $\Tthr{\widehat{N}}^{R \sigma}$ describe the chemo-mechanical couplings due to the concurrent presence of mechanical processes and chemical processes. 
The transient part of the correlation functions can be defined as: $\Delta \Tfor{\widehat{N}}^{\sigma \sigma}(\vec{q},s) \equiv \Tfor{\widehat{N}}^{\sigma \sigma}(\vec{q},s)-\Tfor{\widehat{N}}^{\sigma \sigma}(\vec{q},0)$ and similarly for other correlations.

The procedure analogous to that in the previous sections leads to the constitutive equations in Fourier-Laplace space:
\begin{align} 
\label{eq:CE_general_shear}
\expval{\mat{\widehat{\sigma}}(\vec{q},s)} &= \Tfor{\widehat{\Pi}}(\vec{q},s) :\expval{\mat{\widehat{\gamma}}(\vec{q},s) } +\Tthr{\widehat{\Lambda}}(\vec{q},s)\cdot  \expval{\Delta \vec{\widehat{\mu}}(\vec{q},s)} \,;
\\[2mm]
\expval{\vec{\widehat{r}}(\vec{q},s)} &= \Tthr{\widehat{\Xi}}(\vec{q},s): \expval{\mat{\widehat{\gamma}}(\vec{q},s)} + \mat{\widehat{\Upsilon}}(\vec{q},s)\cdot  \expval{\Delta \vec{\widehat{\mu}} (\vec{q},s)} \,,
\end{align}
where the generalized transport coefficients are given by
\begin{widetext}
\begin{align} 
\Tfor{\widehat{\Pi}}(\vec{q},s) &\equiv 
 \mat{\widehat{Y}}\, \cdot\qty(\Delta  \Tfor{\widehat{N}}^{\sigma \sigma}+ \frac{1}{s\rho_0} \Delta \Tthr{\widehat{N}}^{\sigma R} \cdot \,\mat{\widehat{X}}^{-1}\cdot \,\Delta \Tthr{\widehat{N}}^{R \sigma} )\quad;\label{eq:Pi_hat} \\[3mm]
\Tthr{\widehat{\Lambda}}(\vec{q},s)& \equiv \frac{1}{\rho_0} \, \mat{\widehat{Y}}\, \cdot  \qty(\Delta \Tthr{\widehat{N}}^{\sigma R}+\frac{1}{s\rho_0}\Delta \Tthr{\widehat{N}}^{\sigma R}\cdot \,\mat{\widehat{X}}^{-1}\cdot \,\Delta \mat{\widehat{N}}^{RR})\quad; \label{eq:Lamda_hat} \\[3mm]
\Tthr{\widehat{\Xi}}(\vec{q},s) &\equiv \mat{\kappa}^{-1}\cdot \mat{\widehat{Z}} \, \cdot \qty(\frac{1}{s\rho_0} \qty(\Delta \Tthr{\widehat{N}}^{R \sigma} \cdot \vec{q}) \cdot \, \mat{\widehat{U}}^{-1} \cdot \qty(\vec{q} \cdot \Delta \Tfor{\widehat{N}}^{\sigma \sigma} )+ \Delta \Tthr{\widehat{N}}^{R \sigma} ) \quad; \\[3mm]
\mat{\widehat{\Upsilon}}(\vec{q},s) &\equiv \frac{1}{\rho_0}\mat{\kappa}^{-1} \cdot \mat{\widehat{Z}} \, \cdot\qty(\frac{1}{s\rho_0} \qty(\Delta\Tthr{\widehat{N}}^{R \sigma} \cdot \vec{q})\cdot \, \mat{\widehat{U}}^{-1}\cdot \qty(\vec{q}\cdot \Delta \Tthr{\widehat{N}}^{\sigma R}) + \Delta \mat{\widehat{N}}^{RR} )\label{eq:coefficients_upsilon} \quad,
\end{align}
using the matrices $\mat{U}$\,, $\mat{V}$\,, $\mat{W}$\,, $\mat{X}$\,, $\mat{Y}$\,, and $\mat{Z}$\,:
\begin{align} 
\label{eq:matrix1}
&\mat{\widehat{U}}^{} \equiv \mat{I} - \frac{1}{s\rho_0}\vec{q}\cdot \Delta \Tfor{\widehat{N}}^{\sigma \sigma}\cdot \vec{q}\quad ; \quad \mat{\widehat{V}}^{} \equiv -\frac{1}{s\rho_0}\,i\, \vec{q} \cdot \Delta \Tthr{\widehat{N}}^{\sigma R}\quad ; \nonumber\\[2mm]
&\mat{\widehat{W}}^{} \equiv \frac{1}{s\rho_0}  \Delta \Tthr{\widehat{N}}^{R \sigma}\cdot i\,{\vec{q}}\quad;\quad  \mat{\widehat{X}}^{} \equiv  \mat{I}-\frac{1}{s\rho_0}\Delta \mat{\widehat{N}}^{RR}\quad;  \\[2mm]
&\mat{\widehat{Y}}^{} \equiv \qty(\mat{\widehat{U}}^{}-\mat{\widehat{V}}^{} \cdot \mat{\widehat{X}}^{-1}\cdot \,\mat{\widehat{W}}^{})^{-1} \; ; \; \mat{\widehat{Z}}^{} \equiv \qty(\mat{\widehat{X}}^{}-\mat{\widehat{W}}^{}\cdot\mat{\widehat{U}}^{-1}\cdot \, \mat{\widehat{V}}^{})^{-1}\quad \nonumber.
\end{align}
\end{widetext}
Eqs.~(\ref{eq:CE_general_shear})–(\ref{eq:matrix1}) provide a general set of constitutive relations for active fluids driven by chemical reactions. In Eq.~\eqref{eq:CE_general_shear}, the contribution to the stress tensor that is proportional to the chemical potential difference, $\Delta \vec{\mu}$\,, is known as the active stress~\cite{julicher2018hydrodynamic}. The active stress originates from the chemo-mechanical coupling coefficient $\Tthr{\widehat{\Lambda}}$\,, which is mediated by the tensors, $\Tthr{\widehat{N}}^{\sigma R}$ and $ \Tthr{\widehat{N}}^{R \sigma}$, see Eq.~\eqref{eq:Lamda_hat}. In the limit where this coupling is absent, i.e., $\Tthr{\widehat{N}}^{\sigma R} = \Tthr{\widehat{N}}^{R \sigma} = 0$, it follows that $\Tthr{\widehat{\Lambda}} = \Tthr{\widehat{\Xi}} = 0$, and the active stress consequently vanishes. The system then becomes decoupled, and the equations revert to the separate forms for stress and chemical reaction rates derived previously in Sections \ref{sec:stress_CE} and \ref{sec:chemical_reaction}, respectively.

We find that the effective viscosity, $\Tfor{\widehat{\Pi}}(\vec{q},s)$, is itself modified by chemo-mechanical coupling, see Eq.~\eqref{eq:Pi_hat}. This implies that the memory kernel governing the viscous stress relaxation depends directly on the active chemical processes. We identify this coupling-dependent relaxation, which we call Active Viscoelastic Memory. Consequently, the rheology of the fluid can be fundamentally changed in the presence of chemical processes. In the absence of chemo-mechanical coupling, active stress as well as active viscoelastic memory vanish. 

\subsection{Isotropic fluids driven by single chemical reaction in a two-dimensions \label{sec: simplified_active}}
In this section we derive explicit expressions for the constitutive  Eqs.(\ref{eq:CE_general_shear})–(\ref{eq:coefficients_upsilon}) for isotropic achiral systems in a two‐dimensions with a single chemical reaction—a situation often relevant to biological materials powered by ATP hydrolysis. We denote this chemical reaction as $I$. To simplify further, we use the coordinate frame where the $y$-axis is aligned with the direction of $\vec{q}$, i.e. $\vec{q}=(0,q)$. In this frame, isotropy and achirality impose symmetry constraints on the memory-kernel tensors, which leads to simplified constitutive equations:
\begin{align} 
\label{eq:CE_single_reaction}
&\expval{\widehat{\sigma}_{xy}(\vec{q},s)} = \widehat{\eta}(\vec{q},s)\expval{\widehat{\gamma}_{xy}(\vec{q},s)}\quad; \\[3mm]
&\expval{\widehat{\sigma}_{yy}(\vec{q},s)} =  \widehat{\zeta} (\vec{q},s)\expval{\widehat{\gamma}_{yy}(\vec{q},s)} + \widehat{\Lambda}(\vec{q},s) \expval{\Delta \widehat{\mu}_{I}(\vec{q},s)} ;\\[3mm] 
&\expval{\widehat{r}_I(\vec{q},s)} = \widehat{\Xi} (\vec{q},s)\expval{\widehat{\gamma}_{yy}(\vec{q},s)} + \widehat{\Upsilon} (\vec{q},s)\expval{\Delta \widehat{\mu}_{I} (\vec{q},s)}\label{eq:CE_single_r} \;.
\end{align}
The generalized transport coefficients are given by 
\begin{widetext}
\begin{equation} 
\label{eq:TP1}
\widehat{\eta}(\vec{q},s) = \frac{\Delta \widehat{N}^{\sigma \sigma}_{xyxy}\rho_0 s}{\rho_0 s - q^2\Delta \widehat{N}^{\sigma \sigma}_{xyxy}} \quad;
\end{equation}
\begin{equation} 
\begin{split}
\label{eq:TPzeta}
\widehat{\zeta}(\vec{q},s)= 
 \frac{\rho_0 s(\Delta \widehat{N}^{R \sigma}_{I,yy}\Delta \widehat{N}^{\sigma R}_{yy,I}-\Delta \widehat{N}^{RR}_{II}\Delta \widehat{N}^{\sigma \sigma}_{yyyy} + \Delta \widehat{N}^{\sigma \sigma}_{yyyy}\rho_0 s)}{ (\Delta \widehat{N}^{RR}_{II}-\rho_0 s)(\Delta \widehat{N}^{\sigma \sigma}_{yyyy}q^2 -\rho_0 s) -q^2 \Delta \widehat{N}^{R \sigma}_{I,yy}\Delta \widehat{N}^{\sigma R}_{yy,I} } \quad;
\end{split}
\end{equation}
\begin{equation} 
\begin{split}
\label{eq:TP2}
\widehat{\Lambda}(\vec{q},s)= 
\frac{\widehat{\chi}_{yy,I}\rho_0^2 s^2}{ (\Delta \widehat{N}^{RR}_{II}-\rho_0s)(\Delta \widehat{N}^{\sigma \sigma}_{yyyy}q^2 - \rho_0s) -q^2 \Delta \widehat{N}^{R \sigma}_{I,yy}\Delta \widehat{N}^{\sigma R}_{yy,I}} \quad; \\
\end{split}
\end{equation}
\begin{equation} 
\begin{split}
\label{eq:TP3}
\widehat{\Xi}(\vec{q},s)= 
 \frac{\widehat{\chi}_{I,yy} \rho_0^2 s^2}{(\Delta \widehat{N}^{RR}_{II}-\rho_0 s)(\Delta \widehat{N}^{\sigma \sigma}_{yyyy}q^2-\rho_0s)  - q^2 \Delta \widehat{N}^{R \sigma}_{I,yy}\Delta \widehat{N}^{\sigma R}_{yy,I}} \quad;
\end{split}
\end{equation}
\begin{equation} 
\begin{split}
\label{eq:TP4}
\widehat{\Upsilon}(\vec{q},s) = 
\frac{s}{\kappa_{II}}\frac{  (\Delta \widehat{N}^{R \sigma}_{I,yy}\Delta \widehat{N}^{\sigma R}_{yy,I}-\Delta \widehat{N}^{RR}_{II}\Delta \widehat{N}^{\sigma \sigma}_{yyyy})q^2 + \Delta \widehat{N}^{RR}_{II}\rho_0 s }{(\Delta \widehat{N}^{RR}_{II}-\rho_0 s)(\Delta \widehat{N}^{\sigma \sigma}_{yyyy}q^2-\rho_0s) -q^2 \Delta \widehat{N}^{R \sigma}_{I,yy}\Delta \widehat{N}^{\sigma R}_{yy,I}} \quad;
\end{split}
\end{equation}
with the cross coupling terms,
\begin{equation} 
\label{eq:chi_1}
\widehat{\chi}_{yy,I}(\vec{q},s) \equiv \Delta  \widehat{N}^{\sigma R}_{yy,I}(\vec{q},s)/{\rho_0}\quad  \text{and} \quad \widehat{\chi}_{I,yy}(\vec{q},s) \equiv \Delta  \widehat{N}_{I,yy}^{R \sigma}(\vec{q},s)/\kappa_{II}\quad .
\end{equation}
\end{widetext}
No summation over repeated $I$ is implied. These results should be compared to Eqs.~\eqref{eq:CE_mech_simplified}-\eqref{eq:zeta_hat_def} for the isotropic fluids without chemical reactions. 

Linear irreversible thermodynamics dictates reciprocity relations between cross-coupling transport coefficients, known as Onsager reciprocity~\cite{onsager1931reciprocal,onsager1931reciprocal2}. In this simplified form, we can readily confirm the Onsager reciprocity in equilibrium: $\widehat{\Lambda}=-\widehat{\Xi}$. 
In equilibrium, we have $g^{jj}_{yy}(\vec{q}) = \rho_0 V k_BT$ [Eq.~(\ref{eq:Kubo_j})] and $g^{\mu \mu}_{II}(\vec{q})=Vk_BT\kappa_{II}$ [Eq.~(\ref{eq:kubo_mu})]. Furthermore, we have $\expval{\sigma_{yy}(\vec{q},t) R_I^*(\vec{q})}=- \expval{R_I(\vec{q},t) \sigma_{yy}^*(\vec{q})}$ due to the time reversal symmetry in equilibrium, see Appendix~\ref{sec:symmetry}. The minus sign follows from the distinct time-reversal signatures of the stress and the chemical reaction rate. Substituting these conditions into the definitions of the coupling coefficients [Eq.~\eqref{eq:chi_1}] directly yields the reciprocity relation; see Appendix~\ref{App:Onsager_j_mu} for the detail. In a non-equilibrium steady-state, however, time-reversal symmetry is broken, and these relations do not hold, thus allowing broken Onsager reciprocity.

In this coordinate frame, the chemical reaction does not couple to transverse stress $\expval{\widehat{\sigma}_{xy}}$. 
 Since \ $q_x=0$, momentum conservation contains only $q_\beta \widehat\sigma_{\alpha\beta}=q\,\widehat\sigma_{\alpha y}$; hence $\expval{\widehat\sigma_{xx}}$ cannot feed back into the evolution of momentum density. 
We therefore focus on $\expval{\widehat{\sigma}_{yy}}$ for the specific example in the following section.

\subsection{Simple example of correlation functions \label{sec:example:active}}
The generalized transport coefficients in Eqs.~(\ref{eq:TP1})-(\ref{eq:TP4}) involve correlation functions that must be specified.  For the stress-stress correlation functions, we use again the form of Eq.~\eqref{eq:DN_para}.
For the correlation functions  involving chemical reactions, we  choose functions that satisfy the following conditions: for $s=0$, the transient part of the correlations satisfy $\Delta \widehat{N}^{\sigma \sigma}_{yyyy}(\vec{q},0)=\Delta \widehat{N}^{RR}_{II}(\vec{q},0)= \Delta \widehat{N}^{\sigma R}_{yy,I}(\vec{q},0)=\Delta \widehat{N}^{R \sigma}_{I,yy}(\vec{q},0)=0$\,.  Because the chemical reaction rates are not hydrodynamic variables that become slow in the small $q$ limit, we choose chemical correlations that are independent of $q$\,. The chemical reaction rate is governed by a single relaxation rate $\lambda$ since we consider a single chemical reaction such as ATP hydrolysis. The cross correlations between mechanical and chemical degrees of freedom contain $q$\,. We thus have 
\begin{equation*} 
\begin{split}
\label{eq:}
\Delta \widehat{N}^{\sigma \sigma}_{yyyy}(q,s)=\frac{\eta^\parallel \rho_0 s }{\eta^\parallel q^2 + \rho_0 s}\quad; \quad \Delta \widehat{N}^{RR}_{II}(q,s)= \lambda \frac{  \rho_0 s}{\lambda + s}\, ;
\end{split}
\end{equation*}
\begin{equation} 
\label{eq:N_special}
\Delta \widehat{N}^{\sigma R}_{yy,I}(q,s)=  \nu_{\sigma R} \frac{ \rho_0 s^2}{(s+\lambda)(s+\lambda\, q^2\xi_{\mu}^2)}\quad; 
\end{equation}
\begin{equation*} 
\label{eq:}
\Delta \widehat{N}^{R \sigma}_{I,yy}(q,s)=\nu_{R\sigma}\frac{\kappa_{II} s^2}{(s+\lambda)(s+\lambda\, q^2\xi_{\mu}^2)}\quad.
\end{equation*}
Here $\nu_{\sigma R}$ and $\nu_{R \sigma }$ characterize the chemo-mechanical coupling (with units of inverse-volume), and $\xi_\mu$ is chemo-mechanical coupling length. 
The length $\xi_\mu$ can be interpreted as a reaction-diffusion length of the chemo-mechanical coupling mode. When mechanical transport couples to a chemical process with relaxation rate $\lambda$, a chemo-mechanical coupling mode with effective diffusivity $D_\mu\equiv \lambda\,\xi_\mu^{2}$ naturally emerges. 

For an equilibrium system with time-reversal symmetry, the fluctuation-dissipation theorem and microscopic reversibility hold. From this, Kubo relations [Eq.~\eqref{eq:Kubo_j} and Eq.~\eqref{eq:kubo_mu}] and Onsager reciprocity ($\widehat{\Lambda} = - \widehat{\Xi}$) follow. In the following, we set $\nu \equiv \nu_{\sigma R}=-\nu_{R \sigma}/\varepsilon$ to model the reciprocity breaking where $\varepsilon=1$ respects the reciprocity and $\varepsilon\neq1$ otherwise. 
We note that imposing Onsager reciprocity ($\varepsilon=1$) 
describes small departures from equilibrium.
We substitute these forms of the correlation functions into Eq.~(\ref{eq:TP1})-(\ref{eq:TP4}) to obtain the generalized transport coefficients. The resulting transport coefficients in Laplace space are shown in Eq.~\eqref{eq:zeta_spetial_laplace} in Appendix~\ref{App:Special}.  
We note that when $\varepsilon=0$, the chemo-mechanical feedback contribution to the viscous response vanishes, and $\widehat{\zeta}(q,s)$ reduces to that of the isotropic two-dimensional fluid analyzed in Section~\ref{sec:iso_fluid_2D}.
The parameter values are restricted to ensure that the phenomenological correlations correspond to a realizable stationary process; see Appendix~\ref{app:correlation_realizability}.

In Fig.~\ref{fig:relaxation}a, we show the viscous transport kernel $\zeta(q,t)$ obtained by numerically inverting the Laplace-space expression $\widehat{\zeta}(q,s)$. The instantaneous $\delta(t)$ contribution at $t=0$ is omitted for clarity. At long wavelengths, $q\xi_\mu=0.1$ (left), $-\zeta(q,t)$ relaxes monotonically for all values of $\varepsilon$ shown. Increasing $\varepsilon$ enhances the magnitude of the transient response.
At the intermediate wavelength $q\xi_\mu=1$ (middle), the kernel initially relaxes positively, crosses zero, develops a negative lobe, and subsequently returns to zero. This sign-changing transient becomes more pronounced as $\varepsilon$ increases and represents an emergent reactive, elastic-like contribution to the otherwise viscous relaxation. Its origin is the feedback term
$\Delta\widehat N^{R\sigma}_{I,yy} \Delta\widehat N^{\sigma R}_{yy,I}  \sim \nu_{R\sigma}\nu_{\sigma R}  =-\varepsilon\nu^2$ in Eq.~\eqref{eq:TPzeta},
which allows the chemical dynamics to transiently oppose the direct viscous relaxation.
At short wavelengths, $q\xi_\mu=10$ (right), the kernel decays on a much shorter time scale and becomes effectively local in time. In this regime, the $q$-dependent chemical mode relaxes rapidly and the chemo-mechanical feedback has little influence on the response at finite times.

Figure~\ref{fig:relaxation}b shows $-\zeta(q,t)$ at the fixed scaled time $\lambda t=1$ as a function of the dimensionless wave number $q\xi_\mu$ for representative values of the reciprocity-breaking parameter $\varepsilon$. 
In all cases the curves exhibit a clear crossover near $q\xi_\mu\simeq 1$ (vertical dashed line). 
This crossover follows directly from the chemo-mechanical coupling factor $(s+\lambda q^{2}\xi_\mu^{2})^{-1}$ in Eq.~\eqref{eq:N_special}, which introduces a $q$-dependent chemical relaxation rate $\lambda_q\equiv \lambda q^{2}\xi_\mu^{2}$ and relaxation time scale $\tau_q \equiv 1/\lambda_q$. 
At the observation time $t\sim 1/\lambda$, the chemical mode remains dynamically relevant when $\tau_q\gtrsim 1/\lambda$, i.e.\ $q\xi_\mu\lesssim 1$, and the chemical degrees of freedom significantly renormalize the stress relaxation through the feedback terms $\Delta\widehat N^{R\sigma}_{I,yy}\Delta\widehat N^{\sigma R}_{yy,I}$. 
In contrast, when $\tau_q\ll 1/\lambda$ (i.e.\ $q\xi_\mu\gg 1$), the chemical mode relaxes rapidly and effectively decouples, so that the response approaches the local viscous behavior.

\begin{figure*}
\begin{center}
\includegraphics[width=\textwidth]{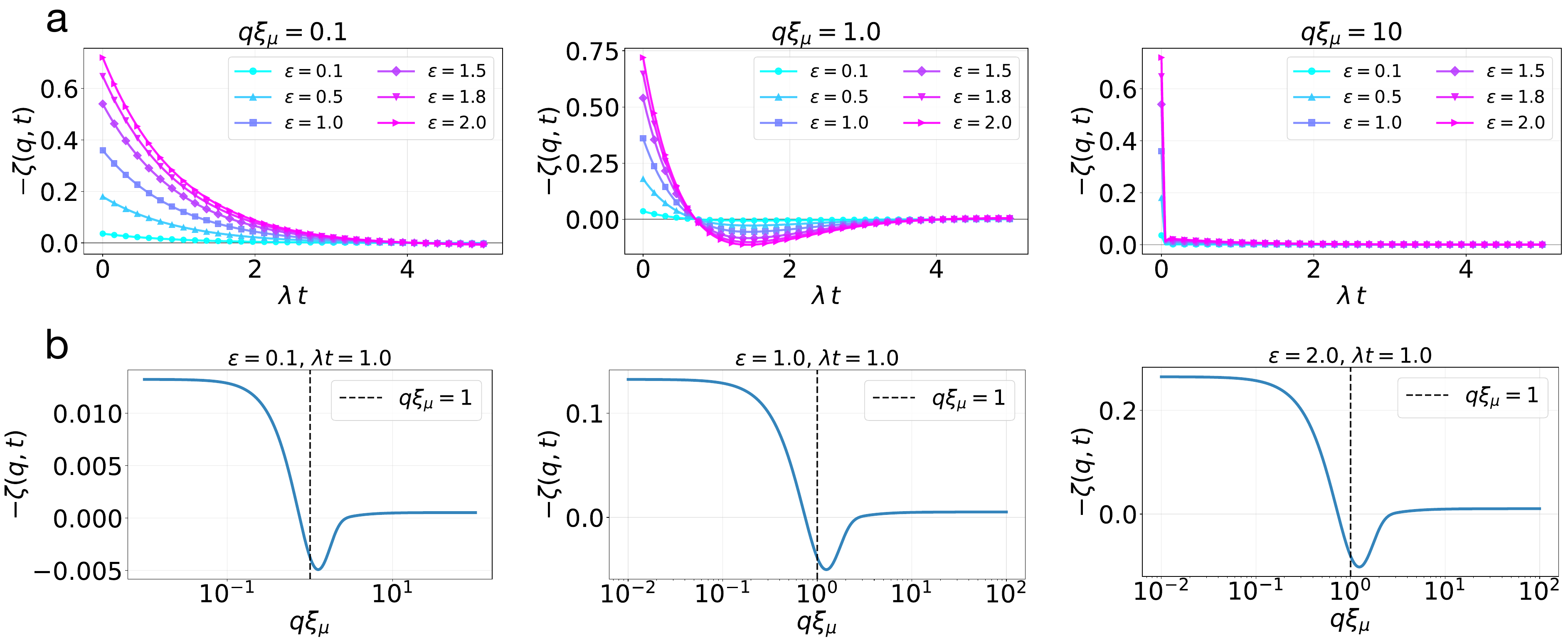}
\caption{\label{fig:relaxation}
{\bf Time-domain viscous transport kernel and its length-scale crossover.}
The time-dependent kernel $-\zeta(q,t)$, obtained by numerical inverse Laplace transform of $\widehat\zeta(q,s)$ (the instantaneous $\delta(t)$ contribution at $t=0$ is omitted for clarity), is shown in scaled variables $\lambda t$ and $q\xi_\mu$.
{\bf (a)} Relaxation curves $-\zeta(q,t)$ versus $\lambda t$ for three representative wavelengths: $q\xi_\mu=0.1$ (left), $q\xi_\mu=1.0$ (middle), and $q\xi_\mu=10$ (right). Each panel compares different reciprocity-breaking parameters $\varepsilon$ (legend). For long wavelengths ($q\xi_\mu\ll 1$) the relaxation is monotonic for all shown $\varepsilon$, whereas near the coupling scale ($q\xi_\mu\simeq 1$) a pronounced oscillatory transient emerges at large $\varepsilon$, reflecting strong chemo-mechanical feedback. At short wavelengths ($q\xi_\mu\gg 1$) the relaxation becomes rapidly decaying and essentially local in time.
{\bf (b)} Snapshot of $-\zeta(q,t)$ at fixed time $\lambda t=1$ as a function of $q\xi_\mu$ for $\varepsilon=0.1$ (left), $\varepsilon=1.0$ (middle), and $\varepsilon=2.0$ (right). The vertical dashed line marks $q\xi_\mu=1$, highlighting a clear crossover between a long-wavelength regime where the chemical mode significantly renormalizes the stress relaxation and a short-wavelength regime where the response approaches local viscous behavior.
Parameters: $\rho_0=5$, $\eta^{\parallel}=1$, $\lambda=1$, $\xi_\mu=1$, $\kappa_{II}=1$ and $\nu=0.6$ (arbitrary units).
}
\end{center}
\end{figure*}

To quantify the rheology of this chemically driven active fluid, we compute the complex modulus, 
\begin{equation}
\label{}
\widetilde{G}(q,\omega) \equiv i \omega \widehat{\zeta}(q,s=i\omega)\quad ,
\end{equation}
where the real and imaginary part of $\widetilde{G}(q,\omega)$, denoted as $\widetilde{G}'(q,\omega)$ and $\widetilde{G}''(q,\omega)$, respectively, is the storage and loss modulus.

\begin{figure*}[]
\begin{center}
\includegraphics[width=\textwidth]{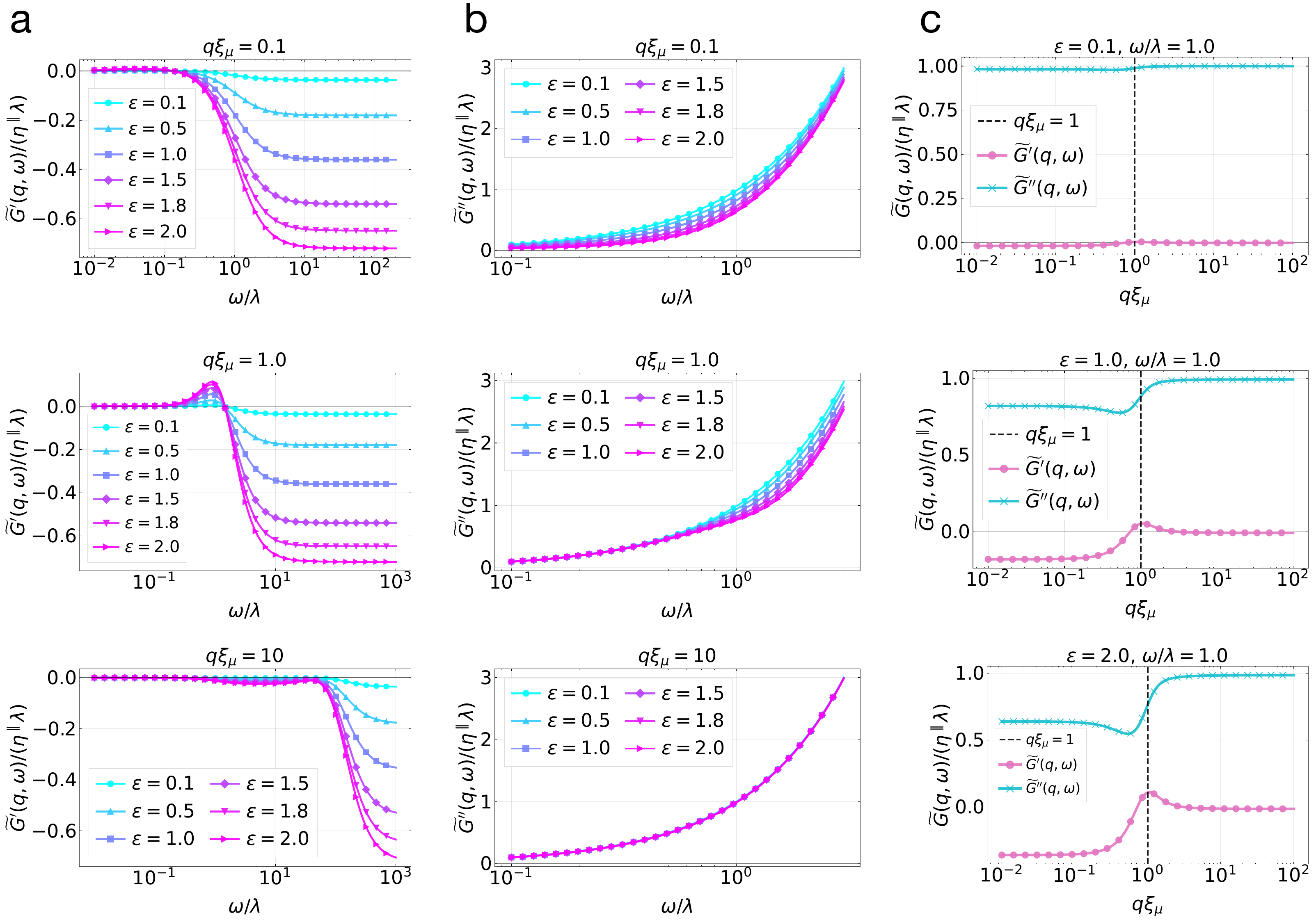}
\end{center}
\caption{\label{fig:Gs}
{\bf Complex moduli in the frequency and wave-number domains.}
Frequency-dependent complex moduli $\widetilde G(q,\omega)=\widetilde G'(q,\omega)+i\,\widetilde G''(q,\omega)$ are shown for varying degrees of reciprocity breaking $\varepsilon$ (color scale).
(a) Storage modulus $\widetilde G'(q,\omega)$ as a function of the scaled frequency $\omega/\lambda$ for three representative length scales $q\xi_\mu=0.1$, $1.0$, and $10$.
For sufficiently large $\omega/\lambda$, $\widetilde G'$ becomes negative, indicating a reactive contribution to the viscoelastic response at finite frequency.
(b) Corresponding loss modulus $\widetilde G''(q,\omega)$ for the same $q\xi_\mu$ values.
(c) Storage and loss moduli at fixed $\omega/\lambda=1$ plotted as functions of the dimensionless wave number $q\xi_\mu$ for three representative values of $\varepsilon=0.1$, $1.0$, and $2.0$.
The strongest $q$-dependence occurs near $q\xi_\mu\simeq 1$ (vertical dashed line), where the $q$-dependent chemo-mechanical relaxation rate $\lambda_q=\lambda q^2\xi_\mu^2$ becomes comparable to the probing frequency ($\lambda_q\sim\omega$).
Model parameters are the same as in Fig.~\ref{fig:relaxation}.}
\end{figure*}

In Fig.~\ref{fig:Gs}a, we show the storage modulus
$\widetilde G'(q,\omega)$ for three representative length scales, $q\xi_\mu=0.1$, $1$, and $10$, and for several values of the reciprocity-breaking parameter $\varepsilon$. In the low-frequency limit, $\widetilde G'(q,\omega)$ vanishes, as expected for a fluid without a static elastic modulus. At finite frequency, the chemo-mechanical coupling generates a
reactive contribution to the response. The corresponding signature is most pronounced around $q\xi_\mu\simeq1$, where the chemical relaxation rate becomes comparable to the probing frequency. In this regime, the storage modulus can first develop a positive maximum and subsequently cross zero and become negative as the frequency increases. The magnitude of this finite-frequency response increases with $\varepsilon$.
The negative storage modulus originates from the phase-delayed response of the chemical mode. When the probing time $1/\omega$ becomes comparable to, or shorter
than, the chemical relaxation time, the chemical variable cannot
adjust instantaneously to the mechanical deformation. The resulting
delayed feedback produces a phase-shifted stress response whose
reactive component can oppose the ordinary elastic response. This can
render the storage modulus negative at non-zero frequencies. The resulting negative storage modulus is a finite-frequency transient feature. It vanishes in the limit $\omega\to0$ and therefore does not correspond to a negative static elastic modulus or a static mechanical instability. 

Figure~\ref{fig:Gs}b shows the corresponding loss modulus
$\widetilde G''(q,\omega)$. Increasing $\varepsilon$ renormalizes the dissipative response over a finite range of frequencies. In particular, for $q\xi_\mu \lesssim 1$, the loss modulus is suppressed relative to the bare viscous response, which is in the regime where the chemical mode is dynamically relevant ($\tau_q\gtrsim1/\lambda$). Therefore strong chemical-feedback can suppress viscous response.
At high frequencies, chemical feedback becomes ineffective and the loss modulus approaches the bare viscous behavior, $\widetilde G''(q,\omega)\simeq\eta^\parallel\omega$. In the low-frequency limit at fixed nonzero $q$, the response remains fluid-like, $\widetilde G''(q,\omega)\simeq\eta_{\mathrm{eff}}\omega$, with a chemically renormalized effective viscosity $\eta_{\mathrm{eff}}=\widehat\zeta(q,0)$.

In Fig.~\ref{fig:Gs}c we plot the storage and loss moduli,
$\widetilde G'(q,\omega)$ and $\widetilde G''(q,\omega)$, at fixed $\omega/\lambda=1$ as functions of the dimensionless wavenumber $q\xi_\mu$ for three representative values of the reciprocity-breaking parameter $\varepsilon$.
For the reciprocity breaking parameter $\varepsilon \gtrsim 1.0$ (Fig.~\ref{fig:Gs}c middle and bottom), the $q$-dependence is most pronounced and becomes non-monotonic around $q\xi_\mu\simeq 1$. 
For small $\varepsilon = 0.1$ (Fig.~\ref{fig:Gs}c top), the chemo-mechanical feedback is small ($\Delta\widehat N^{R\sigma}_{I,yy}\Delta\widehat N^{\sigma R}_{yy,I}\sim \nu_{\sigma R}\nu_{R\sigma}=-\varepsilon\,\nu^{2}$) so that the $q \xi_\mu$-dependence is suppressed.

\section{Discussion}
We summarize our main findings. Eq.~\eqref{eq:dpsi/dt} is a general non-Markovian dynamical equation describing the evolution of correlation functions. 
Eq.~\eqref{eq:dA/dt} provides a model-independent regression equation for steady-state fluctuations and responses. As a consequence, we derive fluctuation-response relations Eq.~\eqref{eq:Harada-sasa_like_physical} and Eq.~\eqref{eq:FDT_chi_physical_main}. With this framework, we derive generalized transport coefficients directly from correlation functions in a non-equilibrium steady state, with time-reversal symmetry breaking entering explicitly, generalizing the equilibrium Green-Kubo relation to non-equilibrium fluids. When chemical reactions are coupled to mechanics, the resulting chemo-mechanical feedback generates a reactive contribution to the viscoelastic response and can produce a negative storage modulus at finite frequency, i.e.\ an effective negative elasticity.

The evolution equation for the correlation function, Eq.~(\ref{eq:dpsi/dt}), is built upon the property of time-translational invariance and a general identity for the steady state correlation functions. The procedure is an exact equivalence transformation of correlation functions, a principle that is shared with the Mori-Zwanzig projection operator formalism~\cite{zwanzig2001nonequilibrium}. 
Our formulation provides a direct link between transport coefficients and correlation functions, consistent with the established relations at equilibrium. Namely, the memory kernel, Eq.~(\ref{eq:K_0}), is expressed explicitly in terms of the flux-flux correlation function, $\mat{\widehat{\phi}}(s)$. These are precisely the correlations that determine transport coefficients via the Green-Kubo relations at equilibrium, thus creating a transparent link between the system's memory, its transport properties, and the underlying microscopic fluxes. When the reactive frequency matrix $\mat{\omega}=0$, our memory-kernel representation, Eq.~\eqref{eq:K_0}, coincides with that of Ref.~\cite{berne1966calculation}. 
Ref.~\cite{berne1966calculation} shows that this representation is equivalent to the memory kernel derived from the Mori-Zwanzig projection-operator, which suggests our formulation is consistent with projection operator approach.

Eq.~\eqref{eq:dA/dt} is a linear response relation with memory of a non-equilibrium steady state, valid for the class of small perturbations of the steady-state distribution that can be expanded in the dynamical variables; see Appendix~\ref{App:correlation_deriv}. Using Eq.~\eqref{eq:dA/dt}, we obtain the fluctuation-response relations, Eq.~\eqref{eq:FDT_chi_physical_main} and Eq.~\eqref{eq:Harada-sasa_like_physical}, derived in Appendix~\ref{sec:FDT}. Our work can be related to early work of Agarwal, who derived a general fluctuation-response relation of a nonequilibrium steady state based on a Fokker-Plank equation~\cite{agarwal1972fluctuation}. This work showed that a response function to a perturbation of the drift of a stochastic process with Gaussian noise can be obtained from a correlation function using the full steady state probability distribution. Both approaches, the one presented in this work and the one of Agarwal are complementary and differ in several ways. We obtained a general expression of fluctuation-response relations without the need to know the full steady state probability distribution. Furthermore, our approach is model free and not based on a Fokker-Planck equation. It is exact when the perturbation $\delta p_h(\vec{X}_{\mkern1mu  0})/p(\vec{X}_{\mkern1mu  0})$ in the function space can be expressed in terms of linear combinations of the components of $\vec{A}(\vec{X}_{\mkern1mu  0})$, independent on any underlying model. By contrast, the approach of Agarwal is exact for perturbations of the drift of an underlying Markovian Langevin dynamics when the full steady-state distribution is known. Both approaches thus have different limitations and different types of generality.

Hydrodynamic descriptions of active matter have been successful in capturing a wide range of biological phenomena~\cite{marchetti2013hydrodynamics,ramaswamy2010mechanics,toner1995long,toner1998flocks,han2021fluctuating,banerjee2017odd,fruchart2023odd,scheibner2020odd}. 
Our formulation extends the applicability of active matter theory to far-from-equilibrium, memory-rich complex fluids.  Furthermore, it makes explicit the connection between fluid transport coefficients and broken time-reversal symmetry: in non-equilibrium fluids, the transport coefficients are renormalized in a manner controlled by the degree of time-reversal breaking, see Eqs.~\eqref{eq:deltaN} and \eqref{eq:eta}. This result helps to bridge active matter physics and the recent progress on non-equilibrium thermodynamics such as stochastic thermodynamics~\cite{falasco2025macroscopic,seifert2012stochastic,peliti2021stochastic,shiraishi2023introduction}.

In this study, we chose the momentum density $\vec j$ and the reaction Gibbs free energy $\Delta \vec{\mu}$ as a choice of variable. This is the minimal set that captures mechanical response and chemical driving. Chemical species densities $n_a$ are not included because we focus on well-mixed chemical backgrounds on the scales of interest: small metabolites typically equilibrate by diffusion faster than the structural degrees of freedom of active materials. Reaction-diffusion effects can be incorporated systematically by enlarging the variable set to include the relevant $n_a$, obeying $\partial_t n_a+\nabla\!\cdot\!\vec {J_a}=-\nu_a^I\, r_I$, where $\vec {J_a}$ is a diffusion flux of species $a$. We also restricted ourselves to isothermal dynamics and neglect heat transport; including temperature and energy as additional variables would extend the theory to thermo-mechanical response.

Our formulation is developed at finite wave number and frequency and is therefore not restricted to strictly hydrodynamic variables. Nevertheless, in the long-wavelength and low-frequency limit, it naturally captures modes whose characteristic frequencies or relaxation rates vanish, including conserved hydrodynamic modes and Nambu-Goldstone modes. By contrast, the reaction Gibbs free-energy difference $\Delta\vec{\mu}$ is generally nonconserved and need not become slow in the long-wavelength limit. This raises the broader question of whether the present formalism could also be useful for applications outside biological and soft matter physics such as the counting of Nambu-Goldstone modes~\cite{hidaka2013counting,minami2018spontaneous} in non-equilibrium steady states.

Our theory is a fluctuation-response framework formulated around a well-defined steady state. Importantly, the reference steady state need not be close to equilibrium: it may be a strongly driven non-equilibrium steady state with large entropy production. 
At a strict glass transition, ergodicity is broken and relaxation to a unique steady state is no longer available. In this regime the assumptions underlying standard linear response cease to apply, and one must adopt an explicitly non-stationary description. By contrast, in the supercooled or glassy regime above the transition where relaxation is slow but finite our approach remains well suited and can be used to characterize non-Markovian transport over the relevant observation window.
Extensions beyond linear response such as higher-order response functions and genuinely nonlinear constitutive behavior are beyond the scope of the present work. We note, however, that the correlation-function evolution equation, Eq.~\eqref{eq:dpsi/dt}, already has an intrinsically nonlinear structure for correlation functions, suggesting a natural starting point for future generalizations to non-ergodic and non-linear response.

Our theory yields several experimentally testable predictions. 
(i) Length-scale crossover set by the chemo-mechanical coupling length. 
Both the stress relaxation kernel $\zeta(q,t)$ and the complex moduli $\widetilde G'(q,\omega)$ and $\widetilde G''(q,\omega)$ exhibit their strongest wave-number dependence around $q\xi_\mu\simeq 1$ (Figs.~\ref{fig:relaxation} and \ref{fig:Gs}). 
Spatially resolved microrheology (e.g.\ tracking tracer particles or applying oscillatory forcing at controlled length scales) should therefore detect a marked change in the viscoelastic response when probing length scales comparable to $\xi_\mu$. 
(ii) Metabolic control of the crossover. 
Changing the chemical driving or reaction kinetics (e.g.\ via nutrient availability, ATP levels, or inhibitors) modifies the effective chemo-mechanical coupling length $\xi_\mu$ and the chemical relaxation rate $\lambda$, thereby shifting the characteristic crossover wave number $q_c\sim 1/\xi_\mu$ and the frequency window where chemo-mechanical feedback is most apparent. 
(iii) Negative storage modulus as a finite-frequency signature of active viscoelastic memory. 
We predict that there can be a regime where the storage modulus becomes negative, $\widetilde G'(q,\omega) < 0$, at non-zero frequencies (Fig.~\ref{fig:Gs}a,c). This indicates a reactive contribution to the viscoelastic response: chemical reaction cycles generate a stress component opposing the imposed strain, producing an apparent ``negative elasticity'' at intermediate and high frequencies.
Verifying these scale- and frequency-dependent signatures would provide a direct experimental probe of Active Viscoelastic Memory.

To conclude, we have developed a general framework for describing non-equilibrium systems, starting from a fundamental identity for steady-state correlation functions. With this approach, we systematically derived the constitutive equations for complex fluids driven by internal chemical reactions. Our analysis uncovers a mechanism beyond the conventional notion of active stress: a direct contribution of chemo–mechanical coupling to the viscous transport coefficients. We believe that this first-principles framework provides a broadly applicable tool for non-equilibrium systems, and in particular for dense active matter, where the interplay of memory and activity gives rise to rich emergent phenomena.

\appendix
\renewcommand{\thefigure}{A\arabic{figure}}
\setcounter{figure}{0} 
\onecolumngrid
\section{Conventions and notations}
\subsection{Fourier and Laplace Transform  \label{sec:convention}}
Throughout this study we consider a $d$-dimensional periodic box of
volume $V$. We adopt the convention of Fourier transform for a field $F(\vec{r})$,
\begin{equation} 
\begin{split}
\label{eq:}
F(\vec{q}) = \int_V d{r}^d F(\vec{r})e^{ i \vec{q}\cdot \vec{r}} \quad ; 
\quad F(\vec{r}) = \frac{1}{V} \sum_{\vec{q}}F(\vec{q}) e^{- i \vec{q}\cdot \vec{r}} \quad.
\end{split}
\end{equation}
For a time-dependent function $F(t)$, the Laplace transform is  
\begin{equation} 
\begin{split}
\label{eq:}
\widehat{F}(s) =  \int_0^\infty dt F(t)e^{ -st} \quad; \quad F(t) =\frac{1}{2 \pi i} \lim_{T\rightarrow \infty} \int ^{a + i T}_{a - i T}e^{st}\widehat{F}(s)ds \quad,
\end{split}
\end{equation}
where $a > a_0$ such that $a_0$ is larger than the real parts of all poles in the complex $s$ plane.  
The Fourier–Laplace transform of a space–time field $F(\vec r,t)$ is written as
\begin{equation} 
\begin{split}
\label{eq:}
\widehat{F}(\vec{q},s) = \int_0^\infty dt e^{ -st} \int_V d{r}^d F(\vec{r},t)e^{ i \vec{q}\cdot \vec{r}} \quad.
\end{split}
\end{equation}
We use the following convention of the Fourier transform in time, 
\begin{equation} 
\begin{split}
\label{eq:}
\widetilde{F}(\omega) =  \int_{-\infty}^\infty dt F(t)e^{ -i\omega t}\quad;\quad F(t) = \frac{1}{2\pi}\int_{-\infty}^\infty d\omega \widetilde F(\omega)e^{ i\omega t }\quad.
\end{split}
\end{equation}

\subsection{Index free notation for tensors \label{App:tcensor_underline}}
Throughout this work we use underlines to visually distinguish the tensorial rank of objects:
a single underline denotes vectors, a double underline denotes matrices (rank-2 tensors), a triple underline denotes rank-3 tensors, and a quadruple underline denotes rank-4 tensors.
Concretely,
\begin{align}
\vec{O} && O_\alpha \ \text{(vector)},\\
\mat{P}  && P_{\alpha\beta}\ \text{(matrix / rank-2 tensor)},\\
\Tthr{Q} && Q_{\alpha\beta\gamma}\ \text{(rank-3 tensor)},\\
\Tfor{R} && R_{\alpha\beta\gamma\delta}\ \text{(rank-4 tensor)}.
\end{align}
For example, in our applications:
\begin{align}
\vec{j}(\vec q,t) &: \text{momentum density (vector)},\\
\mat{\sigma}(\vec q,t) &: \text{stress tensor (matrix)},\\
\mat{\gamma}(\vec q,t) &: \text{non-symmetrized strain-rate (matrix)},\\
\Tthr{N}^{\sigma R}(\vec q,t) &: \text{(Normalized) stress-reaction correlation tensor (rank 3)},\\
\Tfor{N}^{\sigma\sigma}(\vec q,t) &: \text{(Normalized) stress-stress correlation tensor (rank 4)}.
\end{align}

\subsection{Dyadic product \label{App:dyadic}}
Consider a column vector of dynamical variables
\(\vec A(t)=(A_1(t),A_2(t),\ldots) ^\top\) with components $A_i$, $i=1 ,\dots, M$.
We use \(^{\top}\) to denote the transposed and \((\,\cdot\,)^{\dagger}=(\,\cdot\,)^{*\top}\) for Hermitian conjugation.
The dyadic product of two vectors is denoted \(\vec a\,\vec b^{\dagger}\), with components
\([\vec a\,\vec b^{\dagger}]_{ij}=a_i\,b_j^{*}\).

Using this notation, the correlation matrix reads
\begin{equation} 
\begin{split}
\label{eq:}
\mat{\psi}(t)\;\equiv\;\big\langle \vec A(t)\vec A^{\dagger}\big\rangle \quad,
\end{split}
\end{equation}
with components 
\begin{equation} 
\begin{split}
\label{eq:}
\psi_{ij}(t)=\big\langle A_i(t)A^{*}_j\big\rangle \quad.
\end{split}
\end{equation}

\subsection{Tensor contractions \label{App:contractions}}
In the main text we adopt an index-free notation to keep expressions concise. Because this can obscure the order of tensor contractions, we summarize some of the specific contraction conventions used. 
\\ \\ 
\paragraph*{\it Momentum density conservation.} Eq.~\eqref{eq:j_conserv} can be written in index notation as  
\begin{equation} 
\begin{split}
\label{eq:dj/dt_index}
 \partial_t{j}_{\alpha }(\vec{q},t)= -i  q_\beta \sigma_{\alpha \beta}(\vec{q},t) \ .
\end{split}
\end{equation}
We use the convention that repeated indices are summed over.
\\ \\ 
\paragraph*{Strain rate.} 
The definition of the strain rate [Eq.~\eqref{eq:strate}] is expressed in index notation as 
\begin{equation} 
\begin{split}
\label{eq:}
\gamma_{\alpha \beta }(\vec{q},t) = - i q_\beta {j}_\alpha(\vec{q},t)/\rho_0 \ .  
\end{split}
\end{equation}

\paragraph*{Contractions of higher rank tensors.} 
The contraction of rank four tensor in Eq.~\eqref{eq:qNq} is 
\begin{equation}
\begin{split}
q_\gamma\,\widehat{N}_{\alpha\gamma\beta \nu}(\vec q,s)\,q_\nu
=
\big(\vec q\cdot \widehat{\Tfor N}(\vec q,s)\cdot \vec q\big)_{\alpha\beta} \quad,
\end{split}
\end{equation}
where 
\begin{equation}
\begin{split}
\widehat{N}_{\alpha\gamma \beta \nu}(\vec q,s)
\equiv
\rho_0\,
\Big\langle
\widehat{\sigma}_{\alpha\gamma}(\vec q,s)\,
\sigma^*_{\lambda\nu}(\vec q)
\Big\rangle \qty({\mat{g}}^{-1})_{\lambda \beta}\quad.
\end{split}
\end{equation}

The tensor contractions in Eq.~\eqref{eq:phi_j_mu} are defined as follows\,: 
\begin{equation}
\label{eq:qNq_and_qN_contractions_explicit}
\bigl(\vec{q}\cdot \Tfor{N}^{\sigma\sigma}(\vec{q},t)\cdot \vec{q}\bigr)_{\alpha\beta}
=
q_\mu\, N^{\sigma\sigma}_{\alpha\mu\beta \delta}(\vec{q},t)\, q_\delta
\quad,
\end{equation}
\begin{equation}
\bigl(i\,\vec{q}\cdot \Tthr{N}^{\sigma R}(\vec{q},t)\bigr)_{\alpha I}
=
i\,q_\mu\, N^{\sigma R}_{\alpha\mu,I}(\vec{q},t)
\quad,
\end{equation}
\begin{equation}
\bigl(\Tthr{N}^{R\sigma}(\vec{q},t)\cdot i\,\vec{q}\bigr)_{I\beta}
=
N^{R\sigma}_{I,\beta \delta}(\vec{q},t)\, i\,q_\delta
\quad,
\end{equation}
where

\begin{equation}
\label{eq:N_sigmasigma_def_index}
N^{\sigma\sigma}_{\alpha\mu\beta \delta}(\vec q,t)
\equiv
\rho_0\,
\Big\langle
\sigma_{\alpha\mu}(\vec q,t)\,
\sigma^*_{\lambda\delta}(\vec q)
\Big\rangle \qty(\mat{g}^{jj})^{-1}_{\lambda \beta}
\quad,
\end{equation}

\begin{equation}
\label{eq:N_sigmaR_def_index}
N^{\sigma R}_{\alpha\mu,I}(\vec q,t)
\equiv
\rho_0\,
\Big\langle
\sigma_{\alpha\mu}(\vec q,t)\,
R_J^*(\vec q)
\Big\rangle  \qty(\mat{g}^{\mu \mu})^{-1}_{JI}
\quad,
\end{equation}
and
\begin{equation}
\label{eq:N_Rsigma_def_index}
N^{R\sigma}_{I,\beta \delta}(\vec q,t)
\equiv
\rho_0\,
\Big\langle
R_I(\vec q,t)\,
\sigma^*_{\lambda\delta}(\vec q)
\Big\rangle \qty(\mat{g}^{jj})^{-1}_{\lambda \beta}
\quad.
\end{equation}

\paragraph*{Double contractions.}
We use the symbol \(:\) to denote a double contraction between a fourth-rank tensor and a second-rank tensor, defined componentwise as
\[
\qty(\Tfor{A}:\mat{B})_{\alpha\beta}\equiv A_{\alpha\beta\zeta\eta}\,B_{\zeta\eta}\quad.
\]
With this convention, the constitutive relation in Eq.~\eqref{eq:CEQ_fluid},
\[
\Tfor{\widehat{\eta}}(\vec q,s):\expval{\widehat{\mat\gamma}(\vec q,s)} \quad,
\]
means explicitly
\[
\qty(\,\Tfor{\widehat{\eta}}(\vec q,s):\expval{\widehat{\mat\gamma}(\vec q,s)})_{\alpha\beta}
=
\widehat{\eta}_{\alpha\beta\zeta\eta}(\vec q,s)\,
\expval{\widehat{\gamma}_{\zeta \eta}(\vec q,s)}\quad.
\]

Other tensor contractions are explicitly  given in the derivation in Appendices~\ref{App:CE_j} and \ref{App:CE_j_mu}.

\section{Properties of correlation functions}

\subsection{Definition of the correlation functions \label{App:correlation_def}}
We define the correlation of a set of dynamical variables  \(\vec A(t)=(A_1(t),A_2(t),\ldots)^{\top}\), using the time average,
\begin{equation} 
\begin{split}
\label{eq:C_def_1}
\expval{{A}_i(t'')  {A}^*_j(t')}_{} \equiv \lim_{\tau \rightarrow \infty} \frac{1}{\tau} \int_0^\tau {A}_i(t''+t) {A}_j^*(t'+t) dt \quad.
\end{split}
\end{equation}
Here $*$ is complex conjugation. 
Alternatively, the correlation function can be defined through a phase space average as
\begin{equation} 
\begin{split}
\label{eq:C_def_2}
&\expval{{A}_i(t'') {A}^*_j(t')}_{} \equiv  \int d{{\vec{X}_{\mkern1mu t'}}} p(\vec{X}_{\mkern1mu t'}) {A}_j^*({\vec{X}_{\mkern1mu t'}})  \int  d{\vec{X}_{\mkern1mu t''}} p({\vec{X}_{\mkern1mu t''}}|{\vec{X}_{\mkern1mu t'}}){A}_i({\vec{X}_{\mkern1mu t''}}) \quad,
\end{split}
\end{equation}
where $\vec{X}$ denotes the full set of phase space variables. The joint probability distribution, $p(\vec{X}_{\mkern1mu t'}, \vec{X}_{\mkern1mu t''}) = p(\vec{X}_{\mkern1mu t'})p({\vec{X}_{\mkern1mu t''}}|{\vec{X}_{\mkern1mu t'}})$, in Eq.~(\ref{eq:C_def_2}) is a steady state distribution.  Assuming ergodicity in the limit $\tau \rightarrow \infty$, the two definitions yield the same correlation function.

By choosing $t'' = t + s$ and $t' = s$, and invoking time translational invariance, we may set $s = 0$ and define the correlation matrix, 
\begin{equation} 
\begin{split}
\label{eq:}
\mat{\psi}(t) \equiv \expval{\vec{A}(t)  \vec{A}^\dagger} \quad,
\end{split}
\end{equation}
where we used the notation $\vec{A} \equiv \vec{A}(t=0)$. 

\subsection{Time derivatives of correlation functions \label{sec:derivative_correlation}}
We derive general properties of the steady-state correlation functions using the definition in Eq.~\eqref{eq:C_def_1}~\cite{hansen2013theory}. Taking the time derivative of the correlation function with respect to $t'$, we find
\begin{equation} 
\begin{split}
\label{}
\frac{d}{dt'}\expval{\vec{A}(t+t')  \vec{A}^\dagger(t')} = \expval{\vec{\dot{A}}(t+t') \vec{A}^\dagger(t')}_{} + \expval{\vec{A}(t+t') \vec{\dot{A}}^\dagger(t')} =0 \quad.
\end{split}
\end{equation}
Evaluating this expression at $t' = 0$ gives
\begin{equation} 
\begin{split}
\label{eq:dotAiAj=-AidotAj}
\expval{\vec{\dot{A}}(t) \vec{A}^\dagger} =- \expval{\vec{A}(t) {\vec{\dot{A}}^\dagger}} \quad.
\end{split}
\end{equation}
Choosing $t=0$ in Eq.~\eqref{eq:dotAiAj=-AidotAj} shows the static correlation matrix $\mat{\omega}$ is anti-Hermitian. 
Taking the second derivative of $\expval{\vec{A}(t+t')  \vec{A}^\dagger(t')}$ with respect to $t$ and using Eq.~(\ref{eq:dotAiAj=-AidotAj}), we obtain
\begin{equation} 
\begin{split}
\label{eq:d2/dt2psi_SI}
\frac{d^2}{dt^2}\expval{\vec{A}(t) \vec{A}^\dagger} =- \expval{\vec{\dot{A}}(t)  \vec{\dot{A}}^\dagger}_{} \quad.
\end{split}
\end{equation}

\subsection{Symmetry property of correlation functions \label{sec:symmetry}}
When the system possesses time-reversal symmetry, the correlation function satisfies
\begin{equation} 
\begin{split}
\label{eq:time_reveresal}
\expval{A_i(t)A^*_j} =\epsilon_i \epsilon_j \expval{A_i(-t)A^*_j}_{} = \epsilon_i \epsilon_j \expval{A_i A^*_j(t)}_{} = \epsilon_i \epsilon_j  \expval{A_j(t)A^*_i}^* \quad.
\end{split}
\end{equation}
Here, $\epsilon_i (\epsilon_j)$ denotes the time-reversal signature of $A_i (A_j)$. In deriving the above, we have used time translation invariance. This relation implies that, at equilibrium, equal-time ($t = 0$) correlation functions between variables with opposite time-reversal signatures vanish.

When the dynamical variables depend on space and the system exhibits inversion symmetry, the correlation function in Fourier space satisfies
\begin{equation} 
\begin{split}
\label{eq:space_reveresal}
\expval{A_i(\vec{q},t)A^*_j(\vec{q})} =p_i p_j \expval{A_i(-\vec{q},t)A^*_j(-\vec{q})}_{} \quad.
\end{split}
\end{equation}
Here, $p_i$ ($p_j$) denotes the inversion parity of $A_i$ ($A_j$) under the transformation $\vec{q} \rightarrow -\vec{q}$\,.

When the dynamical variables exhibit both time-reversal and spatial inversion symmetry, the correlation function in Fourier space satisfies
\begin{equation} 
\begin{split}
\label{eq:time_reveresal_space}
\expval{A_i(\vec{q},t)A^*_j(\vec{q})} = \epsilon_i \epsilon_j \expval{A_j(\vec{q},t)A^*_i(\vec{q})}^* =\epsilon_i \epsilon_j \expval{A_j(-\vec{q},t)A^*_i(-\vec{q})} =p_i p_j\epsilon_i \epsilon_j \expval{A_j(\vec{q},t)A^*_i(\vec{q})} \quad.
\end{split}
\end{equation}

\section{Fluctuation-Response relations and examples}
\subsection{Derivation of the fluctuation-response relation \label{App:correlation_deriv}}
To derive the equation for the average response of dynamical variables,  we use the definition of the correlation function, Eq.~(\ref{eq:C_def_2}),  applied to Eq.~(\ref{eq:dpsi/dt}).
In element-wise notation, we have
\begin{equation} 
\label{eq:dA/dt_Appc}
\frac{d}{dt} \langle A_i(t)A_j^* \rangle
=
-
\sum_m \int_0^t dt'\,
K_{im}(t-t')\langle A_m(t')A_j^* \rangle
\quad .
\end{equation}

The mean of the dynamical variables can be written in phase-space form as
\begin{equation}
\label{eq:meanA}
\langle A_i(t)\rangle
=
\int d\vec X_{\mkern1mu 0}\,p(\vec X_{\mkern1mu 0})
\int d\vec X_{\mkern1mu  t}\,p(\vec X_{\mkern1mu  t}|\vec X_{\mkern1mu  0})A_i(\vec X_{\mkern1mu  t})
\quad .
\end{equation}

We consider an infinitesimal perturbation of the initial steady-state
probability density,
\begin{equation}
\label{eq:C1_pert_distribution}
p_h(\vec X_{\mkern1mu  0})
=
p(\vec X_{\mkern1mu  0})+\delta p_h(\vec X_{\mkern1mu  0})
\quad ,
\end{equation}
and expand its relative deviation in terms of the chosen dynamical
variables,
\begin{equation}
\delta p_h(\vec X_{\mkern1mu  0})
=
p(\vec X_{\mkern1mu  0})
\sum_j h_j A_j^*(\vec X_{\mkern1mu  0})
\quad ,
\end{equation}
where $\vec h$ denotes a set of small fields conjugate to the variables
$\vec A^*(\vec X_{\mkern1mu  0})$.


The corresponding linear change of the mean is
\begin{align}
\label{eq:meanA_delta}
\delta \langle A_i(t)\rangle
&=
\int d\vec X_{\mkern1mu  0}\,\delta p(\vec X_{\mkern1mu 0})
\int d\vec X_{\mkern1mu  t}\,p(\vec X_{\mkern1mu  t}|\vec X_{\mkern1mu  0})A_i(\vec X_{\mkern1mu  t})
\\
&=
 \sum_j h_j
\int d\vec X_{\mkern1mu  0}\,p(\vec X_{\mkern1mu  0})
\int d\vec X_{\mkern1mu  t}\,p(\vec X_{\mkern1mu  t}|\vec X_{\mkern1mu  0})
A_i(\vec X_{\mkern1mu  t}) A_j^*(\vec X_{\mkern1mu  0})
\nonumber \\
&=
 \sum_j h_j
\langle A_i(t)A_j^*\rangle
\quad . \nonumber
\end{align}

Multiplying Eq.~\eqref{eq:dA/dt_Appc} by \(h_j\), summing over \(j\), and using Eq.~\eqref{eq:meanA_delta}, we obtain
\begin{equation}
\frac{d}{dt}\delta \langle A_i(t)\rangle
=
-
\sum_m \int_0^t dt'\,
K_{im}(t-t')\,
\delta \langle A_m(t')\rangle
\quad .
\end{equation}
Thus the relaxation of the linear perturbation of the mean is governed 
by the same memory kernel that appears in the correlation equation.
The memory kernel $K_{im}(t)$ is evaluated in the reference unperturbed 
steady state, while $\delta\langle A_i(t)\rangle$ denotes the linear response 
of the mean to the perturbed initial ensemble. In the main text, we omit 
$\delta$ for notational simplicity.

\subsection{Fluctuation-Dissipation theorem in non-equilibrium steady states \label{sec:FDT}}
In this section, we derive the fluctuation-response relations for
non-equilibrium steady states using Eq.~\eqref{eq:dA/dt}. 
We first derive the relation for a kinematic perturbation, yielding a form analogous to the Harada-Sasa relation~\cite{harada2005equality}. Next, we derive the form corresponding to the direct generalization of the fluctuation-dissipation theorem to non-equilibrium steady states. We emphasize that these derived relations are fundamentally multivariable in nature. 
The connection between the response functions and dissipation is established in Appendix~\ref{sec:response-dissipation}.

\subsubsection*{Kinematic perturbation: Harada-Sasa form}

We introduce a kinematic perturbation $\vec{\lambda}_{\,\rm kin}(t)$ as an
additive source $\mat g\cdot\vec{\lambda}_{\, \rm kin}(t)$ in the mean dynamics.
With this parametrization, an impulsive perturbation
$\vec{\lambda}(t)=\vec h\,\delta(t)$ produces the initial change $\langle \vec{A}(t=0^+) \rangle=\mat g\cdot\vec h$, identical to the perturbation introduced
in Appendix~\ref{App:correlation_deriv}.  The mean dynamics are then written as
\begin{equation} 
\begin{split}
\label{eq:dA_f}
\frac{d}{dt}  \expval{\vec{A}(t)} = - \int_0^t  dt'  \mat{K}(t-t') \cdot \expval{\vec{A}(t')} + \mat{g}\cdot\vec{\lambda}_{\,\rm kin}(t) \quad.
\end{split}
\end{equation}
Taking the one-sided Laplace transform with $\expval{\vec{A}(t=0)} = 0$, we obtain:
\begin{equation} 
\begin{split}
\label{eq:A_f}
\langle \vec{\widehat{A}}(s)\rangle =\qty( s \mat{I} + \mat{\widehat{K}}(s) ) ^{-1} \cdot \mat{g} \cdot \widehat{\vec{\lambda}}_{\,\rm kin}(s) \quad.
\end{split}
\end{equation}
Equation~(\ref{eq:A_f}) formally defines the kinematic response matrix in Laplace space: 
\begin{equation} 
\begin{split}
\label{eq:Gamma_plus_def}
\mat{\widehat{\Gamma}}(s) \equiv   \qty( s \mat{I} + \mat{\widehat{K}}(s) ) ^{-1}\cdot \mat{g} \quad, 
\end{split}
\end{equation}
where $\mat{\Gamma}(t)$ represents the causal response matrix, defined forward in time ($t\geq0$).

Using the Laplace-space relation corresponding to Eq.~\eqref{eq:dpsi/dt}: $s\mat{\widehat{\psi}}-\mat{g}=-\mat{\widehat{K}}\cdot \mat{\widehat{\psi}}$, the memory kernel $\mat{\widehat{K}}(s)$ can be expanded as:
\begin{equation} 
\begin{split}
\label{eq:Kernel_expansion}
\mat{\widehat{K}}(s) =-s \mat{I} + \mat{g}\cdot \qty(\mat{\widehat{\psi}}(s))^{-1}  \quad . 
\end{split}
\end{equation}
Substituting this result into Eq.~\eqref{eq:Gamma_plus_def}, we obtain a direct relation between the fluctuations and the response in Laplace space:
\begin{equation} 
\begin{split}
\label{eq:FDT_motility}
 \mat{\widehat{\psi}}(s) =  \mat{\widehat{\Gamma}}(s) \quad.
\end{split}
\end{equation}
In the real-time domain, this enforces the causality of the response:
\begin{equation}
\label{eq:FDT_motility_positve_time}
  \mat{\psi}(t)  \theta(t) =  \mat{\Gamma}(t) \quad,
\end{equation}
where $\theta(t)$ is the Heaviside step function. 

To bridge this to the full spectral domain, we decompose the two-sided dynamical correlation matrix $\mat{\psi}(t)$ into positive and negative time components: 
\begin{equation}
\begin{split}
    \mat{\psi}(t) &= \mat{\psi}(t)\theta(t) + \mat{\psi}(t)\theta(-t)  \\[1mm]
     & =\mat{\psi}(t)\theta(t) + \mat{\psi}(-t)^\dagger\theta(-t) \\[1mm]
     &= \mat{\Gamma}(t)\ +  \mat{\Gamma}(-t)^\dagger\quad,
\end{split}
\end{equation}
where we have applied the stationary property of the correlation matrix, $\mat{\psi}(-t)=\mat{\psi}(t)^\dagger$. Moving to Fourier space, this time-domain reconstruction yields:
\begin{equation}
\label{eq:psi_Gamma}
     \mat{\widetilde{\psi}}(\omega) = \mat{\widetilde{\Gamma}}(\omega) +    \mat{\widetilde{\Gamma}}(\omega)^\dagger\quad,
\end{equation}
where we used the Fourier transform property $\mathcal{F}[\mat{\Gamma}(-t)^\dagger](\omega)=\mathcal{F}[\mat{\Gamma}(t)](\omega)^\dagger$. Equation~\eqref{eq:psi_Gamma} is a fluctuation-response relation written in terms of the Hermitian conjugate. To connect with the conventional fluctuation-response relation and to display its non-equilibrium violation more explicitly, we now rewrite Eq.~\eqref{eq:psi_Gamma} in terms of the transpose-symmetric and transpose-antisymmetric parts of the response.

We decomposed the response matrix into its symmetric (S) and antisymmetric (A) parts:
\begin{equation}
\label{eq:Gamma_S_A}
\mat{\Gamma}^{\rm S}(t)  \equiv \frac{1}{2} \qty(\mat{\Gamma}(t)  +  \mat{\Gamma}(t)^\top  )  \quad; \quad \mat{\Gamma}^{\rm A}(t)  \equiv \frac{1}{2} \qty(\mat{\Gamma}(t) -   \mat{\Gamma}(t)^\top  ) \quad.
\end{equation}
Applying this decomposition to the spectral correlation gives:
\begin{equation}
\begin{split}
     \mat{\widetilde{\psi}}(\omega) &= \qty[ \qty(\mat{\widetilde{\Gamma}}^{\rm S}(\omega) +\mat{\widetilde{\Gamma}}^{\rm A}(\omega) ) ] + \qty[ \qty(\mat{\widetilde{\Gamma}}^{\rm S}(\omega) +\mat{\widetilde{\Gamma}}^{\rm A}(\omega))]^\dagger \\[2mm]
     &=2 {\rm Re}\qty[\mat{\widetilde{\Gamma}}^{\rm S}(\omega) ] +2\,i\, {\rm Im}\qty[\mat{\widetilde{\Gamma}}^{\rm A}(\omega) ] \quad.
\end{split}
\end{equation}
This establishes the generalized kinematic fluctuation-response relation:
\begin{equation} 
\begin{split}
\label{eq:Harada-sasa_like_physical}
\mat{\widetilde{\psi}}(\omega) - 2 \,\mathrm{Re}\qty[\mat{\widetilde \Gamma}^{\rm S}(\omega)]=2\,i \,\mathrm{Im}\qty[\mat{\widetilde \Gamma}^{\rm A}(\omega)] \quad.
\end{split}
\end{equation}

In thermal equilibrium, for variables without space-dependence with identical time-reversal
signatures, time-reversal symmetry implies the reciprocity condition $\mat{\psi}(t)=\mat{\psi}(t)^{\top}$, or equivalently $\mat{\widetilde{\psi}}(\omega)=\mat{\widetilde{\psi}}(\omega)^{\top}$; see
Appendix~\ref{sec:symmetry}. This condition becomes the vanishing of the antisymmetric part,
$\mat{\widetilde{\Gamma}}^{\rm A}(\omega)=\mat{0}$. A nonzero $\mat{\widetilde{\Gamma}}^{\rm A}(\omega)$ therefore signals a breakdown of this reciprocity and provides a signature of broken detailed balance for variables with identical time-reversal signatures. The left-hand side of Eq.~\eqref{eq:Harada-sasa_like_physical}
is analogous to the Harada-Sasa relation~\cite{harada2005equality}, quantifying
the violation of the equilibrium fluctuation-dissipation theorem, while the
right-hand side identifies this violation with the emergence of non-reciprocal
response associated with circulating probability currents. For variables with different time-reversal signatures, $\mat{\widetilde{\Gamma}}^{\rm A}(\omega)$ represents equilibrium circulating currents.

\subsubsection*{State perturbation: conventional fluctuation-response form \label{sec:state_pert}}

We introduce a state perturbation $\vec{\lambda}_{\, \rm st}(t)$ such that
$\mat{g}\cdot\vec{\lambda}_{\,\rm st}(t)$ represents the
change of the mean associated with the perturbed
state (Appendix~\ref{App:correlation_deriv}). The mean dynamics are then written as
\begin{equation} 
\begin{split}
\label{eq:dA_h}
\frac{d}{dt}  \langle\vec{A}(t) \rangle   = - \int_0^t dt'  \mat{K}(t-t') \cdot \qty(\expval{\vec{A}(t')} - \mat{g}\cdot\vec{\lambda}_{\,\rm st}(t')) \quad.
\end{split}
\end{equation}
This implies the static response
$\langle A\rangle=\mat{g}\cdot 
\vec\lambda$.
Assuming the system was unperturbed at $t=0$ ($\expval{\vec{A}(t=0)} = 0$), the Laplace transform yields:
\begin{equation}
\label{eq:response_state}
    \expval{ \vec{\widehat{A}}(s)} = (s \mat{I} + \mat{\widehat{K}}(s))^{-1} \cdot \mat{\widehat{K}}(s) \cdot \mat{g}\cdot\vec{\widehat{\lambda}}_{\,\rm st}(s) \quad.
\end{equation}
This  defines the response matrix to a state perturbation in Laplace space:
\begin{equation}
    \mat{\widehat{\chi}}(s) =  \qty(s \mat{I} + \mat{\widehat{K}}(s))^{-1} \cdot \mat{\widehat{K}}(s) \cdot \mat{g}\quad.
    \label{eq:chi_def}
\end{equation}
Here, $\mat{\chi}(t)$ represents the causal response matrix, defined forward in time ($t \geq 0$). 

We obtain the relationship between the kinematic perturbation response matrix $\mat{\widehat{\Gamma}}(s)$ [Eq.~\eqref{eq:Gamma_plus_def}] and the state perturbation response matrix $\mat{\widehat{\chi}}(s)$:
\begin{equation}
    \mat{\widehat{\chi}}(s) = \mat{g}-s \, \mat{\widehat{\Gamma}}(s)\quad.
\end{equation}
This becomes in Fourier space ($s \to i\omega$): 
\begin{equation}
    \mat{\widetilde{\Gamma}}(\omega)  = -\frac{i}{\omega}\qty(\mat{g} - \mat{\widetilde{\chi}}(\omega)) \quad.
\end{equation}
Substituting this into the correlation relation in Fourier space, $\mat{\widetilde{\psi}}(\omega) = \mat{\widetilde{\Gamma}}(\omega)+   \mat{\widetilde{\Gamma}}(\omega)^{\dagger}$ [Eq.~\eqref{eq:psi_Gamma}], we obtain:
\begin{equation}
\label{eq:psi_chi}
    \mat{\widetilde{\psi}}(\omega) = \frac{i}{\omega}\qty(\mat{\widetilde{\chi}}(\omega) -  \mat{\widetilde{\chi}}(\omega)^{ \dagger}) \quad,
\end{equation}
where we have used $\mat{g}=\mat{g}^\dagger$.
Equation~\eqref{eq:psi_chi} is a fluctuation-response relation written in terms of the Hermitian conjugate for state perturbation. To connect with the conventional fluctuation-response relation and to display its non-equilibrium violation more explicitly, we now rewrite Eq.~\eqref{eq:psi_chi} in terms of the transpose-symmetric and transpose-antisymmetric parts of the response.

We decompose the response matrix into its symmetric (S) and antisymmetric (A) parts:
\begin{equation}
\label{eq:cai_S_A}
\mat{\chi}^{\rm S}(t)  \equiv \frac{1}{2} \qty(\mat{\chi}(t)  +    \mat{\chi}(t)^\top   ) \quad; \quad \mat{\chi}^{\rm A}(t)  \equiv \frac{1}{2} \qty(\mat{\chi}(t)  -   \mat{\chi}(t)^\top ) \quad.
\end{equation}
In Fourier space, we obtain
\begin{equation}
\label{eq:chi_dagger_decomp}
    \mat{\widetilde{\chi}}(\omega)
    -
    \mat{\widetilde{\chi}}(\omega)^\dagger
    =
    2i\,{\rm Im}
    \left[
        \mat{\widetilde{\chi}}^{\mkern1mu  \rm S}(\omega)
    \right]
    +
    2\,{\rm Re}
    \left[
        \mat{\widetilde{\chi}}^{\mkern1mu  \rm A}(\omega)
    \right]
    \quad.
\end{equation}
Substituting Eq.~\eqref{eq:chi_dagger_decomp} into
Eq.~\eqref{eq:psi_chi}, we obtain the generalized fluctuation-dissipation
relation,
\begin{equation}
\label{eq:FDT_chi_physical}
    \mat{\widetilde{\psi}}(\omega)
    =
    -
    \frac{2}{\omega}
    {\rm Im}
    \left[
        \mat{\widetilde{\chi}}^{\mkern1mu  \rm S}(\omega)
    \right]+
    \frac{2i}{\omega}
    {\rm Re}
    \left[
        \mat{\widetilde{\chi}}^{\mkern1mu  \rm A}(\omega)
    \right]
    \quad.
\end{equation}

Equation~\eqref{eq:FDT_chi_physical} is the state-perturbation form of the
generalized fluctuation-dissipation relation in a non-equilibrium steady
state. 
For variables without space dependence with identical time-reversal signatures, the antisymmetric part of the response matrix, $\mat{\chi}^{\rm A}(t)$ captures the non-equilibrium response, which is associated with nonequilibrium steady-state probability currents. For variables with different time-reversal signatures, the antisymmetric part of the response, $\mat{\chi}^{\rm A}(t)$, captures the equilibrium reactive coupling.

One can recover the equilibrium fluctuation-dissipation theorem as follows.
At thermal equilibrium, the response matrix is related to the linear response $\mat{R}^{\rm eq}_{\mkern1mu h}(t)$ to a thermodynamic field
$\vec h$ conjugate to $\vec A$ by
\begin{equation}
    \mat{R}^{\rm eq}_{\mkern1mu h}(t)
    =
    \frac{1}{k_BT}\,\mat{\chi}(t)
    \quad.
\end{equation}
For real-valued variables with identical time-reversal signatures, we have
$\mat{\chi}^{\rm A}(t)=\mat{0}$ and
$\mat{\chi}^{\rm S}(t)=\mat{\chi}(t)$. Therefore,
Eq.~\eqref{eq:FDT_chi_physical} reduces to
\begin{equation}
    \mat{\widetilde{\psi}}(\omega)
    =
    -\frac{2k_BT}{\omega}
    \operatorname{Im}
    \left[
        \mat{\widetilde R}^{\mkern1mu  \rm eq}_{\mkern1mu  h}(\omega)
    \right]
    \quad,
\end{equation}
which is the conventional equilibrium fluctuation-dissipation theorem for the Fourier-transform convention used here.


\subsection{From response to dissipation \label{sec:response-dissipation}}

While the fluctuation-response relations, Eqs.~\eqref{eq:FDT_chi_physical_main} and \eqref{eq:FDT_chi_physical}, relate the spontaneous fluctuations of the system to its response to generalized perturbations, they do not by themselves specify the connection between the response functions and dissipation. In general, the generalized fields $\vec{\lambda}_{\,\rm kin}$ [Eq.~\eqref{eq:dA_f}] and $\vec{\lambda}_{\,\rm st}$ [Eq.~\eqref{eq:dA_h}] need not coincide with the force $\vec{f}$ that is energetically conjugate to $\vec{A}$, such that $\vec{A}^{\top}\cdot\vec{f}$ has units of energy. To clarify the bridge between response and dissipation, we consider a perturbation by the energy-conjugate force
$\vec{f}$,
\begin{equation}
\begin{split}
\label{eq:dA_f_diss}
\frac{d}{dt}\expval{\vec{A}(t)}
=
- \int_0^t dt'\,
{\mat{K}}(t-t')
\cdot
\expval{\vec{A}(t')}
+
{\mat{Q}}
\cdot
\vec{f}(t)
\quad ,
\end{split}
\end{equation}
where the constant matrix ${\mat{Q}}$ specifies how the
physical force enters the dynamics. The response function to $\vec{f}$ in Fourier space is given by
\begin{equation}
    \widetilde{\mat{R}}(\omega) = \qty(i \omega \mat{I} + \widetilde{\mat{K}}(\omega))^{-1} \cdot \mat{Q} \quad.
    \label{eq:R_def}
\end{equation}

Using the harmonic components
\begin{equation}
\vec{f}(t)
=
{\rm Re}\qty[
\vec{f}_{\mkern1mu\omega} e^{i\omega t}
]
\quad ; \quad
\expval{\vec{A}(t)}
=
{\rm Re}\qty[
{\expval{\vec{A}}}_{\omega} e^{i\omega t}
]
\quad ,
\end{equation}
where
\begin{equation}
{\expval{\vec{A}}}_{\omega}
=
\widetilde{\mat{R}}(\omega)
\cdot
\vec{f}_{\mkern1mu\omega}
\quad ,
\end{equation}
the cycle-averaged input power at frequency $\omega$ is given by
\begin{equation}
\begin{split}
\bar{P}(\omega)\equiv
\frac{1}{T_\omega}
\int_0^{T_\omega} dt\,
\vec{f}(t)^\top
\cdot
\partial_t\expval{\vec{A}(t)}=
-\frac{\omega}{2}\,
{\rm Im}\qty[
\vec{f}_{\mkern1mu\omega}^{\dagger}
\cdot
\widetilde{\mat{R}}(\omega)
\cdot
\vec{f}_{\mkern1mu\omega}
]
\quad ,
\end{split}
\end{equation}
where $T_\omega=2\pi/\omega$.

We decompose the response matrix into its Hermitian and
anti-Hermitian parts,
\begin{equation}
\widetilde{\mat{R}}(\omega)
=
\widetilde{\mat{H}}_R(\omega)
-
i\,\widetilde{\mat{D}}_R(\omega)
\quad ,
\end{equation}
where
\begin{equation}
\widetilde{\mat{H}}_R
\equiv
\frac{1}{2}
\qty(
\widetilde{\mat{R}}
+
\widetilde{\mat{R}}^\dagger
)
\quad ; \quad
\widetilde{\mat{D}}_R
\equiv
\frac{i}{2}
\qty(
\widetilde{\mat{R}}
-
\widetilde{\mat{R}}^\dagger
)
\quad .
\end{equation}
Since both $\widetilde{\mat{H}}_R$ and
$\widetilde{\mat{D}}_R$ are Hermitian, the cycle-averaged input power
can therefore be written as
\begin{equation}
\bar{P}(\omega)
=
\frac{\omega}{2}\,
\vec{f}_{\mkern1mu\omega}^{\dagger}
\cdot
\widetilde{\mat{D}}_R(\omega)
\cdot
\vec{f}_{\mkern1mu\omega}
\quad .
\label{eq:Power}
\end{equation}
Thus, when $\vec{f}$ is energetically conjugated to $\vec{A}$,
the eigenvalues of $\widetilde{\mat{D}}_R(\omega)$ directly determine
the sign of the cycle-averaged power.

We now express the cycle-averaged input power in
Eq.~\eqref{eq:Power} in terms of the generalized fields
$\vec{\lambda}_{\,\rm kin}$ and $\vec{\lambda}_{\,\rm st}$.
Their harmonic components are written as
\begin{equation}
\vec{\lambda}_{\,\rm kin}(t)
=
{\rm Re}\qty[
\vec{\lambda}^{\rm kin}_{\mkern1mu\omega}
e^{i\omega t}
]
\quad ; \quad
\vec{\lambda}_{\,\rm st}(t)
=
{\rm Re}\qty[
\vec{\lambda}^{\rm st}_{\mkern1mu\omega}
e^{i\omega t}
]
\quad .
\end{equation}

The generalized fields are related to the energy-conjugate force
$\vec{f}$ through
\begin{equation}
\vec{f}_{\mkern1mu\omega}
=
\mat{Q}^{-1}
\cdot
\mat{g}
\cdot
\vec{\lambda}^{\rm kin}_{\mkern1mu\omega}
=
\mat{Q}^{-1}
\cdot
\widetilde{\mat{K}}(\omega)
\cdot
\mat{g}
\cdot
\vec{\lambda}^{\rm st}_{\mkern1mu\omega}
\quad ,
\end{equation}
see Eqs.~\eqref{eq:dA_f} and \eqref{eq:dA_h}.
From Eqs.~\eqref{eq:Gamma_plus_def}, \eqref{eq:chi_def}, and
\eqref{eq:R_def}, we obtain
\begin{equation}
\widetilde{\mat{R}}(\omega)
=
\widetilde{\mat{\Gamma}}(\omega)
\cdot
\mat{g}^{-1}
\cdot
\mat{Q}
=
\widetilde{\mat{\chi}}(\omega)
\cdot
\mat{g}^{-1}
\cdot
\widetilde{\mat{K}}(\omega)^{-1}
\cdot
\mat{Q}
\quad .
\end{equation}

Substituting these relations into Eq.~\eqref{eq:Power}, we obtain
\begin{equation}
\bar{P}(\omega)
=
\frac{\omega}{2}
\left(
\vec{\lambda}^{\rm kin}_{\mkern1mu\omega}
\right)^\dagger
\cdot
\widetilde{\mat{S}}_{\Gamma}(\omega)
\cdot
\vec{\lambda}^{\rm kin}_{\mkern1mu\omega}
=
\frac{\omega}{2}
\left(
\vec{\lambda}^{\rm st}_{\mkern1mu\omega}
\right)^\dagger
\cdot
\widetilde{\mat{S}}_{\chi}(\omega)
\cdot
\vec{\lambda}^{\rm st}_{\mkern1mu\omega}
\quad ,
\end{equation}
where
\begin{equation}
\widetilde{\mat{S}}_{\Gamma}(\omega)
\equiv
\frac{i}{2}
\qty[
\mat{g}
\cdot
\left(\mat{Q}^{-1}\right)^\dagger
\cdot
\widetilde{\mat{\Gamma}}(\omega)
-
\widetilde{\mat{\Gamma}}^\dagger(\omega)
\cdot
\mat{Q}^{-1}
\cdot
\mat{g}
]
\quad ,
\end{equation}
and
\begin{equation}
\widetilde{\mat{S}}_{\chi}(\omega)
\equiv
\frac{i}{2}
\qty[
\mat{g}
\cdot
\widetilde{\mat{K}}^\dagger(\omega)
\cdot
\left(\mat{Q}^{-1}\right)^\dagger
\cdot
\widetilde{\mat{\chi}}(\omega)
-
\widetilde{\mat{\chi}}^\dagger(\omega)
\cdot
\mat{Q}^{-1}
\cdot
\widetilde{\mat{K}}(\omega)
\cdot
\mat{g}
]
\quad .
\end{equation}
Thus, in general, the spectra of the anti-Hermitian parts of
$\widetilde{\mat{\Gamma}}(\omega)$ and
$\widetilde{\mat{\chi}}(\omega)$ do not directly determine the sign of the power. Rather, the relevant dissipative matrices are $\widetilde{\mat{S}}_{\Gamma}$ and
$\widetilde{\mat{S}}_{\chi}$.

\subsection{Example 1 : 2D Overdamped Rotating Harmonic Oscillator \label{App:Harmonic}}
In this section, we illustrate the method presented in Section~\ref{sec:correlation} using the model of a two-dimensional rotating harmonic oscillator. We consider the stochastic differential equation:
\begin{equation} 
\begin{split}
\label{eq:dr}
d \vec{r}(t) = (-\kappa\, \mat{I}+\Omega\, \mat{J} ) \cdot \vec{r}(t) \,dt+ \sqrt{2D}\,\mat{I} \cdot d \vec{W_t} \quad,
\end{split}
\end{equation}
where $\vec{W_t}$ represents a Wiener process. The matrices $\mat{I}$ and $\mat{J}$ are the symmetric identity matrix and the antisymmetric rotation matrix, respectively, given by 
\begin{equation} 
\begin{split}
\label{eq:}
\mat{I} = 
\begin{pmatrix}
1 & 0 \\
0 & 1 
\end{pmatrix} \quad ;\quad
\mat{J} = 
\begin{pmatrix}
0 & -1 \\
1 & 0 
\end{pmatrix}
\end{split}\quad.
\end{equation}
Here, $\kappa$, $\Omega$, and $D>0$ are the trap stiffness, rotation frequency, and diffusion constant, respectively. 
For notational convenience, we define:
\begin{equation} 
\begin{split}
\label{}
\mat{L} \equiv -\kappa\, \mat{I}+\Omega\, \mat{J} \quad ; \quad \mat{B} \equiv \sqrt{2D} \mat{I} \quad.
\end{split}
\end{equation}

\textit{Computation of the static correlation matrix $\mat{g}$\,.}
First, we compute the equal-time (static) correlation function. Using Itô's Lemma, we obtain:
\begin{equation} 
\begin{split}
\label{}
d \expval{\vec{r}(t)\, \vec{r}(t)^\top} &=  \expval{d\vec{r}(t)\, \vec{r}(t)^\top} + \expval{\vec{r}(t)\, d\vec{r}(t)^\top}+ \expval{d\vec{r}(t)\, d\vec{r}(t)^\top}\\[2mm]
&=\mat{L} \cdot  \expval{\vec{r}(t)\, \vec{r}(t)^\top}dt +  \expval{\vec{r}(t)\, \vec{r}(t)^\top} \cdot \mat{L}^\top dt + \mat{B} \cdot \mat{B}^\top dt \quad .
\end{split}
\end{equation}
Therefore, the time evolution of the correlation matrix is given by:
\begin{equation} 
\begin{split}
\label{eq:dt_rr}
\frac{d}{dt}\expval{\vec{r}(t)\, \vec{r}(t)^\top}
&=\mat{L} \cdot  \expval{\vec{r}(t)\, \vec{r}(t)^\top} +  \expval{\vec{r}(t)\, \vec{r}(t)^\top} \cdot \mat{L}^\top  + \mat{B} \cdot \mat{B}^\top =0 \quad ,
\end{split}
\end{equation}
where we used the property that for a stationary state, the time derivative of the equal-time correlation function vanishes. Therefore Eq.~\eqref{eq:dt_rr} leads to the isotropic static correlation,
\begin{equation} 
\begin{split}
\label{}
 \mat{g}  \equiv \expval{\vec{r}(t)\, \vec{r}(t)^\top}
&=\frac{D}{\kappa}\, \mat{I}\quad .
\end{split}
\end{equation}

\textit{Dynamical correlation matrix $\mat{\psi}(\tau)$\,.}
Using Eq.~\eqref{eq:dr}, the equation of motion for the dynamical correlation function is
\begin{equation} 
\begin{split}
\label{}
\frac{d}{d\tau} \expval{\vec{r}(t+\tau)\, \vec{r}(t)^\top} =\mat{L} \cdot  \expval{\vec{r}(t+\tau)\, \vec{r}(t)^\top}\quad ,
\end{split}
\end{equation}
which leads to the solution: 
\begin{equation} 
\begin{split}
\label{}
\expval{\vec{r}(t+\tau)\, \vec{r}(t)^\top} = \exp(\mat{L}\, \tau) \cdot  \expval{\vec{r}(t)\, \vec{r}(t)^\top} \quad .
\end{split}
\end{equation}
Thus, the dynamical correlation function $\mat{\psi}(\tau) \equiv \expval{\vec{r}(t+\tau)\, \vec{r}(t)^\top}$ is given by
\begin{equation} 
\begin{split}
\label{eq:psi_exp}
\mat{\psi}(\tau)
= \exp(\mat{L}\, \tau) \cdot \mat{g} \quad(\tau \geq 0) \quad .
\end{split}
\end{equation}
We note that time-translational invariance leads to the relation:
\begin{equation} 
\begin{split}
\label{}
\mat{\psi}(-\tau)
= \mat{\psi}(\tau)^\top \quad .
\end{split}
\end{equation}
Furthermore, differentiating with respect to $\tau$ gives
\begin{equation} 
\begin{split}
\label{eq:equality_psi_top}
\mat{\dot{\psi}}(-\tau)
= -\mat{\dot{\psi}}(\tau)^\top \quad .
\end{split}
\end{equation}

\textit{Computation of the reactive frequency matrix $\mat{\omega}$\,.}
The static correlation matrix $\mat{\omega}$ is computed symmetrically at $\tau=0$ to account for the discontinuity of the derivative at the time origin: 
\begin{equation} 
\begin{split}
\label{}
\mat{\omega} &\equiv \frac{1}{2}\qty(\mat{\dot{\psi}}(\tau=0^+) +  \mat{\dot{\psi}}(\tau =0^-)) \quad . 
\end{split}
\end{equation}
Using Eq.~\eqref{eq:equality_psi_top} and Eq.~\eqref{eq:psi_exp}, this can be rewritten as: 
\begin{equation} 
\begin{split}
\label{}
\mat{\omega} &= \frac{1}{2}\qty(\mat{\dot{\psi}}(\tau=0^+) -  \mat{\dot{\psi}}^\top(\tau=0^+) ) \\
& = \frac{1}{2}  \qty(\mat{L}\cdot \mat{g} - \qty(\mat{L}\cdot \mat{g})^\top   )\\
& = \frac{\Omega D}{\kappa} \mat{J}\quad .
\end{split}
\end{equation}
Thus, we find that $\mat{\omega}$ is an antisymmetric matrix proportional to $\Omega$: 
\begin{equation} 
\begin{split}
\label{}
\mat{\omega} &= \frac{\Omega D}{\kappa} \mat{J}\quad .
\end{split}
\end{equation}
We consider, in this example, the time reversal signatures of the dynamical variables in $\vec{r}=(x,y)^\top$ to be the same, therefore $\mat{\omega}$ vanishes when the system is in equilibrium ($\Omega= 0$); $\mat{\omega}$ describes a non-reciprocal coupling out of equilibrium.

\textit{Computation of the kernel, $\mat{\widehat{K}}(s)$\,.}
In Laplace space, we have the following relationship for a steady state: 
\begin{equation} 
\begin{split}
\label{}
\Delta \widehat{\phi}(s) \equiv  \widehat{\phi}(s)- \mat{\omega} = -s^2 \mat{\widehat{\psi}}(s) +s \mat{g} \quad . 
\end{split}
\end{equation}
By substituting $\Delta \widehat{\phi}(s)$ into the expression for the kernel $\mat{\widehat{K}}(s)$ [Eq.~\eqref{eq:K}], we obtain:
\begin{equation} 
\begin{split}
\label{}
\mat{\widehat{K}}(s) &= \qty [\mat{I} -  \frac{1}{s} \qty(\Delta \mat{\widehat{\phi}}(s)\cdot \mat{g}^{-1} ) ]^{-1}  \cdot  \qty(\Delta \mat{\widehat{\phi}}(s)\cdot \mat{g}^{-1}) \\
& =-\qty( -\kappa \,  \mat{I} + \Omega \, \mat{J})
\quad . 
\end{split}
\end{equation}
Using Eq.~\eqref{eq:dA/dt} from the main text then leads to:
\begin{equation} 
\begin{split}
\label{}
\frac{d}{dt}\expval{\vec{r}(t)} = \qty( -\kappa \,  \mat{I} + \Omega \, \mat{J}) \cdot  \expval{\vec{r}(t)} \quad,
\end{split}
\end{equation}
which is consistent with the deterministic part of Eq.~\eqref{eq:dr}.

Therefore, although the correlation functions are generated by the underlying dynamics, one does not need to assume or postulate an explicit microscopic equation of motion to extract the linear coefficients. 
Once the relevant time-correlation functions are known (from simulations or experiments), the coefficient matrix can be reconstructed directly through our correlation identities. 

\subsection{Example 2 : 1D Underdamped Harmonic Oscillator \label{App:UDHO}}
In this section, we illustrate the method presented in Section~\ref{sec:correlation} using one-dimensional underdamped harmonic oscillator driven by additive noise. We consider the stochastic differential equation:
\begin{equation}
\begin{split}
\label{eq:udho_sde}
d x(t) &= v(t)\,dt \quad,\\
d v(t) &= \qty(-\kappa\, x(t)-\eta\, v(t))\,dt + \sqrt{2D}\, dW_t \quad,
\end{split}
\end{equation}
where $W_t$ represents a Wiener process and $D>0$ is the diffusion constant. The parameters $\kappa$ and $\eta$ are the trap stiffness and damping coefficient, respectively. 
For notational convenience, we introduce the state vector
\begin{equation}
\begin{split}
\label{eq:udho_state}
\vec{r}(t)\equiv
\begin{pmatrix}
x(t)\\[1mm]
v(t)
\end{pmatrix}\quad,
\end{split}
\end{equation}
so that Eq.~\eqref{eq:udho_sde} can be written in the compact matrix form
\begin{equation}
\begin{split}
\label{eq:udho_dr}
d \vec{r}(t)
=
\mat{L}\cdot \vec{r}(t)\,dt
+
\mat{B}\cdot d\vec{W_t}\quad,
\end{split}
\end{equation}
with
\begin{equation}
\begin{split}
\label{eq:udho_LB}
\mat{L}
\equiv
\begin{pmatrix}
0 & 1\\
-\kappa & -\eta
\end{pmatrix}
\quad;\quad
\mat{B}
\equiv
\sqrt{2D}\,
\begin{pmatrix}
0 & 0 \\
0 & 1
\end{pmatrix}
\quad;\quad
d\vec{W_t}\equiv
\begin{pmatrix}
0 \\
dW_t
\end{pmatrix}\quad.
\end{split}
\end{equation}
A unique stationary state exists for $\eta>0$ and $\kappa>0$.

\textit{Computation of the static correlation matrix $\mat{g}$\,.}
First, we compute the equal-time (static) correlation function. Using It\^{o}'s Lemma, we obtain:
\begin{equation}
\begin{split}
\label{eq:udho_ito}
d \expval{\vec{r}(t)\,\vec{r}(t)^\top}
&=
\expval{d\vec{r}(t)\,\vec{r}(t)^\top}
+\expval{\vec{r}(t)\,d\vec{r}(t)^\top}
+\expval{d\vec{r}(t)\,d\vec{r}(t)^\top}\\[2mm]
&=
\mat{L}\cdot \expval{\vec{r}(t)\,\vec{r}(t)^\top}\,dt
+
\expval{\vec{r}(t)\,\vec{r}(t)^\top}\cdot \mat{L}^\top\,dt
+
\mat{B}\cdot \mat{B}^\top\,dt \quad .
\end{split}
\end{equation}
Therefore, the time evolution of the correlation matrix is given by:
\begin{equation}
\begin{split}
\label{eq:udho_dt_rr}
\frac{d}{dt}\expval{\vec{r}(t)\,\vec{r}(t)^\top}
&=
\mat{L}\cdot \expval{\vec{r}(t)\,\vec{r}(t)^\top}
+
\expval{\vec{r}(t)\,\vec{r}(t)^\top}\cdot \mat{L}^\top
+
\mat{B}\cdot \mat{B}^\top
=0 \quad,
\end{split}
\end{equation}
where we used the property that for a stationary state, the time derivative of the equal-time correlation function vanishes.
Solving Eq.~\eqref{eq:udho_dt_rr} yields
\begin{equation}
\begin{split}
\label{eq:udho_g_result_components}
\expval{x(t)\,v(t)}=\expval{v(t)\,x(t)}=0\quad,\qquad
\expval{v(t)^2}=\frac{D}{\eta}\quad,\qquad
\expval{x(t)^2}=\frac{D}{\eta\,\kappa}\quad.
\end{split}
\end{equation}
Therefore, the static correlation matrix is
\begin{equation}
\begin{split}
\label{eq:udho_g_def}
\mat{g}
\equiv
\expval{\vec{r}(t)\,\vec{r}(t)^\top}
=
\frac{D}{\eta}\,
\begin{pmatrix}
1/\kappa & 0\\
0 & 1
\end{pmatrix}
\quad .
\end{split}
\end{equation}

\textit{Dynamical correlation matrix $\mat{\psi}(\tau)$\,.}
Using Eq.~\eqref{eq:udho_dr}, the equation of motion for the dynamical correlation function is
\begin{equation}
\begin{split}
\label{eq:udho_dpsi}
\frac{d}{d\tau}
\expval{\vec{r}(t+\tau)\,\vec{r}(t)^\top}
=
\mat{L}\cdot
\expval{\vec{r}(t+\tau)\,\vec{r}(t)^\top}
\quad ,
\end{split}
\end{equation}
which leads to the solution:
\begin{equation}
\begin{split}
\label{eq:udho_psi_solution}
\expval{\vec{r}(t+\tau)\,\vec{r}(t)^\top}
=
\exp(\mat{L}\,\tau)
\cdot
\expval{\vec{r}(t)\,\vec{r}(t)^\top}
\quad .
\end{split}
\end{equation}
Thus, the dynamical correlation function $\mat{\psi}(\tau)\equiv \expval{\vec{r}(t+\tau)\,\vec{r}(t)^\top}$ is given by
\begin{equation}
\begin{split}
\label{eq:udho_psi_exp}
\mat{\psi}(\tau)
=
\exp(\mat{L}\,\tau)
\cdot
\mat{g}
\quad(\tau\geq 0)\quad .
\end{split}
\end{equation}
We note that time-translational invariance leads to the relation:
\begin{equation}
\begin{split}
\label{eq:udho_psi_minus}
\mat{\psi}(-\tau)
=
\mat{\psi}(\tau)^\top
\quad .
\end{split}
\end{equation}
Furthermore, differentiating with respect to $\tau$ gives
\begin{equation}
\begin{split}
\label{eq:udho_equality_psi_top}
\mat{\dot{\psi}}(-\tau)
=
-\mat{\dot{\psi}}(\tau)^\top
\quad .
\end{split}
\end{equation}

\textit{Computation of the reactive frequency matrix $\mat{\omega}$\,.}
The reactive frequency matrix $\mat{\omega}$ is computed symmetrically at $\tau=0$ to account for the discontinuity of the derivative at the time origin:
\begin{equation}
\begin{split}
\label{eq:udho_omega_def}
\mat{\omega}
&\equiv
\frac{1}{2}
\qty(
    \mat{\dot{\psi}}(\tau=0^+)
    +
    \mat{\dot{\psi}}(\tau=0^-)
)
\quad .
\end{split}
\end{equation}
Using Eq.~\eqref{eq:udho_equality_psi_top} and Eq.~\eqref{eq:udho_psi_exp}, this can be rewritten as:
\begin{equation}
\begin{split}
\label{eq:udho_omega_calc}
\mat{\omega}
&=
\frac{1}{2}
\qty(
    \mat{\dot{\psi}}(\tau=0^+)
    -
    \mat{\dot{\psi}}^\top(\tau=0^+)
)
\\
&=
\frac{1}{2}
\qty(
    \mat{L}\cdot\mat{g}
    -
    \qty(\mat{L}\cdot\mat{g})^\top
)
\\
&=
\frac{D}{\eta}
\begin{pmatrix}
0 & 1\\
-1 & 0
\end{pmatrix}
\quad .
\end{split}
\end{equation}
Thus, we find that $\mat{\omega}$ is an antisymmetric matrix:
\begin{equation}
\begin{split}
\label{eq:udho_omega_result}
\mat{\omega}
=
\frac{D}{\eta}
\begin{pmatrix}
0 & 1\\
-1 & 0
\end{pmatrix}
\quad .
\end{split}
\end{equation}
We consider, in this example, the time reversal signatures of the dynamical variables in $\vec{r}=(x,v)^\top$ to be different. Therefore, $\mat{\omega}$ does not vanish even in equilibrium and describes the reversible coupling between position and velocity.

\textit{Computation of the kernel, $\mat{\widehat{K}}(s)$\,.}
In Laplace space, we have the following relation for a steady state:
\begin{equation}
\begin{split}
\label{eq:udho_Delphi_relation}
\Delta\mat{\widehat{\phi}}(s)
\equiv
\mat{\widehat{\phi}}(s)-\mat{\omega}
=
-s^2\mat{\widehat{\psi}}(s)+s\mat{g}
\quad .
\end{split}
\end{equation}
From Eq.~\eqref{eq:udho_psi_exp}, the Laplace transform of the correlation function is
\begin{equation}
\begin{split}
\label{eq:udho_psihat}
\mat{\widehat{\psi}}(s)
=
\qty(
    s\mat{I}-\mat{L}
)^{-1}
\cdot
\mat{g}
\quad .
\end{split}
\end{equation}
Therefore,
\begin{equation}
\begin{split}
\label{eq:udho_Delphi_ginv}
\Delta\mat{\widehat{\phi}}(s)
\cdot
\mat{g}^{-1}
&=
-s^2
\qty(
    s\mat{I}-\mat{L}
)^{-1}
+
s\mat{I}
\\
&=
-s\mat{L}
\cdot
\qty(
    s\mat{I}-\mat{L}
)^{-1}
\quad .
\end{split}
\end{equation}
By substituting $\Delta\mat{\widehat{\phi}}(s)$ into the expression for the kernel $\mat{\widehat{K}}(s)$ [Eq.~\eqref{eq:K}], we obtain:
\begin{equation}
\begin{split}
\label{eq:udho_K_calc}
\mat{\widehat{K}}(s)
&=
\qty[
    \mat{I}
    -
    \frac{1}{s}
    \qty(
        \Delta\mat{\widehat{\phi}}(s)
        \cdot
        \mat{g}^{-1}
    )
]^{-1}
\cdot
\qty(
    \Delta\mat{\widehat{\phi}}(s)
    \cdot
    \mat{g}^{-1}
)
\\
&=
\qty[
    \mat{I}
    +
    \mat{L}
    \cdot
    \qty(
        s\mat{I}-\mat{L}
    )^{-1}
]^{-1}
\cdot
\qty[
    -s\mat{L}
    \cdot
    \qty(
        s\mat{I}-\mat{L}
    )^{-1}
]
\\
&=
-\mat{L}
\\
&=
-
\begin{pmatrix}
0 & 1\\
-\kappa & -\eta
\end{pmatrix}
\quad .
\end{split}
\end{equation}
Using Eq.~\eqref{eq:dA/dt} from the main text then leads to:
\begin{equation}
\begin{split}
\label{eq:udho_mean_response}
\frac{d}{dt}
\expval{\vec{r}(t)}
=
\begin{pmatrix}
0 & 1\\
-\kappa & -\eta
\end{pmatrix}
\cdot
\expval{\vec{r}(t)}
\quad ,
\end{split}
\end{equation}
which is consistent with the deterministic part of Eq.~\eqref{eq:udho_sde}.

\section{Constitutive equation for non-equilibrium complex fluids}

\subsection{Derivation of the constitutive equation  \label{App:CE_j}}
In this section, we derive the identity in Eq.~\eqref{eq:main_eqality} leading to the constitutive equation, Eq.~\eqref{eq:CEQ_fluid}, with the generalized viscosity Eq.~\eqref{eq:eta}. 

We first evaluate Eq.~(\ref{eq:dA/dt}) for the choice of the dynamical variable, 
\begin{equation}
\begin{split}
\label{eq:}
\vec{A} =\vec{j}(\vec{q},t)\quad,
\end{split}
\end{equation}

We evaluate the dynamical evolution of $\vec{A}$ using Eq.~\eqref{eq:dA/dt}\,:
\begin{equation}
\begin{split}
\label{eq:eq_motion_ja_SI}
\frac{d}{dt}\expval{\vec{j}(\vec{q},t)}
= -\int_0^t dt' \mat{{K}}(\vec{q},t-t') \cdot \expval{\vec{j}(\vec{q},t')} \quad,
\end{split}
\end{equation}
where the memory kernel in Laplace space is given by 
\begin{equation} 
\begin{split}
\label{eq:K_j}
\mat{\widehat{K}}(\vec{q},s) =  \left[ \mat{I} - \frac{1}{s} \qty(\Delta \mat{\widehat{\phi}}(\vec{q},s) \cdot \mat{g}^{-1})\right]^{-1} \cdot \qty(\Delta \mat{\widehat{\phi}}(\vec{q},s)\cdot \mat{g}^{-1}) \quad;
\end{split}
\end{equation}
with 
\begin{equation} 
\begin{split}
\label{eq:}
\Delta \mat{\widehat{\phi}}(\vec{q},s) = \mat{\widehat{\phi}}(\vec{q},s)-\mat{\widehat{\phi}}(\vec{q},0) \quad. 
\end{split}
\end{equation}

First, we evaluate the memory kernel $\mat{K}(\vec{q},t)$. The static correlation matrix is given by 
\begin{equation}
\begin{split}
\label{eq:g_I}
\mat{g}(\vec{q}) = \expval{\vec{j}(\vec{q})\,\vec{j}^\dagger(\vec{q})} \quad .
\end{split}
\end{equation}
The matrix $\mat{\phi}$ is given by
\begin{equation}
\begin{split}
\label{eq:}
\mat{{\phi}}(\vec{q},t) =\expval{\vec{\dot{j}}(\vec{q},t) \ \vec{\dot{j}}(\vec{q})^\dagger} 
= \expval{\qty(\mat{\sigma}(\vec{q},t) \cdot \vec{q} ) \qty(\vec{q}^\top \cdot \mat{\sigma}^\dagger(\vec{q}) ) } \quad.
\end{split}
\end{equation}
In Laplace space, 
\begin{equation}
\begin{split}
\label{eq:phi_to_N}
\mat{{\widehat{\phi}}}(\vec{q},s)  = \expval{\qty(\mat{\widehat{\sigma}}(\vec{q},s) \cdot \vec{q} ) \qty(\vec{q}^\top \cdot \mat{\sigma}^\dagger(\vec{q}) ) }\quad.
\end{split}
\end{equation}
In the index notation, this reads
\begin{equation}
\begin{split}
\label{eq:}
\widehat{\phi}_{\alpha \beta}(\vec{q},s) =   q_\gamma \expval{\widehat{\sigma}_{\alpha \gamma }(\vec{q},s) {\sigma}_{\beta \epsilon}^*(\vec{q})}  q_\epsilon  \quad.
\end{split}
\end{equation}
We define the normalized stress correlation tensor
\begin{equation}
\begin{split}
\label{eq:}
 \Tfor{ \widehat{N}}^{} (\vec{q},s) \equiv \rho_0 \,  \expval{\mat{\widehat{\sigma}}(\vec{q},s)\, \mat{\sigma}^\dagger(\vec{q})}  \cdot \mat{g}^{-1}\quad,
\end{split}
\end{equation}
or in the index form,
\begin{equation}
\begin{split}
\widehat{N}_{\alpha\gamma\nu\beta}(\vec q,s)
\equiv
\rho_0\,
\Big\langle
\widehat{\sigma}_{\alpha\gamma}(\vec q,s)\,
\sigma^*_{\epsilon \beta}(\vec q)
\Big\rangle\, \qty(\mat{g}^{-1})_{\epsilon \nu}
\quad.
\end{split}
\end{equation}
With this definition, $\mat{\widehat{\phi}}(\vec{q},s)\cdot \mat{g}^{-1}$ can be written in the compact form,
\begin{equation}
\begin{split}
\label{eq:}
 \mat{{\widehat{\phi}}}(\vec{q},s)\cdot \mat{g}^{-1} = \vec{q} \cdot \Tfor{ \widehat{N}}^{} (\vec{q},s)  \cdot \vec{q} /\rho_0\quad. 
\end{split}
\end{equation}
In the index form
\begin{equation}
\begin{split}
\qty[ \widehat{\mat{\phi}}(\vec q,s) \cdot \mat{g}^{-1}]_{\alpha \beta}
=
\frac{1}{\rho_0}\,
q_\gamma\,\widehat{N}_{\alpha\gamma \beta \nu}(\vec q,s)\,q_\nu
=
\frac{1}{\rho_0}\,
\qty(\vec q\cdot \widehat{\Tfor N}(\vec q,s)\cdot \vec q)_{\alpha\beta} \quad.
\end{split}
\end{equation}
Therefore 
\begin{equation}
\begin{split}
\label{eq:Delphi_qNq}
\Delta \mat{\widehat{\phi}}(\vec{q},s)\cdot \mat{g}^{-1} = \vec{q}\cdot \Delta \Tfor{ \widehat{N}}^{} (\vec{q},s)  \cdot \vec{q}/\rho_0 \quad,
\end{split}
\end{equation}
where we have defined the transient part of the stress correlation tensor,
\begin{equation}
\begin{split}
\label{eq:}
 \Delta \Tfor{ \widehat{N}}^{} (\vec{q},s) \equiv  \Tfor{ \widehat{N}}^{} (\vec{q},s)- \Tfor{ \widehat{N}}^{} (\vec{q},0)\quad.
\end{split}
\end{equation}
Substituting Eq.~\eqref{eq:Delphi_qNq} into Eq.~(\ref{eq:K_j}), we obtain
\begin{equation}
\begin{split}
\label{eq:K_middle}
\widehat{K}(\vec{q},s) = \qty[\mat{I} - \frac{1}{\rho_0 s}\vec{q}\cdot \Delta \widehat{\Tfor{N}}(\vec{q},s) \cdot \vec{q} ]^{-1} \cdot \qty(\vec{q}\cdot \Delta \widehat{\Tfor{N}}(\vec{q},s) \cdot \vec{q} /\rho_0) \quad.
\end{split}
\end{equation}
Using the momentum conservation [Eq.~(\ref{eq:j_conserv})] and Eq.~\eqref{eq:K_middle}, Eq.~(\ref{eq:eq_motion_ja_SI}) can be written in the Laplace space as
\begin{equation}
\begin{split}
\label{eq:derive_middle}
-i   \expval{\mat{\widehat{\sigma}}(\vec{q},s)}\cdot \vec{q}     &= -  \mat{\widehat{K}}(\vec{q},s)  \cdot \expval{\vec{\widehat{j}}(\vec{q},s)}\\
&=- \qty[\mat{I}  - \frac{1}{\rho_0 s}\vec{q}\cdot \Delta \widehat{\Tfor{N}}(\vec{q},s) \cdot \vec{q} ]^{-1} \cdot \qty(\vec{q}\cdot \Delta \widehat{ \Tfor{N}}(\vec{q},s) \cdot \vec{q} /\rho_0)  \cdot \expval{\vec{\widehat{j}}(\vec{q},s)}\\
&=- \qty[\mat{I}  - \frac{1}{\rho_0 s}\vec{q}\cdot \Delta \widehat{\Tfor{N}}(\vec{q},s) \cdot \vec{q} ]^{-1} \cdot \qty(i\vec{q}\cdot \Delta \widehat{\Tfor{N}}(\vec{q},s)) : \expval{\mat{\widehat{\gamma}}(\vec{q},s)} \quad.
\end{split}
\end{equation}
Using index notation, Eq.~\eqref{eq:derive_middle} can be written as 
\begin{equation} 
\begin{split}
\label{eq:eqality_index}
q_\beta \qty( \expval{{\widehat{\sigma}_{\alpha \beta }}(\vec{q},s)} -\qty[\mat{I}  - \frac{1}{\rho_0 s}\vec{q}\cdot \Delta \widehat{\Tfor{N}}(\vec{q},s) \cdot \vec{q}]^{-1}_{\alpha \xi }\Delta \widehat{N}_{\xi \beta \eta \epsilon }(\vec{q},s)\expval{{\widehat{\gamma}_{\eta \epsilon}}(\vec{q},s)}  )  =0 \quad.
\end{split}
\end{equation}
We write Eq.~\eqref{eq:eqality_index} using the index-free notation, as given in Eq.~\eqref{eq:main_eqality}, 
\begin{equation} 
\begin{split}
\label{eq:eqality}
  \qty( \expval{{\widehat{\mat{\sigma}}}(\vec{q},s)} - \qty[\mat{I}  - \frac{1}{\rho_0 s}\vec{q}\cdot \Delta \widehat{\Tfor{N}}(\vec{q},s) \cdot \vec{q}]^{-1} \cdot\Delta \Tfor{\widehat{N}}(\vec{q},s): \expval{{\widehat{\mat{\gamma}}}(\vec{q},s)}) \cdot \vec{q}=0 \quad.
\end{split}
\end{equation}

After the gauge fixing procedure in the following Section~\ref{sec:gauge}, the constitutive equation in index form is given by
\begin{equation} 
\begin{split}
\label{eq:}
 \expval{{\widehat{\sigma}_{\alpha \beta }}(\vec{q},s)} = \widehat{\eta}_{\alpha\beta \gamma \delta}(\vec{q},s)\expval{{\widehat{\gamma}_{\gamma \delta}}(\vec{q},s)}    \quad,
\end{split}
\end{equation}
where the generalized viscosity is 
\begin{equation} 
\begin{split}
\label{eq:eta_index}
 \widehat{\eta}_{\alpha \beta \gamma \delta } (\vec{q},s)=\qty[\mat{I}  - \frac{1}{\rho_0 s}\vec{q}\cdot \Delta \widehat{\Tfor{N}}(\vec{q},s) \cdot \vec{q}]^{-1}_{\alpha \xi }\Delta \widehat{N}_{\xi \beta \gamma \delta }(\vec{q},s) \quad.
\end{split}
\end{equation}

\subsection{Gauge freedom in momentum density conservation \label{sec:gauge}}
Our derived expression for the stress response, given in Eq.~\eqref{eq:main_eqality}, does not uniquely determine the stress tensor. This result contains a fundamental ambiguity which stems from the underlying physical constraint that the stress tensor must satisfy: the conservation of momentum, Eq.~\eqref{eq:j_conserv}. This ambiguity can be understood as a gauge freedom in the definitions of the stress and momentum density that leaves the form of the conservation law invariant.

Suppose the stress $\mat{\sigma}^\circ$ and the momentum density $\vec{j}^\circ$ satisfy the conservation law,
\begin{equation} 
\begin{split}
\label{eq:dj/dt_original}
 \partial_t{j}^\circ_{\alpha }(\vec{q},t)= -i  q_\beta \sigma^\circ_{\alpha \beta}(\vec{q},t)\quad.
\end{split}
\end{equation}
The form of this law is preserved under the general transformation:
\begin{equation}
    \label{eq:gauge}
    j_\alpha(\vec{q},t) = j^\circ_\alpha(\vec{q},t) + \mathcal{A}_\alpha(\vec{q},t)\quad, \quad \sigma_{\alpha \beta}(\vec{q},t) = \sigma^\circ_{\alpha \beta}(\vec{q},t) + \mathcal{B}_{\alpha \beta}(\vec{q},t)\quad,
\end{equation}
where the arbitrary fields $\vec{\mathcal{A}}$ and $\mat{\mathcal{B}}$ are linked by the constraint:
\begin{equation}
    \label{eq:gauge_constraint}
    \partial_t \mathcal{A}_\alpha(\vec{q},t) = - i q_\beta \mathcal{B}_{\alpha \beta}(\vec{q},t)\quad.
\end{equation}
To formulate a well-defined constitutive relation, we must first fix this gauge. We adopt the specific choice where the momentum density is not transformed, i.e., $\vec{\mathcal{A}}= 0$. The constraint in Eq.~\eqref{eq:gauge_constraint} then requires the change in the stress tensor, $\mat{\mathcal{B}}$\,, to be divergence-free: $q_\beta \mathcal{B}_{\alpha \beta}=0$. With this gauge choice, the constitutive equations are given by Eq.~\eqref{eq:CEQ_fluid}-\eqref{eq:eta}.

\paragraph*{Gauge invariance of Eq.~\eqref{eq:main_eqality}.}
We demonstrate that Eq.~\eqref{eq:main_eqality} is gauge invariant.  Eq.~\eqref{eq:main_eqality} constrains only the divergence of the stress,
i.e.\ it has the form $(\cdots)\cdot \vec q=0$.  Under the fixed $\vec j$ gauge
($\mathcal A=0$), the stress transforms as $\mat{\sigma}=\mat{\sigma}^{\circ}+\mat{\mathcal B}$
with $q_{\beta}\mathcal B_{\alpha\beta}=0$, hence
\begin{equation}
\langle \widehat{\mat{\sigma}}(\vec q,s)\rangle\cdot \vec q
=
\langle \widehat{\mat{\sigma}}^{\circ}(\vec q,s)\rangle\cdot \vec q \quad.
\end{equation}
Moreover, since $\vec j=\vec j^{\circ}$ in this gauge, the associated objects constructed from
$\vec j$ are unchanged, in particular
\begin{equation}
\mat g(\vec q)=\mat g^{\circ}(\vec q)\equiv \langle \vec j^{\circ}(\vec q)\,\vec j^{\circ\dagger}(\vec q)\rangle,
\qquad
\widehat{\mat{\gamma}}(\vec q,s)=\widehat{\mat{\gamma}}^{\circ}(\vec q,s)\equiv -\,i\,\vec{\widehat j}^{\circ}(\vec q,s)\,\vec q^\top/\rho_0 \quad.
\end{equation}
Therefore Eq.~\eqref{eq:main_eqality} is invariant under the divergence-free stress shift
$\mat{\sigma}= \mat{\sigma}^\circ+\mat{\mathcal B}$ with $q_{\beta}\mathcal B_{\alpha\beta}=0$,
\begin{equation}
    \label{eq:equality_index_original}
   \qty( \expval{{\widehat{\mat{\sigma}}}^\circ(\vec{q},s)} - \qty[\mat{I}  - \frac{1}{\rho_0 s}\vec{q}\cdot \Delta \widehat{\Tfor{N}}^\circ(\vec{q},s) \cdot \vec{q}]^{-1} \cdot\Delta \Tfor{\widehat{N}}^\circ(\vec{q},s):  \expval{{\widehat{\mat{\gamma}}^\circ}(\vec{q},s)}  ) \cdot \vec{q}=0\quad,
\end{equation}
where $\Delta\widehat{\Tfor N}^{\circ}$ is defined from $\mat{\sigma}^{\circ}$ consistently.

\paragraph*{Symmetric gauge.}
We can also choose a gauge that is different from $\vec{\mathcal{A}}=0$.  This gauge is fixed by choosing  
\begin{equation} 
\begin{split}
\label{eq:tr_j}
\mathcal{A}_\alpha = \frac{1}{2} \partial_\beta S_{\alpha \beta} \quad,
\end{split}
\end{equation}
and 
\begin{equation} 
\begin{split}
\label{eq:tr_sigma}
\mathcal{B}_{\alpha \beta}= -\frac{1}{2} \dot{S}_{\alpha \beta} -\partial_\gamma f_{\alpha \beta \gamma} + \partial_\gamma f_{\alpha \gamma \beta} + \partial_\gamma f_{\beta \gamma \alpha} \quad.
\end{split}
\end{equation}
This gauge was analyzed in the context of general relativity by Belinfante~\cite{belinfante1940current} and Rosenfeld~\cite{rosenfeld1940energy} and later applied to fluids by Martin et al.~\cite{martin1972unified}.
Here $S_{\alpha \beta}$ is the spin angular momentum density tensor which is antisymmetric ($S_{\alpha \beta} = - S_{\beta \alpha}$) and $f_{\alpha \beta 
\gamma}$ is defined using the total angular momentum flux $M_{\alpha \beta \gamma}$ as,
\begin{equation} 
\begin{split}
\label{eq:f}
f_{\alpha \beta 
\gamma} \equiv - \frac{1}{2} \qty(M_{\alpha \beta \gamma} - r_\beta \sigma^\circ_{\alpha \gamma} + r_\alpha \sigma^\circ_{\beta \gamma} ) \quad.
\end{split}
\end{equation}
The tensor $f_{\alpha \beta 
\gamma}$ is antisymmetric for the exchange of the first two indices: $f_{\alpha \beta 
\gamma}= -f_{\beta \alpha 
\gamma}$ . 

As detailed in the Appendix A of Martin et al.~\cite{martin1972unified} the transformation for the stress in Eq.~(\ref{eq:tr_sigma}) is chosen to make the stress tensor symmetric: ${\sigma}_{\alpha \beta}={\sigma}_{\beta \alpha}$. This was used for equilibrium fluids to effectively eliminate the spin angular momentum and attribute all angular momentum to its orbital component. However, for non-equilibrium systems like active chiral fluids, the intrinsic spin angular momentum is a crucial physical feature. Therefore, this gauge choice is not suitable for non-equilibrium fluids, which would otherwise conceal the important physics by symmetrizing the stress tensor.

\subsection{\label{sec:Green-Kubo} Green-Kubo relation for viscosity}
The Green–Kubo formula for viscosity in equilibrium can be recovered from the generalized viscosity tensor, Eq.~(\ref{eq:eta}), in the hydrodynamic limit. First, we consider the static momentum density correlation function,
\begin{equation} 
\begin{split}
\label{}
\mat{g}(\vec{q}) &= {\expval{ \frac{1}{d}\sum_{ij} m_i m_j v_\alpha^i v_\alpha^j e^{i\vec{q} \cdot (\vec{r_i} - \vec{r_j})}}}\,  \mat{I}
=\frac{1}{d}{\sum_{i} m_i^2 \expval{(v^i_\alpha)^2}  \,  \mat{I} +\frac{1}{d}\sum_i \sum_{j \neq i} m_i m_j \expval{v^i_\alpha v^{j}_\alpha e^{i\vec{q} \cdot (\vec{r_i} - \vec{r_j})}}}\,  \mat{I} \quad.
\end{split}
\end{equation}
Here $d$ is the spatial dimension and $\mat{I}$ is identity matrix.
In a simple equilibrium fluid the positions and velocities are uncorrelated as well as the velocities of distinct particles, therefore 
\begin{equation} 
\begin{split}
\label{eq:vcoor0}
\sum_i \sum_{j \neq i} m_i m_j \expval{v^i_\alpha v^{j}_\alpha e^{i\vec{q} \cdot (\vec{r_i} - \vec{r_j})}} =\sum_i \sum_{j \neq i} m_i m_j \expval{v^i_\alpha v^{j}_\alpha} \expval{ e^{i\vec{q} \cdot (\vec{r_i} - \vec{r_j})}} =  0 \quad.
\end{split}
\end{equation}
Thus, we obtain 
\begin{equation} 
\begin{split}
\label{eq:Kubo_j_SI}
\mat{g}(\vec{q}) & =\frac{1}{d}\sum_{i} m_i^2 \expval{(v^i_\alpha)^2}\,  \mat{I} = {\sum_i m_i k_BT   } =  {\rho_0 V k_BT} \,  \mat{I}\quad.
\end{split}
\end{equation}
We have used the equipartition theorem.
Therefore, we obtain the following expression for the stress correlation function:  
\begin{equation} 
\begin{split}
\label{}
\widehat{N}_{\alpha \beta \gamma \epsilon}(\vec{q},s)= \frac{\expval{\widehat{\sigma}_{\alpha \beta}(\vec{q},s) \sigma^*_{\gamma \epsilon}(\vec{q})}}{Vk_BT} \quad.
\end{split}
\end{equation}

In equilibrium, the time-reversal symmetry $\mat{\omega}=\mat{0}$ holds, then the hydrodynamically relevant components of the transient correlation tensor simplifies to $\Delta\widehat{N}_{\alpha\beta\gamma\epsilon}(\vec{q},s) = \widehat{N}_{\alpha\beta\gamma\epsilon}(\vec{q},s)$. 
Using this, the generalized viscosity tensor, Eq.~(\ref{eq:eta}), in the hydrodynamic limit ($\vec{q} \rightarrow 0$) can be written as:
\begin{equation} 
\begin{split}
\label{}
\lim_{\vec{q} \rightarrow 0}  \widehat{\eta}_{\alpha \beta \gamma \epsilon}(\vec{q},s) &=\lim_{\vec{q} \rightarrow 0}  \widehat{N}_{\alpha \beta \gamma \epsilon}(\vec{q},s) =   \int_0^\infty  {N}_{\alpha \beta \gamma \epsilon}(\vec{q}=0,t)e^{-s t} dt \quad.
\end{split}
\end{equation}
Taking the limit  $s \rightarrow 0$  subsequently, we obtain
\begin{equation} 
\begin{split}
\label{}
\lim_{s \rightarrow 0} \lim_{q \rightarrow 0} \widehat{\eta}_{\alpha \beta \gamma \epsilon}(\vec{q},s)  =  \frac{1}{V k_BT}\int_0^\infty \expval{{\sigma}_{\alpha \beta}(\vec{q}=0,t) \sigma^*_{\gamma \epsilon}(\vec{q}=0)} dt\quad.
\end{split}
\end{equation}
This is the Green–Kubo formula for the viscosity tensor, which relates the transport coefficients to equilibrium time-correlation functions of the stress tensor.

\section{Kubo formula for chemical potential differences \label{App:CE_mu}}

In equilibrium, we can derive a relationship between the fluctuations of chemical potential differences,  $\expval{\Delta \vec{\mu}(\vec{q})\, \Delta \vec{\mu}^\dagger(\vec{q})}$, and the  susceptibility $\mat{\kappa}$. We begin by expanding the free energy functional $\mathcal{F}$, which is the spatial integration of free energy density $f(\vec{r})$, to second order around its equilibrium state:
\begin{equation} 
\begin{split}
\label{eq:F_chem}
\mathcal{F} \equiv \int dr^d f(\vec{r}) =  F_0 + \frac{1}{2}\int dr^d  n_a(\vec{r})  k_{ab}   n_b(\vec{r})  \quad,
\end{split}
\end{equation}
where 
\begin{equation} 
\begin{split}
\label{eq:}
k_{ab} = \frac{\partial^2   f(\vec{r})}{\partial n_a(\vec{r}) \partial n_b(\vec{r})}
\end{split}
\end{equation}
is the local curvature matrix in composition space.
In Fourier space, the quadratic part of the free energy in Eq.~\eqref{eq:F_chem} reads:
\begin{equation} 
\begin{split}
\label{eq:}
\mathcal{F}_2 \equiv \frac{1}{2V}\sum_\vec{q}    n_a(\vec{q})  k_{ab}  n_b^*(\vec{q}) \quad.
\end{split}
\end{equation}
Applying the equipartition theorem to each mode, we obtain:
\begin{equation} 
\begin{split}
\label{eq:}
\expval{n_a(\vec{q})n_b^*(\vec{q}) } = k^{-1}_{ab} V k_BT \quad.
\end{split}
\end{equation}
The chemical potential is given by $\mu_a(\vec{q}) = k_{ab}n_b(\vec{q})$ in the quadratic approximation, therefore we find the fluctuations of chemical potential differences as:
\begin{equation} 
\begin{split}
\label{eq:Kubo_mu_SI}
\expval{\Delta \mu_I(\vec{q})\Delta \mu^*_J(\vec{q})} = \nu_a^I \nu_b^J k_{a c}k_{bd}\expval{n_c(\vec{q})n_d^*(\vec{q}) } = \kappa_{IJ} V k_BT \quad.
\end{split}
\end{equation}
This establishes the static Kubo formula for chemical potential differences in equilibrium.

\section{Constitutive equations for chemically driven complex fluids \label{App: CE_j_mu_detail} }
\subsection{General form of constitutive equations \label{App:CE_j_mu}}
In this section, we derive the constitutive equation for chemically driven active fluids. Again, our aim is to evaluate Eq.~\eqref{eq:dA/dt} for a choice of dynamical variables. We consider the set of dynamical variables given in Eq.~(\ref{eq:variable_j_mu}): 
\begin{equation} 
\begin{split}
\label{}
\vec{A} =  \qty( \vec{j}(\vec{q},t)\, , \,\Delta \vec{\mu}(\vec{q},t))^\top\quad.
\end{split}
\end{equation}
The static correlation matrix $\mat{g}$ is given by 
\begin{equation} 
\begin{split}
\label{}
\mat{g}(\vec{q}) = \begin{pmatrix}
\expval{\vec{j}(\vec{q})\, \vec{j}^\dagger(\vec{q})} &\expval{\vec{j}(\vec{q})\, \Delta {\vec{\mu}}^{\dagger}(\vec{q})}\\ \\
\expval{\Delta \vec{\mu}(\vec{q})\, \vec{j}^\dagger(\vec{q})} & \expval{\Delta \vec{\mu}(\vec{q}) \, \Delta \vec{\mu}^{\dagger}(\vec{q})}
\end{pmatrix}\quad .
\end{split}
\end{equation}
We consider a steady state in which there is no static cross-correlation
between momentum density and chemical affinities, 
\begin{equation}
\label{eq:no_static_cross}
\expval{\vec{j}(\vec{q})\,\Delta {\vec{\mu}}^{\dagger}(\vec{q})}
=
\expval{\Delta \vec{\mu}(\vec{q})\,\vec{j}^\dagger(\vec{q})}
=\mat{0}\quad.
\end{equation}
Defining the components of $\mat{g}$ as 
\begin{equation}
    \mat{g}^{jj} (\vec{q})\equiv  \expval{\vec{j}(\vec{q})\, \vec{j}^\dagger(\vec{q})} \quad ; \quad \mat{g}^{\mu \mu} (\vec{q}) \equiv \expval{\Delta \vec{\mu}(\vec{q}) \, \Delta \vec{\mu}^{\dagger}(\vec{q})} \quad,
\end{equation}
we have
\begin{equation} 
\begin{split}
\label{}
\mat{g}(\vec{q}) = \begin{pmatrix}
 \mat{g}^{jj}(\vec{q})  &0\\ \\
0 & \mat{g}^{\mu \mu}(\vec{q})
\end{pmatrix}\quad .
\end{split}
\end{equation}
Since the static cross-correlations vanish, the inverse static correlation
matrix is block diagonal:
\begin{equation}
\begin{split}
\label{eq:g_inv_j_mu}
\mat{g}^{-1}(\vec q)
=
\begin{pmatrix}
    \left(\mat{g}^{jj}\right)^{-1}(\vec q) & 0
    \\[2mm]
    0 & \left(\mat{g}^{\mu\mu}\right)^{-1}(\vec q)
\end{pmatrix}
\quad .
\end{split}
\end{equation}

Next, we evaluate the time-dependent correlation matrix $\mat{\phi}(\vec{q},t)$, which takes the form:
\begin{equation} 
\begin{split}
\label{}
\mat{\phi}(\vec{q},t) &=  \begin{pmatrix}
\expval{\dot{\vec{j}}(\vec{q},t)\, \dot{\vec{j}}^{\dagger}(\vec{q})} & \expval{\dot{\vec{j}}(\vec{q},t)\, \Delta \dot{\vec{\mu}}^{\dagger}(\vec{q})}\\ \\
\expval{\Delta \dot{\vec{\mu}}(\vec{q},t)\, \dot{\vec{j}}^\dagger(\vec{q})}& \expval{\Delta \dot{\vec{\mu}}(\vec{q},t)\,  \Delta \dot{\vec{\mu}}^{\dagger}(\vec{q})}
\end{pmatrix} \quad.
\end{split}
\end{equation}
Using the momentum conservation Eq.~\eqref{eq:j_conserv} and Eq.~\eqref{eq:dmu/dt}, the matrix $\mat{\phi}(\vec{q},t)\cdot \mat{g}^{-1}$ can be written as 
\begin{equation} 
\begin{split}
\label{}
\mat{\phi}(\vec{q},t)\cdot \mat{g}^{-1} &= \begin{pmatrix}
\vec{q}\cdot\Tfor{N}^{\sigma \sigma}(\vec{q},t) \cdot \vec{q}/\rho_0&  i\vec{q}\cdot \Tthr{N}^{\sigma R}(\vec{q},t) /\rho_0\\ \\
-\Tthr{N}^{R \sigma}(\vec{q},t)\cdot i\vec{q}/\rho_0 &\mat{N}^{RR}(\vec{q},t)/\rho_0
\end{pmatrix} \quad,
\end{split}
\end{equation}
where the normalized correlation tensors are defined as 
\begin{equation} 
\begin{split}
\label{eq:correlations_SI}
&\Tfor{N}^{\sigma \sigma} (\vec{q},t)  \equiv \rho_0 \,\expval{\mat{\sigma}(\vec{q},t)\, \mat{\sigma}^{\dagger}(\vec{q})} \cdot \qty(\mat{g}^{jj})^{-1}\quad ,  \quad \Tthr{N}^{\sigma R} (\vec{q},t) \equiv \rho_0 \,\expval{\mat{\sigma}(\vec{q},t)\, \vec{R}^{\dagger}(\vec{q})}\cdot \qty(\mat{g}^{\mu \mu})^{-1}  \quad ,\\ \\ 
&\Tthr{N}^{R \sigma} (\vec{q},t)  \equiv\rho_0\, \expval{\vec{R}(\vec{q},t)\, \mat{\sigma}^{\dagger}(\vec{q})} \cdot \qty(\mat{g}^{jj})^{-1}\quad , \quad \mat{N}^{R R} (\vec{q},t)  \equiv  \rho_0\,\expval{\vec{R}(\vec{q},t)\, \vec{R}^{\dagger}(\vec{q})}\cdot \qty(\mat{g}^{\mu \mu})^{-1} \quad . 
\end{split}
\end{equation}
In the index form,
\begin{equation} 
\begin{split}
\label{eq:correlations_SI_index}
&{N}^{\sigma \sigma}_{\alpha \beta \gamma \delta} (\vec{q},t)  \equiv \rho_0 \,\expval{{\sigma}_{\alpha \beta}(\vec{q},t)\, {\sigma}^{*}_{\epsilon \delta}(\vec{q})}  \qty(\mat{g}^{jj})^{-1}_{\epsilon \gamma }\quad ,  \quad {N}^{\sigma R}_{\alpha \beta,I} (\vec{q},t) \equiv \rho_0 \,\expval{{\sigma}_{\alpha \beta}(\vec{q},t)\, {R}_{J}^{*}(\vec{q})} \qty(\mat{g}^{\mu \mu})_{JI}^{-1}  \quad ,\\ \\ 
&{N}^{R \sigma}_{I,\alpha \beta} (\vec{q},t)  \equiv\rho_0\, \expval{{R}_I(\vec{q},t)\, {\sigma}^{*}_{\epsilon \beta}(\vec{q})} \qty(\mat{g}^{jj})_{\epsilon \alpha}^{-1}\quad , \quad {N}^{R R}_{IJ} (\vec{q},t)  \equiv  \rho_0\,\expval{{R}_{I}(\vec{q},t)\, {R}^{*}_K(\vec{q})} \qty(\mat{g}^{\mu \mu})^{-1}_{KJ} \quad . 
\end{split}
\end{equation}
The contractions in the index notation are given by
\begin{equation}
\begin{split}
    \label{eq:qNq_and_qN_contractions}
\bigl(\vec{q}\cdot \Tfor{N}^{\sigma\sigma}(\vec{q},t)\cdot \vec{q}\bigr)_{\alpha\beta}
&=
q_\mu\, N^{\sigma\sigma}_{\alpha\mu \beta \delta}(\vec{q},t)\, q_\delta
\quad;\\ 
\\[1mm]
\bigl(i\,\vec{q}\cdot \Tthr{N}^{\sigma R}(\vec{q},t)\bigr)_{\alpha I}
&=
i\,q_\mu\, N^{\sigma R}_{\alpha\mu,I}(\vec{q},t)
\quad;\\
\\[1mm]
\bigl(\Tthr{N}^{R\sigma}(\vec{q},t)\cdot i\,\vec{q}\bigr)_{I\beta}
&=
i\, N^{R\sigma}_{I, \beta \delta }(\vec{q},t)\, q_\delta
\quad.
\end{split}
\end{equation}
Thus, in Laplace space, $\Delta\mat{\widehat{\phi}}(\vec{q},s)\cdot \mat{g}^{-1}$ is expressed as
\begin{equation}
\label{eq:Delphihat_def}
\Delta \mat{\widehat{\phi}}(\vec{q},s)\cdot \mat{g}^{-1}=
\begin{pmatrix}
\vec{q}\cdot \Delta\Tfor{\widehat{N}}^{\sigma \sigma}(\vec{q},s)\cdot \vec{q}/\rho_0
&
 i\,\vec{q}\cdot \Delta \Tthr{\widehat{N}}^{\sigma R}(\vec{q},s)/\rho_0
\\ \\
-\Delta\Tthr{\widehat{N}}^{R \sigma}(\vec{q},s)\cdot i\,\vec{q}/\rho_0
&
\Delta \mat{\widehat{N}}^{RR}(\vec{q},s)/\rho_0
\end{pmatrix}\quad,
\end{equation}
where  
\begin{equation} 
\begin{split}
\label{}
&\Delta \Tfor{\widehat{N}}^{\sigma \sigma} (\vec{q},s) \equiv  \Tfor{\widehat{N}}^{\sigma \sigma} (\vec{q},s)- \Tfor{\widehat{N}}^{\sigma \sigma} (\vec{q},0)\quad, \quad
\Delta \Tthr{\widehat{N}}^{\sigma R} (\vec{q},s) \equiv  \Tthr{\widehat{N}}^{\sigma R} (\vec{q},s)- \Tthr{\widehat{N}}^{\sigma R} (\vec{q},0)  \quad,\\\\
&\Delta \Tthr{\widehat{N}}^{R \sigma} (\vec{q},s) \equiv \Tthr{\widehat{N}}^{R \sigma} (\vec{q},s)-\Tthr{\widehat{N}}^{R \sigma} (\vec{q},0)\quad, \quad
\Delta \mat{\widehat{N}}^{R R} (\vec{q},s)  \equiv  \mat{\widehat{N}}^{R R} (\vec{q},s)- \mat{\widehat{N}}^{R R} (\vec{q},0) \quad.
\end{split} 
\end{equation}

We compute the contraction $\qty(\Delta \mat{\widehat{\phi}}(\vec{q},s) \cdot \mat{g}^{-1}) \cdot \vec{\widehat{A}}(\vec{q},s)$, which appears inside $\mat{\widehat{K}}(\vec{q},s) \cdot {\vec{\widehat{A}}}(\vec{q},s)$: 
\begin{equation} 
\begin{split}
\label{}
\qty(\Delta \mat{\widehat{\phi}}(\vec{q},s) \cdot \mat{g}^{-1} ) \cdot \vec{\widehat{A}}(\vec{q},s)&=
\begin{pmatrix}
\vec{q}\cdot \Delta \Tfor{\widehat{N}}^{\sigma \sigma}\cdot \vec{q}/\rho_0 & i\vec{q} \cdot \Delta \Tthr{\widehat{N}}^{\sigma R}/\rho_0 \\ \\
- \Delta \Tthr{\widehat{N}}^{R \sigma}\cdot  i {\vec{q}}/\rho_0 & \Delta \mat{\widehat{N}}^{RR}/\rho_0 \\ 
\end{pmatrix}
\cdot
\begin{pmatrix}
\vec{\widehat{j}}\\ \\
{\Delta} \vec{\widehat{\mu}} 
\end{pmatrix}
\\ \\ 
&=
\begin{pmatrix}
\qty(\vec{q}\cdot \Delta \Tfor{\widehat{N}}^{\sigma \sigma}\cdot \vec{q})\cdot \vec{\widehat{j}}/\rho_0+ \qty(i\vec{q}\cdot \Delta \Tthr{\widehat{N}}^{\sigma R})\cdot  \Delta \vec{\widehat{\mu}}/\rho_0  \\ \\
-\qty(i \Delta \Tthr{\widehat{N}}^{R \sigma}\cdot {\vec{q}}) \cdot \vec{\widehat{j}}/\rho_0+ \Delta \mat{\widehat{N}}^{RR}\cdot  {\Delta} \vec{\widehat{\mu}} / \rho_0 
\\
\end{pmatrix}
\\  \\
&=
\begin{pmatrix}
\qty(i\vec{q}\cdot \Delta \Tfor{\widehat{N}}^{\sigma \sigma}): \mat{\widehat{\gamma}}+ \qty(i\vec{q}\cdot \Delta \Tthr{\widehat{N}}^{\sigma R})\cdot \Delta \vec{\widehat{\mu}}/\rho_0   \\ \\ 
\Delta  \Tthr{\widehat{N}}^{R \sigma}:  \mat{\widehat{\gamma} } + \Delta \mat{\widehat{N}}^{RR}\cdot   {\Delta} \vec{\widehat{\mu}} / \rho_0 
\\
\end{pmatrix}\quad.
\end{split}
\end{equation}
We have used $\mat{\gamma}(\vec{q},t) = - i \,\vec{j}(\vec{q},t) \,\vec{q}^\top /\rho_0$\,.

We now make use of the matrix inversion identity for block matrices:
\begin{equation} 
\begin{split}
\label{eq:inverse_block_matrix}
\begin{pmatrix}
\mat{A} & \mat{B} \\ \\
\mat{C} & \mat{D}
\end{pmatrix}
^{-1}
= 
\left[
\begin{matrix}
\qty(\mat{A}-\mat{B}\cdot\mat{D}^{-1}\cdot\mat{C})^{-1} & \mat{0} \\ \\
\mat{0} & \qty(\mat{D}-\mat{C}\cdot\mat{A}^{-1}\cdot\mat{B})^{-1}
\end{matrix}
\right ]
\left[
\begin{matrix}
\mat{I} & - \mat{B}\cdot \mat{D}^{-1} \\ \\
-\mat{C} \cdot\mat{A}^{-1} &\mat{I}
\end{matrix}
\right]\quad, \\
\end{split}
\end{equation}
which holds when $\mat{A}$ and $\mat{D}$ are invertible.  Here $\mat{I}$ is the identity matrix. We use this matrix identity to $\left[\mat{I}-\frac{1}{s} \qty(\Delta \mat{\widehat{\phi}}(\vec{q},s) \cdot \mat{g}^{-1} )\right]^{-1}$ in the kernel $\mat{K}$\ , 
\begin{equation} 
\begin{split}
\label{}
&\left[\mat{I} -\frac{1}{ s} \qty(\Delta \mat{\widehat{\phi}}(\vec{q},s) \cdot \mat{g}^{-1} )\right]^{-1}
=
\begin{pmatrix}
\mat{I}^{} - \vec{q}\cdot \Delta \Tfor{\widehat{N}}^{\sigma \sigma}\cdot \vec{q}/(s\rho_0) & -i\vec{q} \cdot \Delta \Tthr{\widehat{N}}^{\sigma R}/(s\rho_0) \\ \\
 \Delta \Tthr{\widehat{N}}^{R \sigma}\cdot {i\vec{q}}/(s\rho_0) & \mat{I}^{}-\Delta \mat{\widehat{N}}^{RR}/(s\rho_0)\\
\end{pmatrix}^{-1} \\  \\
&\equiv
\begin{pmatrix}
\mat{\widehat{U}}^{ } & \mat{\widehat{V}}^{}\\ \\
\mat{\widehat{W}}^{} & \mat{\widehat{X}}^{}\\
\end{pmatrix}^{-1}
=
\begin{pmatrix}
\mat{\widehat{Y}}^{} & \mat{0}\\ \\ 
\mat{0} & \mat{\widehat{Z}}^{} \\
\end{pmatrix}
\begin{pmatrix}
\mat{I}^{} & -\mat{\widehat{V}}^{} \cdot \mat{\widehat{X}}^{-1}\\ \\
-\mat{\widehat{W}}^{} \cdot \mat{\widehat{U}}^{-1} & \mat{I}^{} \\
\end{pmatrix}\quad.
\end{split}
\end{equation}
We defined the relevant blocks as follows:
\begin{equation} 
\begin{split}
\label{eq:matrix1_SI}
\mat{\widehat{U}}^{} =\mat{I} - \frac{1}{s\rho_0}\vec{q}\cdot \Delta \Tfor{\widehat{N}}^{\sigma \sigma}\cdot \vec{q} \quad; \quad \mat{\widehat{V}}^{} =-\frac{1}{s\rho_0}i\vec{q} \cdot \Delta \Tthr{\widehat{N}}^{\sigma R};
\end{split}
\end{equation}
\begin{equation} 
\begin{split}
\label{eq:matrix2_SI}
\mat{\widehat{W}}^{} =\frac{1}{s\rho_0}  \Delta \Tthr{\widehat{N}}^{R \sigma}\cdot i{\vec{q}}\quad ;\quad  \mat{\widehat{X}}^{} = \mat{I}-\frac{1}{s\rho_0}\Delta \mat{\widehat{N}}^{RR};
\end{split}
\end{equation}
\begin{equation} 
\begin{split}
\label{eq:matrix3_SI}
\mat{\widehat{Y}}^{} = \qty(\mat{\widehat{U}}^{}-\mat{\widehat{V}}^{} \cdot \mat{\widehat{X}}^{-1}\cdot \mat{\widehat{W}}^{})^{-1}; \quad \mat{\widehat{Z}}^{} = \qty(\mat{\widehat{X}}^{}-\mat{\widehat{W}}^{}\cdot\mat{\widehat{U}}^{-1}\cdot\mat{\widehat{V}}^{})^{-1}.
\end{split}
\end{equation}
Combining these, we can now compute $\mat{\widehat{K}}(\vec{q},s)\cdot \vec{\widehat{A}}(\vec{q},s)=\left[\mat{I}-\frac{1}{s}\qty(\Delta \mat{\widehat{\phi}}(\vec{q},s)\cdot \mat{g}^{-1})  \right]^{-1}\cdot
\qty(\Delta \mat{\widehat{\phi}}(\vec{q},s) \cdot \mat{g}^{-1} )\cdot \vec{\widehat{A}}(\vec{q},s)$ as 
\begin{equation} 
\begin{split}
&\mat{\widehat{K}}(\vec{q},s)\cdot \vec{\widehat{A}}(\vec{q},s)=\left[\mat{I}-\frac{1}{s}\qty(\Delta \mat{\widehat{\phi}}(\vec{q},s)\cdot \mat{g}^{-1})  \right]^{-1}\cdot
\qty(\Delta \mat{\widehat{\phi}}(\vec{q},s) \cdot \mat{g}^{-1} )\cdot \vec{\widehat{A}}(\vec{q},s)
\label{}
\\  \\
&=
\begin{pmatrix}
\mat{\widehat{Y}} & \mat{0}\\ \\
\mat{0} & \mat{\widehat{Z}} \\
\end{pmatrix}
\begin{pmatrix}
\mat{I} & -\mat{\widehat{V}} \cdot \mat{\widehat{X}}^{-1}\\ \\
-\mat{\widehat{W}} \cdot \mat{\widehat{U}}^{-1} & \mat{I} \\
\end{pmatrix}
\begin{pmatrix}
\qty(i\vec{q}\cdot \Delta \Tfor{\widehat{N}}^{\sigma \sigma}):\mat{\widehat{\gamma}}+ \qty(i\vec{q}\cdot \Delta \Tthr{\widehat{N}}^{\sigma R})\cdot \Delta \vec{\widehat{\mu}}/\rho_0   \\ \\ 
\Delta  \Tthr{\widehat{N}}^{R \sigma}:   \mat{\widehat{\gamma} } + \Delta \mat{\widehat{N}}^{RR}\cdot   {\Delta} \vec{\widehat{\mu}} / \rho_0 
\\
\end{pmatrix}. 
\end{split}
\end{equation}
The first component is 
\begin{equation}
\begin{split}
 \qty[\mat{\widehat{K}}(\vec{q},s)\cdot \vec{\widehat{A}}(\vec{q},s)]^{(1)} =     \mat{\widehat{Y}}\cdot \Bigg \{ i\vec{q}\cdot \Bigg[ \qty(\Delta  \Tfor{\widehat{N}}^{\sigma \sigma}+ \frac{1}{s\rho_0} \Delta \Tthr{\widehat{N}}^{\sigma R} \cdot \mat{\widehat{X}}^{-1}\cdot \Delta \Tthr{\widehat{N}}^{R \sigma} ) :  \mat{\widehat{\gamma}}  \\[1mm]
 +\qty(\Delta \Tthr{\widehat{N}}^{\sigma R}+\frac{1}{s\rho_0} \Delta \Tthr{\widehat{N}}^{\sigma R}\cdot \mat{\widehat{X}}^{-1}\cdot \Delta \mat{\widehat{N}}^{RR})\cdot{\Delta} \vec{\widehat{\mu}} / \rho_0 \Bigg]  \Bigg \}\quad,
 \end{split}
\end{equation}
and the second component is 
\begin{equation}
\begin{split}
 \qty[\mat{\widehat{K}}(\vec{q},s)\cdot \vec{\widehat{A}}(\vec{q},s)]^{(2)} =  
\mat{\widehat{Z}}\cdot \Bigg \{ \qty(\frac{1}{s\rho_0} \qty(\Delta \Tthr{\widehat{N}}^{R \sigma} \cdot \vec{q}) \cdot \mat{\widehat{U}}^{-1} \cdot \qty(\vec{q} \cdot \Delta \Tfor{\widehat{N}}^{\sigma \sigma} )+ \Delta \Tthr{\widehat{N}}^{R \sigma} ):  \mat{\widehat{\gamma}}    \\[1mm]
+\qty(\frac{1}{s\rho_0} \qty(\Delta \Tthr{\widehat{N}}^{R \sigma} \cdot \vec{q})\cdot \mat{\widehat{U}}^{-1}\cdot \qty(\vec{q}\cdot \Delta \Tthr{\widehat{N}}^{\sigma R}) + \Delta \mat{\widehat{N}}^{RR} ) \cdot {\Delta} \vec{\widehat{\mu}} /\rho_0 \Bigg \} \quad.
 \end{split}
\end{equation}

Combining this with 
\begin{equation}
    \frac{d}{dt}\expval{\vec{A}(\vec{q},t)} = - 
    \begin{pmatrix}
 \expval{\mat{\sigma}(\vec{q},t)}\cdot i \,\vec{q}\\ \\
\expval{\vec{R}(\vec{q},t)}\\
\end{pmatrix}
\quad,
\end{equation}
we obtain in the Laplace space,
\begin{equation}
\begin{split}
\expval{\widehat{\mat{\sigma}}(\vec{q},s)} =\, &   \mat{\widehat{Y}}\cdot\qty(\Delta  \Tfor{\widehat{N}}^{\sigma \sigma} + \frac{1}{s\rho_0} \Delta \Tthr{\widehat{N}}^{\sigma R} \cdot \mat{\widehat{X}}^{-1}\cdot \Delta \Tthr{\widehat{N}}^{R \sigma} ) :  \expval{\mat{\widehat{\gamma}}} \\[2mm]
   &+ \frac{1}{\rho_0} \mat{\widehat{Y}}\cdot \qty(\Delta \Tthr{\widehat{N}}^{\sigma R}+\frac{1}{s\rho_0}\Delta \Tthr{\widehat{N}}^{\sigma R}\cdot \mat{\widehat{X}}^{-1}\cdot \Delta \mat{\widehat{N}}^{RR}) \expval{\Delta \vec{\widehat{\mu}}}\quad,
\end{split}
\end{equation}
and 
\begin{equation}
\begin{split}
\expval{\widehat{\vec{R}}(\vec{q},s)} =&\,  \mat{\widehat{Z}}\cdot \qty(\frac{1}{s\rho_0} \qty(\Delta \Tthr{\widehat{N}}^{R \sigma} \cdot \vec{q}) \cdot \mat{\widehat{U}}^{-1} \cdot \qty(\vec{q} \cdot \Delta \Tfor{\widehat{N}}^{\sigma \sigma}) + \Delta \Tthr{\widehat{N}}^{R \sigma} ): \expval{\mat{\widehat{\gamma}} } \\[2mm]  &+\frac{1}{\rho_0} \mat{\widehat{Z}}\cdot\qty(\frac{1}{s\rho_0}\qty(\Delta \Tthr{\widehat{N}}^{R \sigma} \cdot \vec{q})\cdot \mat{\widehat{U}}^{-1}\cdot \qty(\vec{q}\cdot \Delta \Tthr{\widehat{N}}^{\sigma R}) + \Delta \mat{\widehat{N}}^{RR} ) \cdot \expval{\Delta \vec{\widehat{\mu}}}\quad. 
\end{split}
\end{equation}
Converting $\vec{\widehat{R}}(\vec{q},s)$ to $\vec{\widehat{r}}(\vec{q},s)$ [Eq.~\eqref{eq:R_r}] we obtain the constitutive equations for the stress and chemical reaction rates, 
\begin{equation}
\begin{split}
\expval{\widehat{\mat{\sigma}}(\vec{q},s)} =\, &  \mat{\widehat{Y}}\cdot\qty(\Delta  \Tfor{\widehat{N}}^{\sigma \sigma}+ \frac{1}{s\rho_0} \Delta \Tthr{\widehat{N}}^{\sigma R} \cdot \mat{\widehat{X}}^{-1}\cdot \Delta \Tthr{\widehat{N}}^{R \sigma} ) :  \expval{\mat{\widehat{\gamma}} } \\[2mm]
   &+ \frac{1}{\rho_0}\mat{\widehat{Y}}\cdot \qty(\Delta \Tthr{\widehat{N}}^{\sigma R}+\frac{1}{s\rho_0}\Delta \Tthr{\widehat{N}}^{\sigma R}\cdot \mat{\widehat{X}}^{-1}\cdot \Delta \mat{\widehat{N}}^{RR})\cdot   \expval{\Delta \vec{\widehat{\mu}}}\quad,
\end{split}
\end{equation}
and 
\begin{equation}
\begin{split}
\expval{\widehat{\vec{r}}(\vec{q},s)} =\,& \mat{\kappa}^{-1}\cdot \mat{\widehat{Z}}\cdot \qty(\frac{1}{s\rho_0}\qty( \Delta \Tthr{\widehat{N}}^{R \sigma} \cdot \vec{q} )\cdot \mat{\widehat{U}}^{-1} \cdot \qty(\vec{q} \cdot \Delta \Tfor{\widehat{N}}^{\sigma \sigma}) + \Delta \Tthr{\widehat{N}}^{R \sigma} ): \expval{\mat{\widehat{\gamma}} } \\[2mm]  &+\frac{1}{\rho_0}\mat{\kappa}^{-1} \cdot \mat{\widehat{Z}}\cdot\qty(\frac{1}{s\rho_0} \qty(\Delta \Tthr{\widehat{N}}^{R \sigma} \cdot \vec{q})\cdot \mat{\widehat{U}}^{-1}\cdot \qty(\vec{q}\cdot \Delta \Tthr{\widehat{N}}^{\sigma R} )+ \Delta \mat{\widehat{N}}^{RR} ) \cdot   \expval{\Delta \vec{\widehat{\mu}}}\quad. 
\end{split}
\end{equation}
They are explicitly written using index as follows: 
\begin{equation}
\begin{split}
\label{eq:CE_sigma_index_literal_nofunc}
\expval{\widehat{\sigma}_{\alpha\beta}(\vec{q},s)}
=\;&
\widehat{Y}_{\alpha \xi}\;
\Biggl[
\Delta \widehat{N}^{\sigma\sigma}_{\xi\beta \gamma \zeta }
+\frac{1}{s\rho_0}\,
\Delta \widehat{N}^{\sigma R}_{\xi\beta,I}\;
(\widehat{X}^{-1})_{IJ}\;
\Delta \widehat{N}^{R\sigma}_{J, \gamma \zeta }
\Biggr]\;
\expval{\widehat{\gamma}_{\gamma\zeta}}
\\[1mm]
&\;+\frac{1}{\rho_0}\,
\widehat{Y}_{\alpha \xi}\;
\Biggl[
\Delta \widehat{N}^{\sigma R}_{\xi\beta,K}
+\frac{1}{s\rho_0}\,
\Delta \widehat{N}^{\sigma R}_{\xi\beta,I}\;
(\widehat{X}^{-1})_{IJ}\;
\Delta \widehat{N}^{RR}_{JK}
\Biggr]\;
\expval{\Delta \widehat{\mu}_{K}}\quad.
\end{split}
\end{equation}

\begin{equation}
\begin{split}
\label{eq:CE_r_index_literal_nofunc}
\expval{\widehat{r}_{I}(\vec{q},s)}
=\;&
(\kappa^{-1})_{IJ}\;
\widehat{Z}_{JK}\;
\Biggl[
\frac{1}{s\rho_0}\,
\qty(\Delta \widehat{N}^{R\sigma}_{K,\alpha \beta}\;
q_{\beta})\;
(\widehat{U}^{-1})_{\alpha\gamma}\;
\qty(q_{\delta}\;
\Delta \widehat{N}^{\sigma\sigma}_{\gamma\delta \zeta \epsilon})
+\Delta \widehat{N}^{R\sigma}_{K, \zeta \epsilon}
\Biggr]\;
\expval{\widehat{\gamma}_{\zeta \epsilon}}
\\[1mm]
&\;+\frac{1}{\rho_0}\,
(\kappa^{-1})_{IJ}\;
\widehat{Z}_{JK}\;
\Biggl[
\frac{1}{s\rho_0}\,
\qty(\Delta \widehat{N}^{R\sigma}_{K,\alpha \beta}\;
q_{\beta})\;
(\widehat{U}^{-1})_{\alpha\gamma}\;
\qty(q_{\delta}\;
\Delta \widehat{N}^{\sigma R}_{\gamma\delta,L})
+\Delta \widehat{N}^{RR}_{KL}
\Biggr]\;
\expval{\Delta \widehat{\mu}_{L}}\quad.
\end{split}
\end{equation}
The generalised transport coefficients in the index forms are given as 
\begin{align}
\label{eq:coefficients_index}
&\widehat{\Pi}_{\alpha\beta \gamma \zeta}(\vec q,s)
\equiv
\widehat{Y}_{\alpha\xi}\,
\Biggl[
\Delta \widehat{N}^{\sigma\sigma}_{\xi\beta  \gamma \zeta}
+\frac{1}{s\rho_0}\,
\Delta \widehat{N}^{\sigma R}_{\xi\beta,I}\,
(\widehat{X}^{-1})_{IJ}\,
\Delta \widehat{N}^{R\sigma}_{J, \gamma \zeta }
\Biggr]\,
\quad,
\\[2mm]
&\widehat{\Lambda}_{\alpha\beta,I}(\vec q,s)
\equiv
\frac{1}{\rho_0}\,
\widehat{Y}_{\alpha \xi}\,
\Biggl[
\Delta \widehat{N}^{\sigma R}_{\xi\beta,I}
+\frac{1}{s\rho_0}\,
\Delta \widehat{N}^{\sigma R}_{\xi\beta,J}\,
(\widehat{X}^{-1})_{JK}\,
\Delta \widehat{N}^{RR}_{KI}
\Biggr]\,
\quad,
\\[2mm]
&\widehat{\Xi}_{I,\zeta \gamma}(\vec q,s)
\equiv
(\kappa^{-1})_{IJ}\,
\widehat{Z}_{JK}\,
\Biggl[
\qty(\Delta \widehat{N}^{R\sigma}_{K,\alpha \beta}\;
q_{\beta})\;
(\widehat{U}^{-1})_{\alpha\delta}\;
\qty(q_{\epsilon}\;
\Delta \widehat{N}^{\sigma\sigma}_{\delta\epsilon \zeta \gamma})\;
\frac{1}{s\rho_0}
+\Delta \widehat{N}^{R\sigma}_{K,\zeta \gamma}
\Biggr]\,
\quad,
\\[2mm]
&\widehat{\Upsilon}_{I J}(\vec q,s)
\equiv
\frac{1}{\rho_0}\,
(\kappa^{-1})_{I K}\,
\widehat{Z}_{K L}\,
\Biggl[
\frac{1}{s\rho_0}\,
\qty(\Delta \widehat{N}^{R\sigma}_{L,\alpha \beta}\;
q_{\beta})\;
(\widehat{U}^{-1})_{\alpha\delta}\;
\qty(q_{\epsilon}\;
\Delta \widehat{N}^{\sigma R}_{\delta\epsilon,J})
+\Delta \widehat{N}^{RR}_{LJ}
\Biggr]\,
\quad.
\end{align}

Eq.~(\ref{eq:CE_sigma_index_literal_nofunc})-(\ref{eq:CE_r_index_literal_nofunc}) are the constitutive equations for active fluids, driven by arbitrary number of chemical reactions. The index-free representation of the constitutive equations are shown in Eqs.~(\ref{eq:CE_general_shear}-\ref{eq:coefficients_upsilon}) in the main text.

\subsection{Simplified constitutive equations: Isotropic achiral fluids driven by single chemical reaction \label{App:simplified}}
We consider the scenario where a single chemical reaction such as ATP hydrolysis drives the system. When the single chemical reaction $I$ drives the system,  the above constitutive equations take simpler form,
\begin{equation}
\begin{split}
\label{eq:single_I_CE_gamma}
\expval{\widehat{\sigma}_{\alpha\beta}(\vec{q},s)}
=\;&
\widehat{Y}_{\alpha\xi}\;
\Biggl[
\Delta \widehat{N}^{\sigma\sigma}_{\xi\beta \gamma \zeta}
+\frac{1}{s\rho_0}\,
\Delta \widehat{N}^{\sigma R}_{\xi\beta,I}\;
\Delta \widehat{N}^{R\sigma}_{I, \gamma \zeta}/\widehat{X}_{II}
\Biggr]\;
\expval{\widehat{\gamma}_{\gamma\zeta}}
\\[1mm]
&\;+\frac{1}{\rho_0}\,
\widehat{Y}_{\alpha \xi}\;
\Biggl[
\Delta \widehat{N}^{\sigma R}_{\xi\beta,I}
+\frac{1}{s\rho_0}\,
\Delta \widehat{N}^{\sigma R}_{\xi\beta,I}\;
\Delta \widehat{N}^{RR}_{II}/\widehat{X}_{II}
\Biggr]\;
\expval{\Delta \widehat{\mu}_{I}}\quad.
\end{split}
\end{equation}

\begin{equation}
\begin{split}
\label{eq:single_I_CE_r}
\expval{\widehat{r}_{I}(\vec{q},s)}
=\;& \frac{\widehat{Z}_{II}}{\kappa_{II} }
\Biggl[
\frac{1}{s\rho_0}\,
\qty(\Delta \widehat{N}^{R\sigma}_{I,\alpha \beta}\;
q_{\beta})\;
(\widehat{U}^{-1})_{\alpha\gamma}\;
\qty(q_{\delta}\;
\Delta \widehat{N}^{\sigma\sigma}_{\gamma\delta \zeta \epsilon})
+\Delta \widehat{N}^{R\sigma}_{I,\zeta \epsilon}
\Biggr]\;
\expval{\widehat{\gamma}_{\zeta \epsilon}}
\\[1mm]
&\;+\frac{1}{\rho_0}
\frac{\widehat{Z}_{II}}{\kappa_{II} }
\Biggl[
\frac{1}{s\rho_0}\,
\qty(\Delta \widehat{N}^{R\sigma}_{I,\alpha \beta}\;
q_{\beta})\;
(\widehat{U}^{-1})_{\alpha\gamma}\;
\qty(q_{\delta}\;
\Delta \widehat{N}^{\sigma R}_{\gamma\delta,I})
+\Delta \widehat{N}^{RR}_{II}
\Biggr]\;
\expval{\Delta \widehat{\mu}_{I}}\quad.
\end{split}
\end{equation}
There is no summation over the repeated $I$ implied.

To further simplify the constitutive equations, we obtain the expressions for achiral isotropic fluids in two dimensions in the coordinate frame where $y$-axis points along $\vec{q}$\,. In this frame, $\vec{q} = (0,q)$, therefore the non-symmetrized strain $\widehat{\gamma}_{yx}(\vec{q},s)=\widehat{\gamma}_{xx}(\vec{q},s) = 0$. We further exploit the symmetry constraints of isotropic achiral tensor fields, for which only combinations with an even number of identical spatial indices are non-zero in even-rank tensors~\cite{wittmer2023correlations}. We apply this symmetry constraint to the material properties, i.e., generalized transport coefficients.  Then Eqs.~(\ref{eq:single_I_CE_gamma}-\ref{eq:single_I_CE_r}) become

\begin{equation}
\begin{split}
\label{eq:CE_singleI_shear_new}
\expval{\widehat{\sigma}_{xy}}
=\;& 
\widehat{Y}_{xx}\;
\Delta \widehat{N}^{\sigma\sigma}_{xyxy}
\expval{\widehat{\gamma}_{xy}}\quad.
\end{split}
\end{equation}

\begin{equation}
\begin{split}
\label{eq:CE_singleI_xx_new_correct}
\expval{\widehat{\sigma}_{xx}}
=\;&
\widehat{Y}_{xx}\;
\Biggl(
\Delta \widehat{N}^{\sigma\sigma}_{x x y y}
+\frac{1}{s\rho_0}\,
\frac{\Delta \widehat{N}^{\sigma R}_{x x,I}\;
      \Delta \widehat{N}^{R\sigma}_{I, y y}}{\widehat{X}_{II}}
\Biggr)\;
\expval{\widehat{\gamma}_{yy}}
\\[1mm]
&\;+\frac{1}{\rho_0}\,
\widehat{Y}_{xx}\;
\Delta \widehat{N}^{\sigma R}_{x x,I}\Biggl(1
+\frac{1}{s\rho_0}\,\Delta \widehat{N}^{RR}_{II}/\widehat{X}_{II}
\Biggr)\;
\expval{\Delta \widehat{\mu}_{I}}\quad.
\end{split}
\end{equation}

\begin{equation}
\begin{split}
\label{eq:CE_singleI_yy_new}
\expval{\widehat{\sigma}_{yy}}
=\;&
\widehat{Y}_{yy}\;
\Biggl(
\Delta \widehat{N}^{\sigma\sigma}_{y y y y}
+\frac{1}{s\rho_0}\,
\frac{\Delta \widehat{N}^{\sigma R}_{y y,I}\;
      \Delta \widehat{N}^{R\sigma}_{I, y y}}{\widehat{X}_{II}}
\Biggr)
\expval{\widehat{\gamma}_{yy}}
\\[1mm]
&\;+\frac{1}{\rho_0}\,
\widehat{Y}_{yy}\;
\Delta \widehat{N}^{\sigma R}_{y y,I}\Biggl(1
+\frac{1}{s\rho_0}\,\Delta \widehat{N}^{RR}_{II}/\widehat{X}_{II}
\Biggr)
\expval{\Delta \widehat{\mu}_{I}}\;.
\end{split}
\end{equation}

\begin{equation}
\begin{split}
\label{eq:CE_singleI_r_new}
\expval{\widehat{r}_{I}}
=\;&
\frac{\widehat{Z}_{II}}{\kappa_{II}}
\Biggl[
\frac{q^2}{s\rho_0}\,
\Delta \widehat{N}^{R\sigma}_{I, y y}\;
(\widehat{U}^{-1})_{y y}\;
\Delta \widehat{N}^{\sigma\sigma}_{y y y y}
+\Delta \widehat{N}^{R\sigma}_{I, y y}
\Biggr]
\expval{\widehat{\gamma}_{yy}}
\\[1mm]
&\;+\frac{1}{\rho_0}\,
\frac{\widehat{Z}_{II}}{\kappa_{II}}
\Biggl[
\frac{q^2}{s\rho_0}\,
\Delta \widehat{N}^{R\sigma}_{I, y y}\;
(\widehat{U}^{-1})_{y y}\;
\Delta \widehat{N}^{\sigma R}_{y y,I}
+\Delta \widehat{N}^{RR}_{II}
\Biggr]
\expval{\Delta \widehat{\mu}_{I}}\quad.
\end{split}
\end{equation}
Substituting $({\widehat{U}}^{-1})_{yy}$, $\widehat{X}_{II}$, $\widehat{Y}_{xx}$, $\widehat{Y}_{yy}$, and $\widehat{Z}_{II}$
into Eqs.~\eqref{eq:CE_singleI_shear_new}-\eqref{eq:CE_singleI_r_new} yields the simplified constitutive equations quoted in Eqs.~\eqref{eq:CE_single_reaction}-\eqref{eq:CE_single_r} with the memory kernels in Eqs.~\eqref{eq:TP1}-\eqref{eq:TP4}.

\subsection{Onsager reciprocity \label{App:Onsager_j_mu}}
A fundamental consequence of microscopic time-reversal symmetry is the Onsager reciprocal relations, which state that the matrix of phenomenological coefficients linking thermodynamic fluxes to forces must be related near equilibrium. Here we show that, in equilibrium, the chemo-mechanical coupling coefficients in Eqs.~\eqref{eq:TP2}-\eqref{eq:TP3} follows the Onsager reciprocal relation: $\widehat{\Lambda}(\vec{q},s)=-\widehat{\Xi}(\vec{q},s)$. Note that the minus sign stems from the distinct time reversal signatures for the momentum density and chemical potential difference (see Appendix~\ref{sec:symmetry}). 
Using the Kubo formulas in equilibrium, Eq.~(\ref{eq:Kubo_j}) and Eq.~(\ref{eq:kubo_mu}), to the cross-coupling coefficients, Eq.~(\ref{eq:chi_1}),
we obtain
\begin{equation} 
\begin{split}
\label{eq:codition1}
\widehat{\chi}_{yy,I}(\vec{q},s) 
=\frac{\expval{\widehat{\sigma}_{y y}(\vec{q},s)R^*_{I}(\vec{q})}-\expval{\widehat{\sigma}_{y y}(\vec{q},0)R^{*}_{I}(\vec{q})}}{\kappa_{II} V k_BT} \qquad,
\end{split}
\end{equation}
and
\begin{equation} 
\begin{split}
\label{eq:codition2}
\widehat{\chi}_{I,yy}(\vec{q},s) 
=\frac{\expval{\widehat{R}_{I}(\vec{q},s)\sigma_{y y}^*(\vec{q})}-\expval{\widehat{R}_{I}(\vec{q},0)\sigma_{y y}^*(\vec{q})}}{\kappa_{II} V k_BT} \quad.
\end{split}
\end{equation}
The time-reversal signature of $\sigma_{yy}(\vec{q},t)$ and $R_I(\vec{q},t)$ are distinct, therefore the correlation satisfies (see Appendix~\ref{sec:symmetry}), 
\begin{equation} 
\begin{split}
\label{eq:codition3}
\expval{\widehat{\sigma}_{y y}(\vec{q},s)R^{*}_{I}(\vec{q})}=- \expval{\widehat{R}_{I}(\vec{q},s)\sigma_{y y}^*(\vec{q})} \quad.
\end{split}
\end{equation}
Combining Eqs.~(\ref{eq:codition1})-(\ref{eq:codition3}) leads to $\widehat{\chi}_{yy,I}(\vec{q},s)=-\widehat{\chi}_{I,yy} (\vec{q},s)$ and thus $\widehat{\Lambda}(\vec{q},s)=-\widehat{\Xi} (\vec{q},s)$. This proves Onsager reciprocal relation. In a non-equilibrium situation in general, however, the chemo-mechanical coupling coefficients $\widehat{\Lambda}$ and $\widehat{\Xi}$ need not respect Onsager reciprocity.

\subsection{Memory kernels using specific forms of correlation functions \label{App:Special}}
In this section, we list the generalized transport coefficients evaluated for the choice of the correlation functions given in Eq.~(\ref{eq:N_special}). Substituting the chosen form of the correlation functions into the generalized transport coefficients, Eqs.~\eqref{eq:TPzeta}-\eqref{eq:TP4}, we obtain the following expressions in Fourier-Laplace space:
\begin{equation} 
\begin{split}
\label{eq:zeta_spetial_laplace}
&\widehat{\zeta}(q,s)=\frac{\rho_0\left(-\,\varepsilon \kappa_{II}\nu^2 s\, (\eta^\parallel q^{2}+\rho_0 s)
	+	\eta^\parallel \rho_0(\lambda+s)\big(s+\lambda q^{2}\xi_\mu^{2}\big)^{2}\right)}
{\;\varepsilon \kappa_{II} \nu^2 q^{2}\,(\eta^\parallel q^{2}+\rho_0 s)
	+	\rho^2_0(\lambda+s)\big(s+\lambda q^{2}\xi_\mu^{2}\big)^{2}}\quad; \\ \\
&\widehat{\Lambda}(q,s)=\frac{\nu \rho_0(\lambda+s)(\eta^{\parallel} q^{2}+\rho_0 s)(s+\lambda q^{2}\xi_\mu^{2})}
{\;\varepsilon \kappa_{II}\nu^2 q^{2}\,(\eta^{\parallel} q^{2}+\rho_0 s)
+\rho^2_0(\lambda+s)(s+\lambda q^{2}\xi_\mu^{2})^{2}}\quad;  \\ \\ 
&\widehat{\Xi}(q,s)=-\frac{\varepsilon\,\kappa_{II}\,\nu \rho_0\,(\lambda+s)\big(\eta^{\parallel} q^{2}+\rho_0 s\big)\big(s+\lambda q^{2}\xi_\mu^{2}\big)}
{\;\kappa_{II}\left(\varepsilon\,\kappa_{II}\nu^{2} q^{2}\big(\eta^{\parallel} q^{2}+\rho_0 s\big)
+\rho^2_0(\lambda+s)\big(s+\lambda q^{2}\xi_\mu^{2}\big)^{2}\right)} \quad; \\ \\ 
&\widehat{\Upsilon}(q,s)= \frac{-\varepsilon \, \kappa_{II} \, \nu^2 \, q^{2}\,  s \big(\eta^{\parallel} q^{2}+\rho_{0} s\big)
+\lambda \rho^2_{0}(\lambda+s)\big(s+\lambda q^{2}\xi_{\mu}^{2}\big)^{2}}
{\kappa_{II}\left(\varepsilon\,\kappa_{II}\,\nu^{2} q^{2}\big(\eta^{\parallel} q^{2}+\rho_{0} s\big)
+\rho^2_{0}(\lambda+s)\big(s+\lambda q^{2}\xi_{\mu}^{2}\big)^{2}\right)}\quad.
\end{split}
\end{equation}

\subsection{Physical parameter regimes for the phenomenological correlation functions}
\label{app:correlation_realizability}

In this section we derive the parameter regimes where the phenomenological correlation functions introduced in Eq.~\eqref{eq:N_special} stay physically consistent. We use the longitudinal momentum and chemical-potential fluctuations considered in the Section~\ref{sec: simplified_active},
\begin{equation}
    \vec{A}
    =
    \left(
        j_y,
        \Delta\mu_I
    \right)^{\top}
    \quad .
\end{equation}

The correlation functions introduced in Eq.~\eqref{eq:N_special} must correspond to a time-translationally invariant process. This implies a positivity condition on the correlation spectrum:
\begin{equation}
    {\mat{\widetilde{\psi}}}(q,\omega) \succeq0 \quad.
    \label{eq:psiposi}
\end{equation}
Let
\begin{equation}
    \mat {\widehat P}(q,s)
    \equiv
    \mat{g}^{-1/2}
    \cdot
    {\mat{\widehat{\psi}}}(q,s)
    \cdot
    \mat{g}^{-1/2}
\end{equation}
denote the one-sided correlation matrix normalized by the static covariance. Since $\mat g$ is Hermitian positive definite, $\mat g^{-1/2}$ is Hermitian. This transform does not alter the positive semidefiniteness of ${\mat{\widetilde{\psi}}}(q,\omega)$.
Using the stationary correlation identity, Eq.~\eqref{eq: SecondD_in_Laplace}, together with
the functions in Eq.~\eqref{eq:N_special} with $\nu_{\sigma R}=\nu$ and
$\nu_{R\sigma}=-\varepsilon\nu$, we obtain
\begin{equation}
    \widehat{\mat{P}}(q,s)
    =
    \begin{pmatrix}
        \displaystyle
        \frac{1}{s+\eta^{\parallel}q^2/\rho_0}
        &
        \displaystyle
        -\frac{
            i q\nu\sqrt{\kappa_{II}/\rho_0}\sqrt{\bar{\rho}}
        }{
            (s+\lambda)(s+\lambda q^2\xi_\mu^2)
        }
        \\[5mm]
        \displaystyle
        -\frac{
            i q\varepsilon\nu\sqrt{\kappa_{II}/\rho_0}/\sqrt{\bar{\rho}}
        }{
            (s+\lambda)(s+\lambda q^2\xi_\mu^2)
        }
        &
        \displaystyle
        \frac{1}{s+\lambda}
    \end{pmatrix} \quad,
    \quad 
    \label{eq:P_matrix_note}
\end{equation}
where we have defined non-dimensional parameter
\begin{equation}
    \bar{\rho}
    \equiv
    \frac{\rho_0 g_{\mu\mu}}
         {\kappa_{II}g_{jj}}
    \quad .
    \label{eq:rho_bar_definition}
\end{equation}
The equilibrium Kubo relations $g_{jj}=\rho_0Vk_{\mathrm B}T$ and
$g_{\mu\mu}=\kappa_{II}Vk_{\mathrm B}T$ correspond to
$\bar{\rho}=1$.

\begin{figure}[h]
    \centering
    \includegraphics[width=\linewidth]{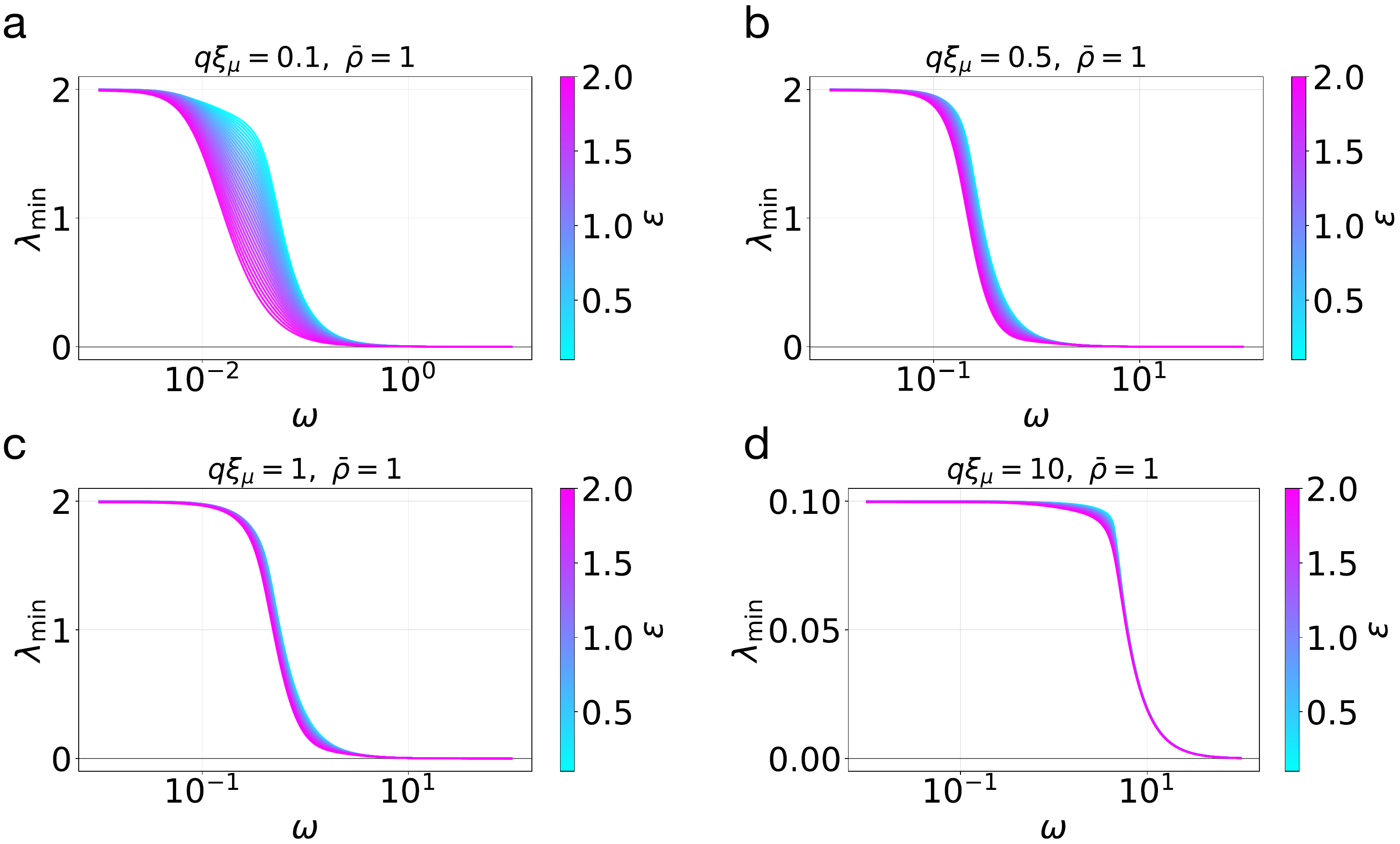}
    \caption{Minimum eigenvalue of the two-sided spectrum in Eq.~\eqref{eq:Sqw} as a function of frequency $\omega$ for $\bar{\rho}=1$, shown for several wavelengths $q$ and different degrees of Onsager-reciprocity breaking $\varepsilon$. A non-negative minimum eigenvalue at all
frequencies is required for the spectrum to be positive semidefinite and hence physically realizable.
Parameters: $\rho_0=5$, $\eta^{\parallel}=1$, $\lambda=1$, $\xi_\mu=1$, $\kappa_{II}=1$ and $\nu=0.6$
(arbitrary units). }
    \label{fig:positive_definite}
\end{figure}

We define the two-sided power spectrum,
\begin{equation}
\label{eq:Sqw}
    \mat{\widetilde{S}}(q,\omega)
    \equiv
    \mat{\widehat{P}}
    \left(q,i\omega\right)
    +
    \mat{\widehat{P}}^\dagger
    \left(q,i\omega\right)
    \quad .
\end{equation}
This must be Hermitian positive semidefinite at every $q$ and $\omega$,
\begin{equation}
    {\mat{\widetilde{S}}}(q,\omega)
    \succeq
    \mat{0}
    \quad .
    \label{eq:spectral_realizability}
\end{equation}
The diagonal elements of ${\mat{\widetilde{S}}}$ are non-negative by construction. The nontrivial requirement is therefore
\begin{equation}
    \left|
        \widetilde S_{j\mu}(q,\omega)
    \right|^2
    \leq
    \widetilde S_{jj}(q,\omega)
    \widetilde S_{\mu\mu}(q,\omega)
    \quad .
    \label{eq:spectral_cauchy_schwarz}
\end{equation}
For this condition to be satisfied for all $\omega$ and $q$, the parameters have to follow the condition: 
\begin{equation}
    \frac{\kappa_{II}\nu^2}{4\eta^\parallel\lambda}
    \frac{(\bar{\rho}+\varepsilon)^2}{\bar{\rho}}
    \max\left[
        1,
        \left(
            \frac{\eta^\parallel}
            {\rho_0\lambda\xi_\mu^2}
        \right)^2
    \right]
    \leq1
    \quad .
    \label{eq:correlation_realizability_condition}
\end{equation}
Figure~\ref{fig:positive_definite} shows the minimum eigenvalue of $\mat{\widetilde{S}}(q,\omega)$ in Eq.~\eqref{eq:Sqw} for $\bar{\rho}=1$ with the parameters used in the figures \ref{fig:relaxation}-\ref{fig:Gs}. The plot stays non-negative for all frequency regimes therefore satisfies the condition Eq.~\eqref{eq:spectral_realizability}.

 \section{Mass-Momentum Density fluctuations \label{App:mass_momentum}}
\subsection{Summary of the results}
In this section we present the stress constitutive equation obtained when the momentum density and mass density are included simultaneously as dynamical variables. The detailed derivations are given in Section~\ref{App:CE_j_rho}. This extended choice captures the compressional mode explicitly, and thereby allows the static bulk-response contribution to enter the stress constitutive relation. We choose the variables
\begin{equation}
\begin{split}
\label{eq:A_j_rho_def}
\vec A(\vec q,t)
=
\qty(\vec j(\vec q,t)\,,\,\rho(\vec q,t))^{\top}\quad,
\end{split}
\end{equation}
where $\vec j$ and $\rho$ denote the momentum density and mass density, respectively. The equal-time correlation matrix is
\begin{equation}
\begin{split}
\label{eq:C_j_rho_def}
\mat g(\vec q)=
\begin{pmatrix}
\expval{\vec j(\vec q)\,\vec j^{\dagger}(\vec q)}
&
\expval{\vec j(\vec q)\,\rho^{\dagger}(\vec q)}
\\\\
\expval{\rho(\vec q)\,\vec j^{\dagger}(\vec q)}
&
\expval{\rho(\vec q)\,\rho^{\dagger}(\vec q)}
\end{pmatrix}\quad.
\end{split}
\end{equation}
For simplicity we consider 
\begin{equation}
\label{eq:no_static_cross_rho}
\expval{\vec{j}(\vec{q})\, {\rho}^{\dagger}(\vec{q})}
=
\expval{\rho(\vec{q})\,\vec{j}^\dagger(\vec{q})}
=\vec{0}\quad.
\end{equation}

We have two conservation laws: 
\begin{equation}
\label{eq:cons_laws_j_rho}
\partial_t \vec j(\vec q,t)= -i\,\mat\sigma(\vec q,t)\cdot \vec q\quad,
\qquad
\partial_t \rho(\vec q,t)= i\,\vec q\cdot \vec j(\vec q,t)\quad. 
\end{equation}
These conservation laws with $\gamma_{\alpha\beta}=-iq_\beta j_\alpha/\rho_0$ in Laplace space gives 
\begin{equation}
\label{eq:rho_trace_gamma_relation_concise}
\widehat{\rho}(\vec q,s)= -\frac{\rho_0}{s}\,\mathrm{tr}\,\widehat{\mat{\gamma}}(\vec q,s)\quad,
\end{equation}
where the initial mean density fluctuation vanishes,
$\langle\rho(\vec q)\rangle=0$.
Therefore, stress contribution mediated by $\rho$ appears proportional to $\mathrm{tr}\,\widehat{\mat{\gamma}}$\,, i.e.\ the compressional mode.

After the calculation analogous to the previous sections, we obtain the constitutive relation for the stress in Fourier-Laplace space,
\begin{equation}
\label{eq:CE_sigma_j_rho_compact}
\expval{\widehat{\mat{\sigma}}(\vec q,s)}
=
\Tfor{\widehat{\Pi}}(\vec q,s):\expval{\widehat{\mat{\gamma}}(\vec q,s)}
+\mat{\widehat{\zeta}}(\vec q,s)\,\expval{\mathrm{tr}\,\widehat{\mat{\gamma}}(\vec q,s)}\quad,
\end{equation}
where the generalized viscosity tensor $\Tfor{\widehat{\Pi}}$ and the bulk coupling tensor $\mat{\widehat{\zeta}}$ are
\begin{equation}
\label{eq:eta_def_j_rho}
\Tfor{\widehat{\Pi}}(\vec q,s)
\equiv
\mat{\widehat Y}\cdot
\Biggl(
\Delta\Tfor{\widehat N}^{\sigma\sigma}
+\frac{1}{s\rho_0}\,\widehat X^{-1}\,\qty[\qty(\Delta\Tthr{\widehat N}^{\sigma j}\cdot \vec{q})\qty( \vec{q}\cdot \Delta\Tthr{\widehat N}^{j\sigma} )]
\Biggr)\quad,
\end{equation}
\begin{equation}
\label{eq:zeta_def_j_rho}
\mat{\widehat{\zeta}}(\vec q,s)
\equiv
-\frac{i}{s}\,
\mat{\widehat Y}\cdot
\qty(\Delta\Tthr{\widehat N}^{\sigma j}\cdot \vec q)\,
\widehat{X}^{-1}\quad.
\end{equation}
The constitutive equation is written in the index form, 
\begin{equation}
\begin{split}
\label{eq:CE_sigma_j_rho_compact_index}
\big\langle \widehat{\sigma}_{\alpha\beta}(\vec q,s)\big\rangle
=\;&
\widehat{\Pi}_{\alpha \beta \zeta \gamma}(\vec{q},s) \big\langle \widehat{\gamma}_{\zeta \gamma}(\vec q,s)\big\rangle + \widehat{\zeta}_{\alpha \beta}(\vec{q},s) \big\langle \widehat{\gamma}_{\kappa\kappa}(\vec q,s)\big\rangle \quad;
\end{split}
\end{equation}
\begin{equation}
\begin{split}
\label{eq:eta_def_j_rho_index}
\widehat{\Pi}_{\alpha \beta \zeta \gamma}(\vec{q},s)  \equiv \widehat{Y}_{\alpha \xi}\;
\Biggl[
\Delta \widehat N^{\sigma\sigma}_{\xi\beta \zeta \gamma}(\vec q,s)
+\frac{1}{s\rho_0}\,
\widehat X^{-1}\,
\Delta \widehat N^{\sigma j}_{\xi\beta,\mu}(\vec q,s)q_\mu
q_\nu\Delta \widehat N^{j\sigma}_{\nu, \zeta \gamma }(\vec q,s)
\Biggr]\quad ;
\end{split}
\end{equation}
\begin{equation}
\begin{split}
\label{eq:zeta_def_j_rho_index}
\widehat{\zeta}_{\alpha \beta}(\vec{q},s)\equiv -\frac{i}{s}\,
\widehat{Y}_{\alpha\xi}\;
\Bigl(
\Delta \widehat N^{\sigma j}_{\xi\beta,\nu}(\vec q,s)\,q_\nu
\Bigr)\;
\widehat{X}^{-1}\quad.
\end{split}
\end{equation}

The second term in Eq.~\eqref{eq:CE_sigma_j_rho_compact},
$\mat{\widehat{\zeta}}(\vec q,s)\,\mathrm{tr}\,\widehat{\mat{\gamma}}(\vec q,s)$,
arises from the coupling between stress and momentum density. 
In the absence of this coupling, $\Delta \Tthr{\widehat{N}}^{\sigma j}=\Delta \Tthr{\widehat{N}}^{j\sigma}=0$,
the off-diagonal blocks vanish so that $\widehat{\mat Y}=\widehat{\mat U}^{-1}$, see Eq.~\eqref{eq:matrix1_j_rho}-\eqref{eq:matrix3_j_rho}, and
Eq.~\eqref{eq:eta_def_j_rho}-\eqref{eq:zeta_def_j_rho} reduces to the expression for the viscous kernel, Eq.~\eqref{eq:eta}, given in the main text.

As detailed in the section~\ref{App:compress}, we show that the trace of the bulk coupling kernel, $\widehat{\zeta}_{\kappa\kappa}(\vec{q},s)$, captures the static, pressure-like contribution to the fluid response. From this, the $\vec{q}$-dependent static compressibility $c_0(\vec{q})$ is obtained and can be written in proportional to the static density structure factor,
\begin{equation}
\label{eq:c0_main}
c_{0}(\vec q) = A(\vec q)\,g^{\rho\rho}(\vec q)\quad.
\end{equation}
In thermal equilibrium, the proportionality coefficient $A(\vec{q})$ is fixed by fluctuation-dissipation theorem and becomes a thermodynamic constant independent of the wave vector, given by
\begin{equation}
\label{eq:A_eq_main}
A_{\mathrm{eq}}= \frac{1}{\rho_0^2 \, V \, k_B T} \quad.
\end{equation}

\subsection{Derivation of the constitutive equations \label{App:CE_j_rho}}
In this section, we detail the derivation of the constitutive  equations.
We first compute the matrix \(\mat g\)\,. We consider a steady state in which there is no static cross-correlation between momentum density and mass density,
\begin{equation}
\label{eq:no_static_cross_j_rho}
\expval{\vec j(\vec q)\,\rho^{\dagger}(\vec q)}
=
\expval{\rho(\vec q)\,\vec j^{\dagger}(\vec q)}
=\vec 0\quad,
\end{equation}
therefore,
\begin{equation}
\begin{split}
\label{eq:g_j_rho_identity}
\mat g(\vec q)
=
\begin{pmatrix}
\mat{g}^{jj}  & \vec 0\\
\vec 0 & g^{\rho \rho}
\end{pmatrix}\quad,
\end{split}
\end{equation}
where the components are given by
\begin{equation}
    \mat{g}^{jj} \equiv \expval{\vec{j}(\vec{q})\, \vec{j}^\dagger(\vec{q})} \quad ; \quad
    g^{\rho \rho} \equiv \expval{\rho(\vec q)\,\rho^{\dagger}(\vec q)} \quad.
\end{equation}
The inverse of $\mat{g}$ becomes diagonal,
\begin{equation}
\begin{split}
\label{eq:}
\mat g^{-1}(\vec q)
=
\begin{pmatrix}
 \qty(\mat{g}^{jj})^{-1} & \vec 0\\
\vec 0 & \qty(g^{\rho \rho})^{-1}
\end{pmatrix}\quad.
\end{split}
\end{equation}

Next, we evaluate the time-dependent correlation matrix \(\mat\phi(\vec q,t)\), which takes the form
\begin{equation}
\begin{split}
\label{eq:phi_j_rho_def}
\mat\phi(\vec q,t)
&=
\begin{pmatrix}
\expval{\dot{\vec j}(\vec q,t)\,\dot{\vec j}^{\dagger}(\vec q)}
&
\expval{\dot{\vec j}(\vec q,t)\,\dot\rho^{\dagger}(\vec q)}
\\\\
\expval{\dot\rho(\vec q,t)\,\dot{\vec j}^{\dagger}(\vec q)}
&
\expval{\dot\rho(\vec q,t)\,\dot\rho^{\dagger}(\vec q)}
\end{pmatrix}\quad.
\end{split}
\end{equation}

We define the normalized correlation tensors,
\begin{equation}
\begin{split}
\label{eq:N_defs_j_rho}
&\Tfor N^{\sigma\sigma}(\vec q,t)
\equiv
\rho_0\expval{\mat{\sigma}(\vec q,t)\,\mat{\sigma}^\dagger(\vec q) } \cdot \qty(\mat{g}^{jj})^{-1}\quad, \quad
\Tthr N^{\sigma j}(\vec q,t)
\equiv
\rho_0\,\expval{\mat{\sigma}(\vec q,t)\, \vec{j}^\dagger(\vec{q})}\qty(g^{\rho \rho})^{-1}\quad,\\[2mm]
&\Tthr N^{j\sigma}(\vec q,t)
\equiv
\rho_0\, \expval{\vec{j}(\vec{q},t)\,\mat{\sigma}^\dagger(\vec{q})} \cdot\qty(\mat{g}^{jj})^{-1} \quad, \quad
\mat N^{jj}(\vec q,t)
\equiv
\rho_0\, \expval{\vec{j}(\vec{q},t)\,\vec{j}^\dagger(\vec{q})} \qty(g^{\rho \rho})^{-1}\quad.
\end{split}
\end{equation}
They are written in the index notation: 
\begin{align}
\label{eq:N_index_defs_j_rho}
&N^{\sigma\sigma}_{\alpha \beta \gamma \delta}(\vec q,t)
\equiv
\rho_0\expval{{\sigma}_{\alpha \beta}(\vec q,t)\,{\sigma}_{\epsilon \delta}^*(\vec q) } \qty(\mat{g}^{jj})_{\epsilon \gamma}^{-1}\quad, \quad
 N^{\sigma j}_{\alpha \beta,\gamma}(\vec q,t)
\equiv
\rho_0\,\expval{{\sigma}_{\alpha \beta}(\vec q,t)\, {j}_{\gamma}^*(\vec{q})}\qty(g^{\rho \rho})^{-1}\quad,\\[2mm]
& N^{j\sigma}_{\alpha,\beta \gamma}(\vec q,t)
\equiv
\rho_0\, \expval{{j}_{\alpha}(\vec{q},t)\,{\sigma}_{\epsilon \gamma}^*(\vec{q})\,} \qty(\mat{g}^{jj})_{\epsilon \beta}^{-1} \quad, \quad
 N^{jj}_{\alpha \beta}(\vec q,t)
\equiv
\rho_0\, \expval{{j}_\alpha(\vec{q},t)\,{j}^*_\beta(\vec{q})} \qty(g^{\rho \rho})^{-1}\quad.
\end{align}

Using these normalized correlation tensors, the matrix product \(\mat\phi(\vec q,t)\cdot \mat{g}^{-1}\) becomes
\begin{equation}
\begin{split}
\label{eq:phi_in_terms_of_N_j_rho}
\mat\phi(\vec q,t)\cdot \mat{g}^{-1}
=
\begin{pmatrix}
\vec q\cdot \Tfor N^{\sigma\sigma}(\vec q,t)\cdot \vec q/\rho_0
&
-\vec q\cdot \Tthr N^{\sigma j}(\vec q,t)\cdot \vec q/\rho_0
\\\\
-\vec q^{}\cdot \Tthr N^{j\sigma}(\vec q,t)\cdot \vec q/\rho_0
&
\vec q^{}\cdot \mat N^{jj}(\vec q,t)\cdot \vec q/\rho_0
\end{pmatrix}\quad.
\end{split}
\end{equation}
The contractions are explicitly given as 
\begin{align}
\label{eq:q_contractions_j_rho}
&\bigl(\vec q\cdot \Tfor N^{\sigma\sigma}(\vec q,t)\cdot \vec q\bigr)_{\alpha\beta}=
q_\mu\,N^{\sigma\sigma}_{\alpha\mu \beta \delta}(\vec q,t)\,q_\delta
\quad,\quad
\bigl(\vec q\cdot \Tthr N^{\sigma j}(\vec q,t)\cdot \vec q\bigr)_{\alpha}
=
q_\mu\,N^{\sigma j}_{\alpha\mu,\nu}(\vec q,t)\,q_\nu
\quad,
\\[2mm]
&\bigl(\vec q^{}\cdot \Tthr N^{j\sigma}(\vec q,t)\cdot \vec q\bigr)_{\beta}
=
q_\nu\,N^{j\sigma}_{\nu,\beta \delta}(\vec q,t)\,q_\delta
\quad, \quad
\vec q^{}\cdot \mat N^{jj}(\vec q,t)\cdot \vec q
=
q_\alpha\,N^{jj}_{\alpha\beta}(\vec q,t)\,q_\beta
\quad.
\end{align}
In Laplace space, the transient part \(\Delta\mat{\widehat\phi}(\vec q,s) \cdot \mat{g}^{-1}\) is defined analogously to Eq.~\eqref{eq:delphi}:
\begin{equation}
\label{eq:Delphihat_def_j_rho}
\Delta \mat{\widehat\phi}(\vec q,s)\cdot \mat{g}^{-1}
=
\begin{pmatrix}
\vec q\cdot \Delta\Tfor{\widehat N}^{\sigma\sigma}(\vec q,s)\cdot \vec q/\rho_0
&
-\vec q\cdot \Delta\Tthr{\widehat N}^{\sigma j}(\vec q,s)\cdot \vec q/\rho_0
\\\\
-\vec q^{}\cdot \Delta\Tthr{\widehat N}^{j\sigma}(\vec q,s)\cdot \vec q/\rho_0
&
\vec q^{}\cdot \Delta\mat{\widehat N}^{jj}(\vec q,s)\cdot \vec q/\rho_0
\end{pmatrix}\quad,
\end{equation}
where, for example,
\(\Delta\Tfor{\widehat N}^{\sigma\sigma}(\vec q,s)\equiv \Tfor{\widehat N}^{\sigma\sigma}(\vec q,s)-\Tfor{\widehat N}^{\sigma\sigma}(\vec q,0)\),
and similarly for the other blocks.

We compute the contraction \( \qty(\Delta\mat{\widehat\phi}(\vec q,s)\cdot \mat{g}^{-1}) \cdot  \vec{\widehat A}(\vec q,s)\), which appears inside
\(\mat{\widehat K}(\vec q,s)\cdot \vec{\widehat A}(\vec q,s)\):
\begin{equation}
\begin{split}
\label{eq:Delphi_Ahat_j_rho_step1}
\qty(\Delta\mat{\widehat\phi}(\vec q,s)\cdot \mat{g}^{-1} ) \cdot \vec{\widehat A}(\vec q,s)
&=
\begin{pmatrix}
\vec q\cdot \Delta\Tfor{\widehat N}^{\sigma\sigma}\cdot \vec q/\rho_0
&
-\vec q\cdot \Delta\Tthr{\widehat N}^{\sigma j}\cdot \vec q/\rho_0
\\\\
-\vec q^{}\cdot \Delta\Tthr{\widehat N}^{j\sigma}\cdot \vec q/\rho_0
&
\vec q^{}\cdot \Delta\mat{\widehat N}^{jj}\cdot \vec q/\rho_0
\end{pmatrix}
\begin{pmatrix}
\vec{\widehat j}\\\\
\widehat\rho
\end{pmatrix}
\\\\
&=
\begin{pmatrix}
\qty(\vec q\cdot \Delta\Tfor{\widehat N}^{\sigma\sigma}\cdot \vec q)\cdot \vec{\widehat j}/\rho_0
-\qty(\vec q\cdot \Delta\Tthr{\widehat N}^{\sigma j}\cdot \vec q)\,\widehat\rho/\rho_0
\\\\
-\qty(\vec q^{}\cdot \Delta\Tthr{\widehat N}^{j\sigma}\cdot \vec q)\cdot  \vec{\widehat j}/\rho_0
+\qty(\vec q^{}\cdot \Delta\mat{\widehat N}^{jj}\cdot \vec q)\,\widehat\rho/\rho_0
\end{pmatrix}\quad.
\end{split}
\end{equation}
From the continuity relation
$\widehat\rho(\vec q,s)=i\,q_\alpha \widehat j_\alpha(\vec q,s)/s$
and strain-rate definition
$\widehat\gamma_{\alpha\beta}(\vec q,s)=-i\,q_\beta\,\widehat j_\alpha(\vec q,s)/\rho_0$, we have $\widehat{\rho}(\vec q,s) = -\rho_0\widehat{\gamma}_{\alpha \alpha}(\vec q,s)/s$.
Thus we obtain in index notation
\begin{equation}
\begin{split}
\label{eq:Delphi_Ahat_j_rho_index}
\bigl[\qty(\Delta\mat{\widehat\phi}(\vec q,s)\cdot \mat{g}^{-1}) \cdot \vec{\widehat A}(\vec q,s)\bigr]^{(1)}_{\alpha}
&=
i\,q_\mu\,
\Delta \widehat N^{\sigma\sigma}_{\alpha\mu \lambda \nu}(\vec q,s)\,
\widehat\gamma_{\lambda\nu}(\vec q,s)
\;+\;
\frac{1}{s}\,
q_\mu\,\Delta \widehat N^{\sigma j}_{\alpha\mu,\nu}(\vec q,s)\,q_\nu\;
\widehat\gamma_{\kappa\kappa}(\vec q,s)\quad,
\\[1mm]
\bigl[\qty(\Delta\mat{\widehat\phi}(\vec q,s)\cdot \mat{g}^{-1}) \cdot \vec{\widehat A}(\vec q,s)\bigr]^{(2)}
&=
-\,i\,q_\nu\,
\Delta \widehat N^{j\sigma}_{\nu,\lambda \mu}(\vec q,s)\,
\widehat\gamma_{\lambda\mu}(\vec q,s)
\;-\;
\frac{1}{s}\,
q_\mu\,\Delta \widehat N^{jj}_{\mu\nu}(\vec q,s)\,q_\nu\;
\widehat\gamma_{\kappa\kappa}(\vec q,s)\quad,
\end{split}
\end{equation}
Equivalently, in index-free notation we write
\begin{equation}
\begin{split}
\label{eq:Delphi_Ahat_j_rho_indexfree}
\qty(\Delta\mat{\widehat\phi}(\vec q,s)\cdot \mat{g}^{-1})\cdot  \vec{\widehat A}(\vec q,s)
=
\begin{pmatrix}
\qty(i\,\vec q\cdot \Delta\Tfor{\widehat N}^{\sigma\sigma}):\widehat{\mat\gamma}
\;+\;\dfrac{1}{s}\,\qty(\vec q\cdot \Delta\Tthr{\widehat N}^{\sigma j}\cdot \vec q)\,\mathrm{tr}\,\widehat{\mat\gamma}
\\\\
-\qty(\,i\,\vec q\cdot \Delta\Tthr{\widehat N}^{j\sigma}):\widehat{\mat\gamma}
\;-\;\dfrac{1}{s}\,\qty(\vec q\cdot \Delta\mat{\widehat N}^{jj}\cdot \vec q)\,\mathrm{tr}\,\widehat{\mat\gamma}
\end{pmatrix}\quad.
\end{split}
\end{equation}

We now make use of the matrix inversion identity for block matrices using Eq.~\eqref{eq:inverse_block_matrix}.
We first compute \(\qty[\mat I-\frac{1}{s} \qty(\Delta\mat{\widehat\phi}(\vec q,s)\cdot \mat{g}^{-1})]^{-1}\) appearing in the memory kernel
\(\mat{\widehat K}(\vec q,s)=\qty[\mat I-\frac{1}{s}\qty(\Delta\mat{\widehat\phi}(\vec q,s)\cdot \mat{g}^{-1})]^{-1}\cdot \qty(\Delta\mat{\widehat\phi}(\vec q,s)\cdot \mat{g}^{-1})\)\,:
\begin{equation}
\begin{split}
\label{eq:ginv_structure_j_rho}
\qty[\mat I-\frac{1}{s} \qty(\Delta\mat{\widehat\phi}(\vec q,s)\cdot \mat{g}^{-1})]^{-1}
&=
\begin{pmatrix}
\mat I-\dfrac{1}{s\rho_0}\,\vec q\cdot \Delta\Tfor{\widehat N}^{\sigma\sigma}\cdot \vec q
&
\dfrac{1}{s\rho_0}\,\vec q\cdot \Delta\Tthr{\widehat N}^{\sigma j}\cdot \vec q
\\\\
\dfrac{1}{s\rho_0}\,\vec q^{}\cdot \Delta\Tthr{\widehat N}^{j\sigma}\cdot \vec q
&
1-\dfrac{1}{s\rho_0}\,\vec q^{}\cdot \Delta\mat{\widehat N}^{jj}\cdot \vec q
\end{pmatrix}^{-1}
\\[2mm]
&\equiv
\begin{pmatrix}
\mat{\widehat U} & \vec{\widehat V}\\[1mm]
\vec{\widehat W} & \widehat X
\end{pmatrix}^{-1}
=
\begin{pmatrix}
\mat{\widehat Y} & \vec 0\\[1mm]
\vec 0 & \widehat Z
\end{pmatrix}
\begin{pmatrix}
\mat I & -\vec{\widehat V}\,\widehat X^{-1}\\[1mm]
-\vec{\widehat W}\cdot \mat{\widehat U}^{-1} & 1
\end{pmatrix}\quad,
\end{split}
\end{equation}
where the density block is scalar, hence $\widehat X$, $\widehat Z$ are scalars and the bottom-right identity is $1$.
We defined the relevant blocks as follows:
\begin{equation}
\begin{split}
\label{eq:matrix1_j_rho}
\mat{\widehat U}
&\equiv
\mat I-\frac{1}{s\rho_0}\,\vec q\cdot \Delta\Tfor{\widehat N}^{\sigma\sigma}\cdot \vec q\quad; \qquad
\vec{\widehat V}
\equiv
\frac{1}{s\rho_0}\,\vec q\cdot \Delta\Tthr{\widehat N}^{\sigma j}\cdot \vec q\quad,
\end{split}
\end{equation}
\begin{equation}
\begin{split}
\label{eq:matrix2_j_rho}
\vec{\widehat W}
&\equiv
\frac{1}{s\rho_0}\,\vec q^{}\cdot \Delta\Tthr{\widehat N}^{j\sigma}\cdot \vec q\quad; \qquad
\widehat X
\equiv
1-\frac{1}{s\rho_0}\,\vec q^{}\cdot \Delta\mat{\widehat N}^{jj}\cdot \vec q\quad,
\end{split}
\end{equation}
and
\begin{equation}
\begin{split}
\label{eq:matrix3_j_rho}
\mat{\widehat Y}
&\equiv
\qty(\mat{\widehat U}-\vec{\widehat V}\,\widehat X^{-1}\, \vec{\widehat W})^{-1}\quad; \qquad
\widehat Z
\equiv
\qty(\widehat X- \vec{\widehat W}\cdot\mat{\widehat U}^{-1}\cdot \vec{\widehat V})^{-1}\quad.
\end{split}
\end{equation}

Combining these, 
\begin{equation}
\begin{split}
\label{eq:K_Ahat_j_rho_blockfactored}
\mat{\widehat K}(\vec q,s)\cdot \vec{\widehat A}(\vec q,s)
&=
\qty[\mat I-\frac{1}{s} \qty(\Delta\mat{\widehat\phi}(\vec q,s)\cdot \mat{g}^{-1})]^{-1}\cdot
\qty(\Delta\mat{\widehat\phi}(\vec q,s)\cdot \mat{g}^{-1}) \cdot \vec{\widehat A}(\vec q,s)
\\[1mm]
&=
\begin{pmatrix}
\mat{\widehat Y} & \vec 0\\[1mm]
\vec 0 & \widehat Z
\end{pmatrix}
\begin{pmatrix}
\mat I & -\vec{\widehat V}\,\widehat X^{-1}\\[1mm]
-\vec{\widehat W}\cdot\mat{\widehat U}^{-1} & 1
\end{pmatrix}
\begin{pmatrix}
\qty(i\,\vec q\cdot \Delta\Tfor{\widehat N}^{\sigma\sigma}):\widehat{\mat\gamma}
\;+\;\dfrac{1}{s}\,\qty(\vec q\cdot \Delta\Tthr{\widehat N}^{\sigma j}\cdot \vec q)\,\mathrm{tr}\,\widehat{\mat\gamma}
\\\\
-\,\qty(i\,\vec q\cdot \Delta\Tthr{\widehat N}^{j\sigma}):\widehat{\mat\gamma}
\;-\;\dfrac{1}{s}\,\qty(\vec q\cdot \Delta\mat{\widehat N}^{jj}\cdot \vec q)\,\mathrm{tr}\,\widehat{\mat\gamma}
\end{pmatrix}\quad.
\end{split}
\end{equation}

This gives the first component, 
\begin{equation}
\begin{split}
 \qty[\mat{\widehat K}(\vec q,s)\cdot \vec{\widehat A}(\vec q,s)   ]^{(1)} =\,& \mat{\widehat{Y}}\cdot i\,\vec q\cdot \Bigg[\Biggl(
\Delta\Tfor{\widehat N}^{\sigma\sigma}
+\frac{1}{s\rho_0}\,\widehat X^{-1}\,\Big[\qty(\Delta\Tthr{\widehat N}^{\sigma j}\cdot  \vec{q}) \qty(\vec{q} \cdot\Delta\Tthr{\widehat N}^{j\sigma})\Big]\Biggr)
:\widehat{\mat\gamma} \\ 
&- \frac{i}{s}\,\qty(\Delta\Tthr{\widehat N}^{\sigma j}\cdot \vec q) \widehat{X}^{-1}\,\mathrm{tr}\,\widehat{\mat\gamma} \, \Bigg]\quad,
\end{split}
\end{equation}
where we have simplified the term proportional to $\mathrm{tr}\,\widehat{\mat\gamma}$ using the form of $\widehat{X}$ [Eq.~\eqref{eq:matrix2_j_rho}]. The second component is given as 
\begin{equation}
\begin{split}
 \qty[\mat{\widehat K}(\vec q,s)\cdot \vec{\widehat A}(\vec q,s)   ]^{(2)} =&  -\widehat{Z} \, i\, \vec{q} \cdot \bigg[\bigg(\frac{1}{s\rho_0}  \qty(\Delta \Tthr{\widehat N}^{j \sigma}\cdot \vec{q}) \cdot  \mat{\widehat U}^{-1} \cdot \qty( \vec{q}\cdot\Delta \Tfor{\widehat N}^{\sigma \sigma}) + \Delta \Tthr{\widehat N}^{j \sigma}\bigg):\mat{\widehat \gamma}  \\[2mm]
&-i \,\frac{1}{s}\bigg\{\frac{1}{s\rho_0}\qty(\Delta \Tthr{\widehat N}^{j \sigma} \cdot q)\cdot \mat{\widehat U}^{-1}\cdot \qty(\vec{q} \cdot \Delta \Tthr{\widehat N}^{\sigma j} \cdot \vec{q}) +\qty( \Delta \mat{\widehat N}^{jj}\cdot \vec{q})\bigg\} \mathrm{tr}\,\widehat{\mat\gamma} \, \bigg]\quad.
\end{split}
\end{equation}

Combining this with 
\begin{equation}
    \frac{d}{dt}\expval{\vec{A}(\vec{q},t)} = -
    \begin{pmatrix}
 \expval{\mat{\sigma}(\vec{q},t)}\cdot i \,\vec{q}\\ \\
-i\vec{q}\cdot \expval{\vec{j}(\vec{q},t)}\\
\end{pmatrix}
\quad,
\end{equation}
we obtain the constitutive equations in Laplace space, 
\begin{equation}
\begin{split}
    \expval{\mat{\widehat{\sigma}}(\vec{q},s)} =\, &\mat{\widehat{Y}} \cdot \Biggl(
\Delta\Tfor{\widehat N}^{\sigma\sigma}
+\frac{1}{s\rho_0}\,\widehat X^{-1}\, \Big[\qty(\Delta\Tthr{\widehat N}^{\sigma j}\cdot \vec{q} )\qty(\vec{q}\cdot \Delta\Tthr{\widehat N}^{j\sigma}) \Big]\Biggr)
:\expval{\widehat{\mat\gamma}} \\[2mm]
&-\frac{i}{s}\,  \cdot \mat{\widehat{Y}} \cdot \,\qty( \Delta\Tthr{\widehat N}^{\sigma j}\cdot \vec q)\,\widehat{X}^{-1}\,
\expval{\mathrm{tr}\,\widehat{\mat\gamma}}\quad; 
\end{split}
\end{equation}
\begin{equation}
\begin{split}
    \expval{\vec{\widehat{j}}(\vec{q},s)} =\,&  \widehat{Z}\,\qty(\frac{1}{s\rho_0}  \qty(\Delta \Tthr{\widehat N}^{j \sigma}\cdot \vec{q}) \cdot  \mat{\widehat U}^{-1} \cdot \qty(\vec{q}\cdot \Delta \Tfor{\widehat{N}}^{\sigma \sigma}) + \Delta \Tthr{\widehat N}^{j \sigma}):\expval{\mat{\widehat \gamma}}\\[2mm]
&-\frac{i}{s}\, \widehat{Z}\,\bigg[\frac{1}{s\rho_0}\qty( \Delta \Tthr{\widehat N}^{j \sigma} \cdot q)\cdot \mat{\widehat U}^{-1}\cdot \qty(\vec{q} \cdot \Delta \Tthr{\widehat N}^{\sigma j} \cdot \vec{q}) +\qty( \Delta \mat{\widehat N}^{jj}\cdot \vec{q})\bigg]\, \expval{\mathrm{tr}\,\widehat{\mat\gamma}} \quad.
\end{split}
\end{equation}

The index forms are written as 
\begin{equation}
\begin{split}
\label{eq:CE_sigma_j_rho_index}
\big\langle \widehat{\sigma}_{\alpha\beta}(\vec q,s)\big\rangle
=\;&
\widehat{Y}_{\alpha \xi}\;
\Biggl[
\Delta \widehat N^{\sigma\sigma}_{\xi\beta \gamma \zeta}(\vec q,s)
+\frac{1}{s\rho_0}\,
\widehat X^{-1}\,
\Delta \widehat N^{\sigma j}_{\xi\beta,\mu}(\vec q,s)q_\mu
q_\nu\Delta \widehat N^{j\sigma}_{\nu,\gamma \zeta }(\vec q,s)
\Biggr]\;
\big\langle \widehat{\gamma}_{\gamma\zeta}(\vec q,s)\big\rangle
\\[1mm]
&\;
-\frac{i}{s}\,
\widehat{Y}_{\alpha\xi}\;
\Bigl(
\Delta \widehat N^{\sigma j}_{\xi\beta,\nu}(\vec q,s)\,q_\nu
\Bigr)\;
\widehat{X}^{-1}\, 
\big\langle \widehat{\gamma}_{\kappa\kappa}(\vec q,s)\big\rangle
\quad.
\end{split}
\end{equation}

\begin{equation}
\begin{split}
\label{eq:CE_j_j_rho_index}
\big\langle \widehat{j}_{\alpha}(\vec q,s)\big\rangle
=\;&
\widehat Z\;
\Biggl[
\Delta \widehat N^{j\sigma}_{\alpha,\zeta \eta}(\vec q,s)
+\frac{1}{s\rho_0}\,
\Delta \widehat N^{j\sigma}_{\alpha,\mu \beta}(\vec q,s)\,q_\beta\,
(\widehat U^{-1})_{\mu\gamma}q_\delta\;
\Delta \widehat N^{\sigma\sigma}_{\gamma\delta\zeta \eta}(\vec q,s)
\Biggr]\;
\big\langle \widehat{\gamma}_{\zeta \eta}(\vec q,s)\big\rangle
\\[1mm]
&\;
-\frac{i}{s}\;
\widehat Z\;
\Biggl[
\frac{1}{s\rho_0}\,
\Bigl(\Delta \widehat N^{j\sigma}_{\alpha,\mu \beta}(\vec q,s)\,q_\beta\Bigr)\;
(\widehat U^{-1})_{\mu\gamma}\;
\Bigl(q_\delta\,\Delta \widehat N^{\sigma j}_{\gamma\delta,\nu}(\vec q,s)\;q_\nu\Bigr)
\;+\;
\Delta \widehat N^{jj}_{\alpha \nu}(\vec q,s)\,q_\nu
\Biggr]\;
\big\langle \widehat{\gamma}_{\kappa\kappa}(\vec q,s)\big\rangle
\quad.
\end{split}
\end{equation}

\subsection{Identification of pressure-like term and connection to the compressibility \label{App:compress}}
We show in this section that the kernel $\mat{\widehat{\zeta}}(\vec{q},s)$ is related to the static compressibility. We define a compressibility as the linear-response kernel of the density to an isotropic pressure-like scalar $\widehat p(\vec q,s)$,
\begin{equation}
\label{eq:def_compressibility_qs}
\widehat c(\vec q,s)
\;\equiv\;
\frac{1}{\rho_0}\,
\frac{\partial \langle \widehat\rho(\vec q,s)\rangle}{\partial \langle \widehat p(\vec q,s)\rangle}\quad.
\end{equation}
Using Eq.\eqref{eq:rho_trace_gamma_relation_concise}, we obtain
\begin{equation}
\label{eq:rho_trace_gamma_relation_used}
\langle \widehat\rho(\vec q,s)\rangle
= -\,\frac{\rho_0}{s}\,\langle \widehat\gamma_{\kappa\kappa}(\vec q,s)\rangle
\quad,
\end{equation}
so that
\begin{equation}
\label{eq:def_compressibility_via_tracegamma}
\widehat c(\vec q,s)
=
-\,\frac{1}{s}\,
\frac{\partial \langle \widehat\gamma_{\kappa\kappa}(\vec q,s)\rangle}
{\partial \langle \widehat p(\vec q,s)\rangle}\quad.
\end{equation}

Next we decompose the strain-rate tensor into deviatoric and trace parts,
\begin{equation}
\label{eq:dev_trace_decomposition}
\widehat\gamma_{\alpha\beta}
=
\widehat\gamma^{\rm dev}_{\alpha\beta}
+\frac{1}{d}\,\delta_{\alpha\beta}\,\widehat\gamma_{\kappa\kappa}\quad,
\qquad
\widehat\gamma^{\rm dev}_{\kappa\kappa}=0\quad.
\end{equation}
With this, the stress constitutive relation can be written as
\begin{equation}
\label{eq:stress_dev_trace_form}
\langle \widehat\sigma_{\alpha\beta}(\vec q,s)\rangle
=
\widehat\Pi_{\alpha\beta\gamma\zeta}(\vec q,s)\,
\langle \widehat\gamma^{\rm dev}_{\gamma\zeta}(\vec q,s)\rangle
+
\widehat B_{\alpha\beta}(\vec q,s)\,
\langle \widehat\gamma_{\kappa\kappa}(\vec q,s)\rangle \quad,
\end{equation}
where we defined the bulk coupling tensor
\begin{equation}
\label{eq:def_bulk_tensor_B}
\widehat B_{\alpha\beta}(\vec q,s)
\;\equiv\;
\frac{1}{d}\,\widehat\Pi_{\alpha\beta\kappa\kappa}(\vec q,s)
+\widehat\zeta_{\alpha\beta}(\vec q,s)\quad.
\end{equation}
The isotropic pressure-like scalar is identified with the trace of the stress,
\begin{equation}
\label{eq:def_pressure_trace}
\langle \widehat p(\vec q,s)\rangle
\equiv
-\frac{1}{d}\,\langle \widehat\sigma_{\kappa\kappa}(\vec q,s)\rangle \quad.
\end{equation}
For a purely dilatational deformation, $\langle \widehat\gamma^{\rm dev}_{\alpha\beta}(\vec q,s)\rangle=0$, the constitutive law reduces to
\begin{equation}
\label{eq:def_pressure_dilation}
\langle \widehat p(\vec q,s)\rangle
=
-\frac{1}{d}\,\widehat B_{\kappa\kappa}(\vec q,s)\,
\langle \widehat\gamma_{\lambda\lambda}(\vec q,s)\rangle \quad.
\end{equation}
Combining Eqs.~\eqref{eq:def_compressibility_via_tracegamma} and \eqref{eq:def_pressure_dilation}, we obtain the generalized isotropic compressibility
\begin{equation}
\label{eq:compressibility_final}
\widehat c(\vec q,s)
=
\frac{1}{s}\,
\frac{d}{\widehat B_{\kappa\kappa}(\vec q,s)}\quad.
\end{equation}

The static compressibility may be obtained as
\begin{equation}
\label{eq:static_compressibility}
c_{0}(\vec q) \equiv \lim_{s\rightarrow 0} \widehat c(\vec q,s)  =  \lim_{s\rightarrow 0} \frac{1}{s}\frac{d}{\widehat B_{\kappa\kappa}(\vec q,s)} \quad .
\end{equation}
To compute this explicitly, we evaluate the small-$s$ scaling of the bulk kernel
$\widehat{B}_{\kappa\kappa}(\vec q,s)$.
This requires a careful analysis of the matrix blocks in Eqs.~\eqref{eq:matrix1_j_rho}-\eqref{eq:matrix3_j_rho}.
From Eq.~\eqref{eq: SecondD_in_Laplace}, we have
\begin{equation}
    \mat{g} - \frac{1}{s}\Delta \mat{\widehat{\phi}}(s) = s \,\mat{\widehat{\psi}}(s) \quad .
\end{equation}
Since $\lim_{s \to 0} \int_0^\infty dt e^{-st} \mat{\psi}(t)=\mat{\widehat{\psi}}(0)+ \mathcal{O}(s)$, we find that as $s \to 0$,
\begin{equation}
    s\, \mat{\widehat{\psi}}(s) \sim \mathcal{O}(s)  \quad.
\end{equation}
Because the static correlation matrix $\mat{g}$ is block-diagonal, the identity $\Delta \mat {\widehat{\phi}}(s) = s\mat{g} - s^2\mat{\widehat{\psi}}(s)$ enforces distinct scalings on the transient correlation tensors.

For the $jj$ block,
\begin{equation}
    \Delta \mat{\widehat{\phi}}^{jj}(\vec{q},s) = s \mat{g}^{jj} - s^2 \mat{\widehat{\psi}}^{jj}(\vec{q},s) \sim \mathcal{O}(s) \quad,
\end{equation}
which leads to
\begin{equation}
    \Delta \Tfor{\widehat{N}}^{\sigma \sigma}(\vec{q},s)  \sim \mathcal{O}(s)  \quad.
\end{equation}
The matrix $\mat{\widehat{U}}$ is thus
\begin{equation}
\begin{split}
\label{eq:U_scaling}
\mat{\widehat U}
&\equiv
\mat I-\frac{1}{s\rho_0}\,\vec q\cdot \Delta\Tfor{\widehat N}^{\sigma\sigma}\cdot \vec q = \mat{I} - \frac{1}{s}\Delta \mat{\widehat{\phi}}^{jj}(\vec{q},s) \cdot \qty(\mat{g}^{jj})^{-1} = s \, \mat{\widehat{\psi}}^{jj}(\vec{q},s) \cdot \qty(\mat{g}^{jj})^{-1} \sim \mathcal{O}(s) \quad.
\end{split}
\end{equation}

For the $\rho\rho$ block,
\begin{equation}
    \Delta {\widehat{\phi}}^{\rho\rho}(\vec{q},s) = sg^{\rho \rho}  - s^2 {\widehat{\psi}}^{\rho \rho }(\vec{q},s) \sim \mathcal{O}(s) \quad,
\end{equation}
which leads to
\begin{equation}
    \Delta \mat{\widehat{N}}^{jj}(\vec{q},s)  \sim \mathcal{O}(s) \quad.
\end{equation}
Further, we obtain
\begin{equation}
  \widehat X
\equiv
1-\frac{1}{s\rho_0}\,\vec q\cdot \Delta\mat{\widehat N}^{jj}\cdot \vec q = 1-\frac{1}{s}\Delta\widehat{\phi}^{\rho \rho}(\vec{q},s)  \qty({g}^{\rho \rho})^{-1} = s \,\widehat{\psi}^{\rho \rho}(\vec{q},s)\qty({g}^{\rho \rho})^{-1}  \sim \mathcal{O}(s) \quad.
\end{equation}

For the cross-coupling $j\rho$ and $\rho j$ blocks, the static correlation is zero, yielding
\begin{equation}
    \Delta \vec{\widehat{\phi}}^{j\rho}(\vec{q},s) = - s^2 \vec{\widehat{\psi}}^{j \rho}(\vec{q},s) \sim \mathcal{O}(s^2) \quad,
\end{equation}
which leads to
\begin{equation}
    \Delta \Tthr{\widehat{N}}^{\sigma j}(\vec{q},s)  \sim \mathcal{O}(s^2) \quad.
\end{equation}
This scaling ensures that the vector $\vec{\widehat V}$ scales as
\begin{equation}
    \vec{\widehat V}
\equiv
\frac{1}{s\rho_0}\,\vec q\cdot \Delta\Tthr{\widehat N}^{\sigma j}\cdot \vec q = -\frac{1}{s}\Delta \vec{\widehat{\phi}}^{j \rho}(\vec{q},s)\qty({g}^{\rho \rho})^{-1} = s \vec{\widehat{\psi}}^{j \rho}(\vec{q},s)\qty({g}^{\rho \rho})^{-1} \sim \mathcal{O}(s)\quad.
\end{equation}
Similarly, $\Delta \Tthr{\widehat{N}}^{j \sigma }(\vec{q},s) \sim \mathcal{O}(s^2)$ and $\vec{\widehat{W}} \sim \mathcal{O}(s)$.

Using these results, we evaluate the scaling of inverse matrix $\mat{\widehat Y}$\,, which reads
\begin{equation}
\mat{\widehat Y}
\equiv
\qty(\mat{\widehat U}-\vec{\widehat V}\,\widehat X^{-1}\, \vec{\widehat W})^{-1}  \sim \mathcal{O}(s^{-1}) \quad.
\end{equation}

Substituting these scalings into Eqs.~\eqref{eq:eta_def_j_rho} and \eqref{eq:zeta_def_j_rho}, we obtain
\begin{equation}
\Tfor{\widehat{\Pi}}(\vec q,s)\sim\mathcal{O}(1)\quad; \quad \mat{\widehat{\zeta}}(\vec{q},s) \sim\mathcal{O}(s^{-1}) \quad.
\end{equation}
Since $\widehat{B}_{\kappa\kappa}(\vec q,s)=\widehat{\Pi}_{\kappa\kappa\lambda\lambda}(\vec q,s)/d+\widehat{\zeta}_{\kappa\kappa}(\vec q,s)$,
the viscous contribution $\widehat{\Pi}_{\kappa\kappa\lambda\lambda}/d$ vanishes when multiplied by $s$. The leading elastic contribution for $s\to 0$ ($\vec{q} \ne 0$) is thus governed entirely by the $1/s$ pole in $\widehat{\zeta}_{\kappa\kappa}(\vec q,s)$, which yields a finite static compressibility,
\begin{equation}
\label{eq:c0_from_zeta}
c_{0}(\vec q)= \lim_{s\to 0}\frac{d}{s\,\widehat{\zeta}_{\kappa\kappa}(\vec q,s)}
\equiv A(\vec q)\,g^{\rho\rho}(\vec q)\quad,
\end{equation}
where $A(\vec q)$ is a finite prefactor that depends on the details of the correlation kernels.
In equilibrium, the density structure factor is related to the isothermal compressibility ($\kappa_T$) via
\begin{equation}
\label{eq:rho_rho_eq}
\lim_{\vec{q}\to 0}g^{\rho\rho}_{\mathrm{eq}}(\vec q)=\lim_{\vec{q}\to 0}\langle \rho(\vec q)\rho^\dagger(\vec q)\rangle_{\mathrm{eq}}
=\rho_0^{2}\,V\,k_B T\,\kappa_T \quad.
\end{equation}
By the static fluctuation-dissipation theorem, this macroscopic relation generalizes to finite wavevectors such that $g^{\rho\rho}_{\mathrm{eq}}(\vec q) = \rho_0^2 \, V \, k_B T \, c_0(\vec q)$. 
Comparing this with Eq.~\eqref{eq:c0_from_zeta}, we can identify the prefactor in thermal equilibrium as a constant independent of $\vec q$\,:
\begin{equation}
\label{eq:A_prefactor_eq}
A_{\mathrm{eq}} = \frac{1}{\rho_0^2 \, V \, k_B T} \quad.
\end{equation}
Out of equilibrium, $A(\vec q)$ may acquire a $\vec q$-dependence and deviate from this thermal value, reflecting the presence of active, non-thermal density fluctuations.

\twocolumngrid
\bibliography{ref.bib}

\end{document}